\documentclass[]{aastex63}
\usepackage{tabularx}
\usepackage{amsmath}
\usepackage{soul}
\renewcommand\hl[1]{#1}

\begin{document}

% Title
\title{Origin of Life Molecules in the Atmosphere After Big Impacts on the Early Earth}
\author{Nicholas F. Wogan}
\affiliation{Department of Earth and Space Sciences, University of Washington, Seattle, WA 98195}
\affiliation{Virtual Planetary Laboratory, University of Washington, Seattle, WA98195}

\author{David C. Catling}
\affiliation{Department of Earth and Space Sciences, University of Washington, Seattle, WA 98195}
\affiliation{Virtual Planetary Laboratory, University of Washington, Seattle, WA98195}

\author{Kevin J. Zahnle}
\affiliation{Space Science Division, NASA Ames Research Center, Moffett Field, CA 94035}
\affiliation{Virtual Planetary Laboratory, University of Washington, Seattle, WA98195}

\author{Roxana Lupu}
\affiliation{Eureka Scientific, Inc., Oakland, CA 94602}

\begin{abstract}

The origin of life on Earth would benefit from a prebiotic atmosphere that produced nitriles, like HCN, which enable ribonucleotide synthesis. However, geochemical evidence suggests that Hadean air was relatively oxidizing with negligible photochemical production of prebiotic molecules. These paradoxes are resolved by iron-rich asteroid impacts that transiently reduced the entire atmosphere, allowing nitriles to form in subsequent photochemistry. Here, we investigate impact-generated reducing atmospheres using new time-dependent, coupled atmospheric chemistry and climate models, which account for gas-phase reactions and surface-catalysis. The resulting H$_2$-, CH$_4$- and NH$_3$-rich atmospheres persist for millions of years, until hydrogen escapes to space. HCN and HCCCN production and rainout to the surface can reach $10^9$ molecules cm$^{-2}$ s$^{-1}$ in hazy atmospheres with a mole ratio of $\mathrm{CH_4} / \mathrm{CO_2} > 0.1$. Smaller $\mathrm{CH_4} / \mathrm{CO_2}$ ratios produce HCN rainout rates $< 10^5$ molecules cm$^{-2}$ s$^{-1}$, and negligible HCCCN. The minimum impactor mass that creates atmospheric $\mathrm{CH_4} / \mathrm{CO_2} > 0.1$ is $4 \times 10^{20}$ to $5 \times 10^{21}$ kg (570 to 1330 km diameter), depending on how efficiently iron reacts with a steam atmosphere, the extent of atmospheric equilibration with an impact-induced melt pond, and the surface area of nickel that catalyzes CH$_4$ production. Alternatively, if steam permeates and deeply oxidizes crust, impactors $\sim 10^{20}$ kg could be effective. Atmospheres with copious nitriles have $> 360$ K surface temperatures, perhaps posing a challenge for RNA longevity, although cloud albedo can produce cooler climates. Regardless, post-impact cyanide can be stockpiled and used in prebiotic schemes after hydrogen has escaped to space.

\end{abstract}

\section{Introduction}

Two essential aspects of life are a genome and catalytic reactions, so the presence of ribonucleotide molecular ``fossils'' in modern biochemistry \citep{White_1976,Goldman_2021} and the ability of RNAs to store genetic information and catalyze reactions have led to the hypothesis that RNA-based organisms originated early \hl{\mbox{\citep{Cech_2012,Gilbert_1986}}}. This hypothesis proposes a stage of primitive life with RNA as a self-replicating genetic molecule that evolved by natural selection, which, at some point, became encapsulated in a cellular membrane and may have interacted with peptides from the beginning in the modified hypothesis of the RNA-Peptide World \citep[e.g.][]{Di_1997,Muller_2022}. In any case, RNA must be produced abiotically on early Earth for such scenarios. Chemists have proposed several prebiotic schemes that require nitriles - hydrogen cyanide (HCN), cyanoacetylene (HCCCN), cyanamide (H$_2$NCN), and cyanogen (NCCN) - to synthesize ribonucleobases, which are building blocks of RNA \citep{Benner_2020,Sutherland_2016,Yadav_2020}.

Abiotic synthesis of nitriles in nature is known to occur efficiently from photochemistry in reducing N$_2$-CH$_4$ atmospheres \citep{Zahnle_1986,Tian_2011}. Indeed, Titan's atmosphere, composed of mostly N$_2$ and CH$_4$, makes HCN, HCCCN and NCCN \citep{Strobel_2009}.

Geochemical evidence does not favor a volcanic source for a CH$_4$-rich prebiotic atmosphere. Redox proxies in old rocks indicate that Earth's mantle was only somewhat more reducing 4 billion years ago \citep{Aulbach_2016,Nicklas_2019}. Therefore, volcanoes would have mostly produced relatively oxidized gases like H$_2$O, CO$_2$ and N$_2$ instead of highly reduced equivalents, H$_2$, CH$_4$ and NH$_3$ \citep{Holland_1984,Catling_2017,Wogan_2020}. Thus, steady-state volcanism would have likely produced Hadean \hl{(4.56 - 4.0 Ga)} air with CO$_2$ and N$_2$ as bulk constituents, whereas reducing gases, such as CH$_4$, would have been minor or very minor.

However, \citet{Urey_1952} suggested that the prebiotic atmosphere was transiently reduced by large asteroid impacts. In more detail, \citet{Zahnle_2020} argued that iron-rich impact ejecta could react with an impact-vaporized ocean to generate H$_2$ ($\mathrm{Fe} + \mathrm{H_2O} \leftrightarrow \mathrm{FeO} + \mathrm{H_2}$). As the H$_2$O- and H$_2$-rich atmosphere cools, their chemical equilibrium modeling with parameterized quenching finds that H$_2$ can combine with atmospheric CO or CO$_2$ to generate CH$_4$. After several thousand years of cooling, the steam condenses to an ocean, leaving a H$_2$ dominated atmosphere containing CH$_4$. \citet{Zahnle_2020} used a photochemical box model to show that such a reducing atmosphere would have generated prebiotic molecules like HCN. The reducing atmospheric state terminates when H$_2$ escapes to space after millions of years.

Model simplicity in \citet{Zahnle_2020} left critical questions unanswered. Their model of a cooling steam post-impact atmosphere did not explicitly simulate chemical kinetics pertinent to Earth, which may inaccurately estimate the generated CH$_4$. Additionally, their photochemical box model did not include all relevant reactions or distinguish between different prebiotic nitriles (e.g. HCN and HCCCN). Finally, \citet{Zahnle_2020} only crudely computed the climate of post-impact atmospheres, yet surface temperature is important for understanding the possible fate of prebiotic feedstock molecules. These molecules are needed to initiate prebiotic synthesis and must be available in the prebiotic environment.

Here, we improve upon the calculations made in \citet{Zahnle_2020} using more sophisticated and accurate models of post-impact atmospheres. We estimate post-impact H$_2$ production by considering reactions between the atmosphere and delivered iron, and equilibration between the atmosphere and impact-generated melt. Our model explicitly simulates the 0-D chemical kinetics of a cooling steam atmosphere, considering gas-phase reactions, as well as reactions occurring on nickel surfaces which catalyze CH$_4$ production given that nickel is expected to be delivered by big impactors. After post-impact steam condenses to an ocean, we simulate the long-term evolution of a reducing atmosphere with a 1-D photochemical-climate model, quantifying HCN and HCCCN production and the climate in which they are deposited on Earth's surface. Additionally, we discuss the possible fate and preservation of prebiotic molecules in ponds or lakes on Hadean land. Finally, we discuss how ``lucky'' primitive life was if created by post-impact molecules, given a need to not be subsequently annihilated by further impactors.

\section{Methods}

We organize our investigation of post-impact Hadean atmospheres in three phases of atmospheric evolution depicted in Figure \ref{fig:impact_diagram}. Below, we briefly describe our numerical models for each phase and complete descriptions can be found in the Appendix.

In Phase 1, an impactor collides with Earth, vaporizing the ocean, and H$_2$ is generated by reactions between the atmosphere and iron-rich impact ejecta, and atmospheric reactions with an impact-produced melt pond. Our model of this phase (Appendix \ref{sec:phase1_appendix}) accounts for H$_2$ generation from impactor iron by assuming each mole of iron delivered to the atmosphere removes one mole of oxygen. For example, Fe can sequester O atoms from steam:

\begin{equation}
  \mathrm{Fe} + \mathrm{H_2O} \rightarrow \mathrm{FeO} + \mathrm{H_2}
\end{equation}
Simulations that consider reactions between the atmosphere and impact-melted crust follow a similar procedure to the one described in \citet{Itcovitz_2022}. Our model requires that the atmosphere and melt have the same oxygen fugacity. The oxygen fugacity of the melt is governed by relative amounts of ferric and ferrous iron \citep{Kress_1991}:

\begin{equation} 
  \label{eq:kress_carmichael}
  0.5 \mathrm{O_2} + 2 \mathrm{FeO} \leftrightarrow \mathrm{Fe_2O_3}
\end{equation}
We assume that oxygen atoms can flow from the atmosphere into the melt (or visa-versa), and use an equilibrium constant for Reaction \ref{eq:kress_carmichael} from \citet{Kress_1991}. Finally, we compute a chemical equilibrium state of the atmosphere (or atmosphere-melt system) at 1900 K using thermodynamic data from NIST for 96 gas-phase species (Appendix \ref{sec:photochem_reactions}). The result gives the estimated amount of H$_2$ generated by an impact.

In Phase 2 of Figure \ref{fig:impact_diagram}, the steam atmosphere cools for thousands of years generating CH$_4$ and NH$_3$, and eventually, the steam condenses to an ocean. We simulate these events with the 0-D kinetics-climate box model fully described in Appendix \ref{sec:kinetics_climate}. The gas-phase model tracks 96 species connected by 605 reversible reactions (Appendix \ref{sec:photochem_reactions}), but we do not account for photolysis. The model also optionally accounts for reactions that occur on nickel surfaces using the chemical network described in \citet{Schmider_2021}. As discussed later in Section \ref{sec:phase2}, nickel is potentially delivered to Earth's surface by impacts which may catalyze methane production. In the model, atmospheric temperature changes as energy is radiated to space and is modulated by latent heat released from water condensation. We estimate the net energy radiated to space by using a parameterization of calculations performed with our radiative transfer code (Appendix \ref{sec:clima}).

\begin{table}
  \caption{\hl{Opacities used in climate modeling}}
  \label{tab:used_opacities}
  \begin{center}
  \begin{tabularx}{0.9\linewidth}{p{0.25\linewidth} | p{0.4\linewidth} | p{0.25\linewidth}}
    \hline \hline
    Line absorption & Continuum CIA absorption & Rayleigh Scattering \\
    \hline
    H$_2$O, CO$_2$, CH$_4$ & CO$_2$-CO$_2$, N$_2$-N$_2$, CH$_4$-CH$_4$, H$_2$-CH$_4$, H$_2$-H$_2$, H$_2$O-H$_2$O, H$_2$O-N$_2$ & N$_2$, CO$_2$, H$_2$O, H$_2$ \\
    \hline
  \end{tabularx}
  \end{center}
\end{table}

During Phase 3, photochemistry generates HCN and other prebiotic molecules. Hydrogen in the H$_2$ dominated atmosphere escapes to space over millions of years, ushering in the return of a CO$_2$ and N$_2$ atmosphere. We use our time-dependent photochemical-climate model, \emph{Photochem} (Appendix \ref{sec:photochem}), to simulate this phase of atmospheric evolution. The model solves a system of partial differential equations approximating molecular transport in the vertical direction and the effect of chemical reactions, photolysis, condensation, rainout in droplets of water, and \hl{hydrogen} atmospheric escape. \hl{Specifically, the model rains out haze particles and HCN among a few other atmospheric species listed in Appendix \mbox{\ref{sec:photochem_reactions}}. We simulate diffusion-limited and hydrodynamic hydrogen escape using Equation (47) in \mbox{\citet{Zahnle_2020}}.} Our reaction network (Appendix \ref{sec:photochem_reactions}) acceptably reproduces the steady-state composition of Earth and Titan (Appendix Figure \ref{fig:earth_titan_valid}). \hl{When reproducing the chemistry of Earth and Titan we fix the temperature profile to measured values, rather than self-consistently compute the climate.} We evolve the model equations accurately over time using the CVODE Backward Differential Formula (BDF) method \citep{Hindmarsh_2005}. As the atmosphere evolves, we compute self-consistent temperature structures using the radiative transfer code described and validated in Appendix \ref{sec:clima}. \hl{Unless otherwise noted in the text, our climate calculations use the opacities in Table \mbox{\ref{tab:used_opacities}}, which is a subset of the opacities available in our radiative transfer code (Appendix Table \mbox{\ref{tab:climate_opacities}}). Climate calculations do not account for the radiative effects of clouds or hazes. However, our UV radiative transfer for computing photolysis rates do account for haze absorption and scattering.}

\section{Results}

\begin{figure}
  \centering
  \includegraphics[width=1.0\textwidth]{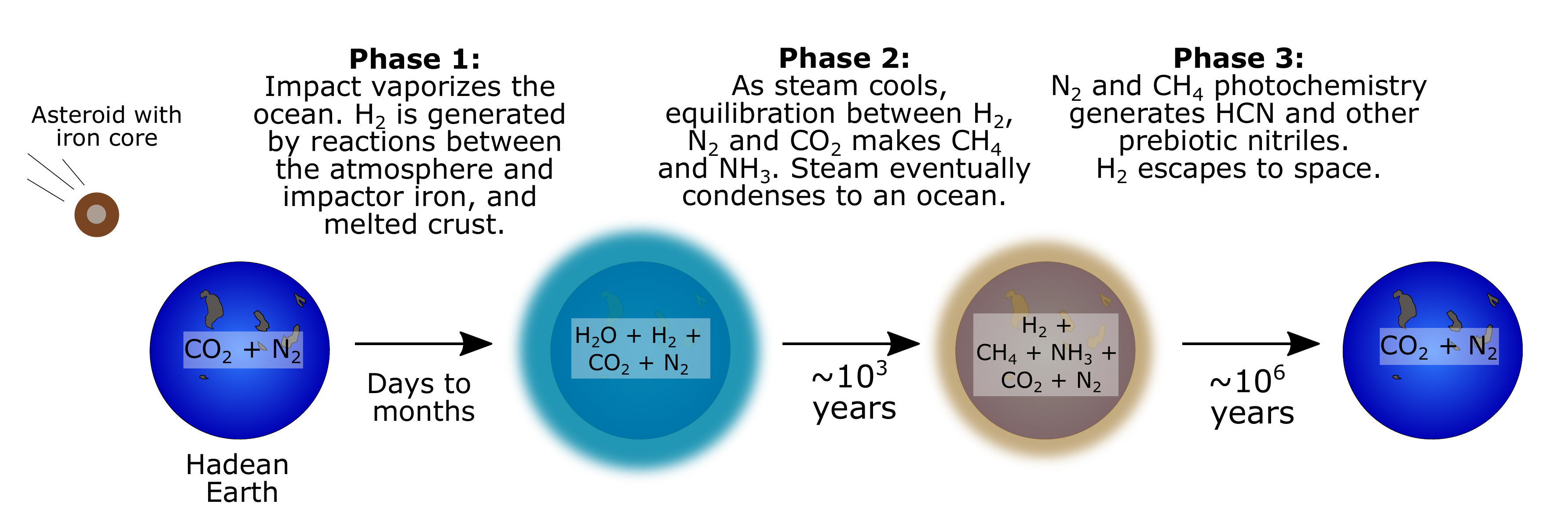}
  \caption{The three phases of atmospheric evolution after a large asteroid impact on the Hadean Earth. In Phase 1, the impactor vaporizes the ocean and heats up the atmosphere. Iron delivered by the impactor reacts with hot steam to make H$_2$. H$_2$ is also modulated by equilibration between the atmosphere and an impact-generated melt pond. In Phase 2, as the steam-rich atmosphere cools for thousands of years, H$_2$ reacts with CO$_2$ to make atmospheric CH$_4$. Ultimately, the steam condenses to an ocean. Finally, in Phase 3, N$_2$ and CH$_4$ photochemistry generates HCN and other prebiotic nitriles. The H$_2$ dominated atmosphere escapes to space over millions of years, causing the return of a more oxidizing N$_2$ and CO$_2$ atmosphere.}
  \label{fig:impact_diagram}
\end{figure}

The following sections simulates the three post-impact phases of atmospheric evolution shown in Figure \ref{fig:impact_diagram} for impactor masses between $10^{20}$ and $10^{22}$ kg (360 to 1680 km diameter) under various modeling assumptions.

\subsection{Phase 1: Reducing the steam-generated atmosphere with impactor iron} \label{sec:phase1}

Within days, a massive asteroid impact would leave the Hadean Earth with a global $\sim$2000 K rock and iron vapor atmosphere, the iron derived from the impactor's core \citep{Itcovitz_2022}. In the following months to years, energy radiated downward from the silicates would vaporize a large fraction of the ocean, adding steam to the atmosphere \citep{Sleep_1989}. At this point, steam should rapidly react with iron to generate H$_2$. Eventually, the iron vapor and then rock would rain out leaving behind a steam-dominated atmosphere containing H$_2$, as well as CO$_2$ and N$_2$ from the pre-impact atmosphere. The sequence of metal followed by silicate condensation with falling temperature is loosely analogous to that of the well-known condensation sequence of the solar nebula.

Furthermore, the massive impact would generate a melt pool on Earth's surface inside the impact crater, which may contain reducing impact-derived iron. The atmosphere and melt pool could react to a redox-equilibrium state. This could add or sequester H$_2$ from the atmosphere, depending on whether the melt was more or less reducing than the atmosphere \citep{Itcovitz_2022}.

Recently, \citet{Itcovitz_2022} used a smoothed-particle hydrodynamics (SPH) code with $0.5 \times 10^6$ - $3 \times 10^6$ particles of 150 - 250 km diameter to estimate the amount of H$_2$ generated as these processes unfold under several different impact scenarios on the Hadean Earth. In their fiducial case (i.e. their ``Model 1A''), they assume that 100\% of iron delivered by an impactor is available to react and reduce a post-impact steam atmosphere. In another scenario, they assume that only $\sim$15 - 30\% of impactor iron reacts with the steam atmosphere based on their SPH simulations (their ``Model 1B'') \citep{Citron_2022}. For both cases, they also consider equilibration between the atmosphere and a melt pool (their ``Model 2'', ``Model 3A'' and ``Model 3B''). In their simulations, the melt pool is extremely reducing or more oxidizing depending on whether they assume it contains a fraction of the impactor's iron, and they use SPH models to predict the amount of iron accreted to the melt pool \citep{Citron_2022}. Overall, they conclude that melt-atmosphere equilibration generates about as much H$_2$ as their fiducial case as long as the iron delivered to the melt-atmosphere system can equilibrate. However, if iron delivered to the melt pool sinks into Earth and cannot react with the atmosphere, then approximately 2 - 10 times less H$_2$ is produced compared to their fiducial scenario (see the erratum in \citet{Itcovitz_2022}).

\begin{figure}
  \centering
  \includegraphics[width=0.9\textwidth]{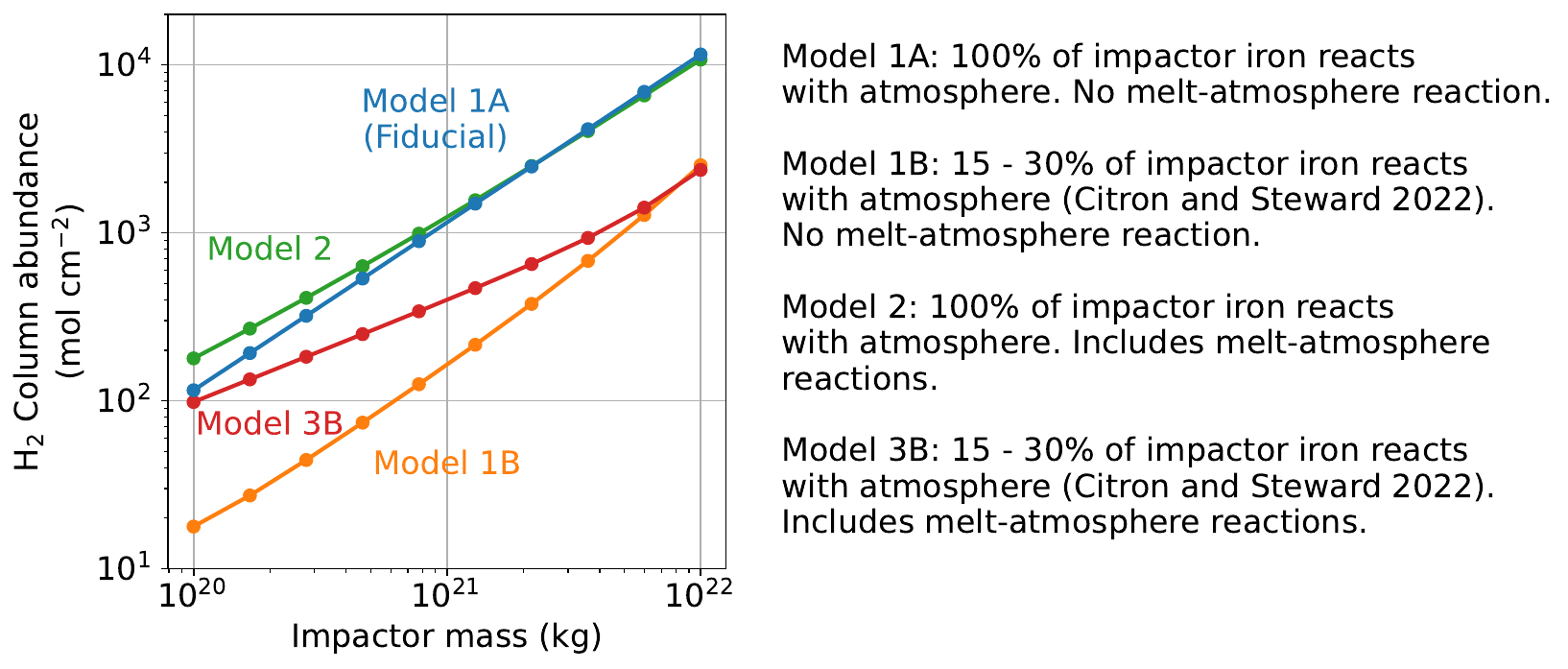}
  \caption{Post-impact H$_2$ generation as a function of impactor mass under different modeling assumptions. Models 1A, 1B, 2 and 3B are identical to those describe in Figure 1 of \citet{Itcovitz_2022}. The simulation's pre-impact volatile inventories, impact angle, and impact velocity are listed in Table \ref{tab:fiducial_parameters}. In Model 1A, all iron delivered by an impact reacts with steam to produce H$_2$. The resulting atmosphere does not equilibrate with a impact-generated melt pool. Model 1B assumes that a fraction ($\sim 15\%$ to $\sim 30\%$) of impactor iron reduces the steam atmosphere based on SPH simulations \citep{Citron_2022}, and that the atmosphere is chemically isolated from a melt pool. Model 2 is like Model 1A while also including post-impact equilibration with a melt pool with a redox state of $\Delta$FMQ-2.3 to represent peridotite \citep{Itcovitz_2022}. Model 3B assumes that a fraction ($\sim 15\%$ to $\sim 30\%$) of impactor iron reacts with the steam atmosphere based on SPH simulations, and includes melt-atmosphere redox equilibration with a pool magma initially at $\Delta$FMQ-2.3. As stated in Section \ref{sec:phase1}, we nominally assume Model 1A throughout the main text calculations. We also include simulations in the appendix that instead adopt Model 1B, which we consider to be a plausible lower bound for post-impact H$_2$ generation.}
  \label{fig:melt_reaction_sup}
\end{figure}

\citet{Itcovitz_2022} considers impactors between $2 \times 10^{21}$ and $2 \times 10^{22}$ kg, and assumes the pre-impact Earth has 1.85 oceans of water, 100 bars CO$_2$ and 2 bars of N$_2$. However, we investigate impacts as small as $10^{20}$ kg, and our nominal model (Table \ref{tab:fiducial_parameters}) assumes only 0.5 bars of pre-impact CO$_2$ motivated by models of the Hadean carbonate-silicate cycle \citep{Kadoya_2020} and assuming little mantle-hosted carbonate is vaporized. Therefore, we use a similar model (Appendix \ref{sec:phase1_appendix}) to the one described in \citet{Itcovitz_2022} to predict the post-impact H$_2$ for our alternative model assumptions (Table \ref{tab:fiducial_parameters}) and impactor sizes. Figure \ref{fig:melt_reaction_sup} shows the results.

Our calculations give two end-member scenarios for impact H$_2$ production which we consider for subsequent calculations in this article. The more optimistic case assumes that 100\% of the impactor's iron reacts with an atmosphere that is chemically isolated from a melt pool (``Model 1A'' in Figure \ref{fig:melt_reaction_sup}). Following \citet{Zahnle_2020}, we adopt this scenario as our nominal model throughout the main text. This assumption produces a similar amount of H$_2$ as an atmosphere-melt system that retains most of the impactor's iron (e.g. ``Model 2'' in Figure \ref{fig:melt_reaction_sup}), which is consistent with \citet{Itcovitz_2022}. The ``Model 2'' calculation assumes the melt pool has an initial oxygen fugacity of $\Delta$FMQ-2.3 which is appropriate for a peridotite melt \citep{Itcovitz_2022}.\footnote{FMQ is the fayalite-magnetite-quartz redox buffer. See Chapter 7 in \citet{Catling_2017} for a discussion of redox buffers.} However, our results are not sensitive to this assumption because, for ``Model 2'', initial melt oxygen fugacities between $\Delta$FMQ and $\Delta$FMQ-4 changes the generated H$_2$ by a factor of at most $\sim 1.3$.

The less-optimistic case for H$_2$ production is ``Model 1B'' in Figure \ref{fig:melt_reaction_sup}, which assumes that only a fraction of the impactor iron reacts with an atmosphere \hl{($\sim 15\%$ to $\sim 30\%$)}, and that the latter does not react with a melt pool. We compute the fraction of available iron by extrapolating SPH simulations of impacts traveling at twice Earth's escape velocity and colliding with Earth at a 45$^{\circ}$ angle (Appendix Figure \ref{fig:citron_interpolations}), which is the most probable angle \citep{Citron_2022}. Most simulations shown in the main text have a complementary figure in the Appendix that makes this alternative pessimistic assumption regarding post-impact H$_2$ generation.

\begin{table}
  \caption{Nominal model assumptions}
  \label{tab:fiducial_parameters}
  \begin{center}
  \begin{tabularx}{0.9\linewidth}{p{0.3\linewidth} | p{0.2\linewidth} | p{0.3\linewidth}}
    \hline \hline
    Parameter & symbol & value \\
    \hline
    Pre-impact ocean inventory & $N_\mathrm{H_2O}$ & $1.5 \times 10^4$ mol cm$^{-2}$ (i.e. 1 ocean)$^\text{a}$ 
    \\
    Pre-impact CO$_2$ inventory & $N_\mathrm{CO_2}$ & 12.5 mol cm$^{-2}$ (i.e. ``0.5 bars'')$^\text{b}$  
    \\
    Pre-impact N$_2$ inventory & $N_\mathrm{N_2}$ & 36 mol cm$^{-2}$ (i.e. ``1 bar'')$^\text{c}$  
    \\
    Impactor mass & $M_\text{imp}$ & $10^{20}$ - $10^{22}$ kg
    \\
    Iron mass fraction of the impactor & $m_\text{Fe,imp}$ & 0.33 
    \\
    Fraction of iron that reacts with atmosphere & $X_\mathrm{Fe,atmos}$ & 1.0$^\text{d}$ 
    \\
    Impact angle & - & 45$^{\circ}$
    \\
    Impact velocity relative to Earth & - & $20.7$ km s$^{-1}$
    \\
    Eddy diffusion coefficient$^\text{e}$ & $K_{zz}$ & $10^6$ cm$^{2}$ s$^{-1}$ 
    \\
    Aerosol particle radius$^\text{e}$ & - & 0.1 $\mu$m
    \\
    Troposphere relative humidity & $\phi$ & 1 
    \\
    Surface Albedo & $A_s$ & 0.2 
    \\
    Temperature of the stratosphere & T$_\mathrm{strat}$ & 200 K 
    \\
    Rainfall rate & $R_\mathrm{rain}$ & $1.1 \times 10^{17}$ molecules cm$^{-2}$ s$^{-1}$  (Modern Earth's value) 
    \\
    HCN deposition velocity$^\text{f}$ & $v_\mathrm{dep,HCN}$ & $7 \times 10^{-3}$ cm s$^{-1}$
    \\
    HCCCN deposition velocity$^\text{g}$ & $v_\mathrm{dep,HCCCN}$ & $7 \times 10^{-3}$ cm s$^{-1}$
    \\
    \hline
    \multicolumn{3}{p{0.9\linewidth}}{

    $^\text{a}$ \hl{The source and inventory of surface H$_2$O throughout the Hadean is debated \mbox{\citep{Miyazaki_2022,Korenaga_2021,Johnson_2020}} and even how much water is present on the modern Earth (e.g., \mbox{\citet{Lecuyer_1998}} estimates 0.3-3 oceans in Earth's mantle). Our nominal case of one modern ocean is one possibility among several.}

    $^\text{b}$ \hl{Based on Hadean carbon cycle modeling in \mbox{\citet{Kadoya_2020}}.}

    $^\text{c}$ \hl{Based on Figure 5 in \mbox{\citet{Catling_2020}}}.

    $^\text{d}$ This is the ``Model 1A'' scenario for H$_2$ production described near the end of Section \ref{sec:phase1} and in Figure \ref{fig:melt_reaction_sup}.

    $^\text{e}$ Assumed to be constant as a function of altitude.

    $^\text{f}$ Estimated based on the HCN hydrolysis rate in the ocean (Appendix \ref{sec:hcn_vdep}).

    $^\text{g}$ Assumed to the the same as HCN.
  }
  \end{tabularx}
  \end{center}
\end{table}

\subsection{Phase 2: The cooling post-impact steam atmosphere} \label{sec:phase2}

After reactions between impact-derived iron and steam produce H$_2$, the atmosphere would radiate at a rate determined by the optical properties of water vapor \citep{Zahnle_2020}. Chemical reactions would initially be rapid, forcing the whole atmosphere to chemical equilibrium. Methane is thermodynamically preferred at lower temperatures \hl{(e.g., more methane is prefered in a gas at 1000 K than a gas at 1500 K)}, so it should become more abundant as the atmosphere cools. Eventually the atmosphere would reach a temperature where the reactions producing methane would be extremely sluggish compared to the rate of atmospheric cooling. At this point, the methane abundance would freeze, or quench. Ammonia would exhibit the same behavior as methane by initially rising in abundance then quenching when kinetics become slow. After several thousand years, water vapor condenses and rains out of the atmosphere to form an ocean.

We use the 0-D kinetics-climate box model described in Appendix \ref{sec:kinetics_climate} to simulate these events. By simulating each elementary chemical reaction, the model automatically computes methane and ammonia quenching as the atmosphere cools and temperature-dependent reactions slow. We first consider gas-phase kinetics, and later we will also consider nickel-surface kinetics.

\begin{figure}
  \centering
  \includegraphics[width=0.6\textwidth]{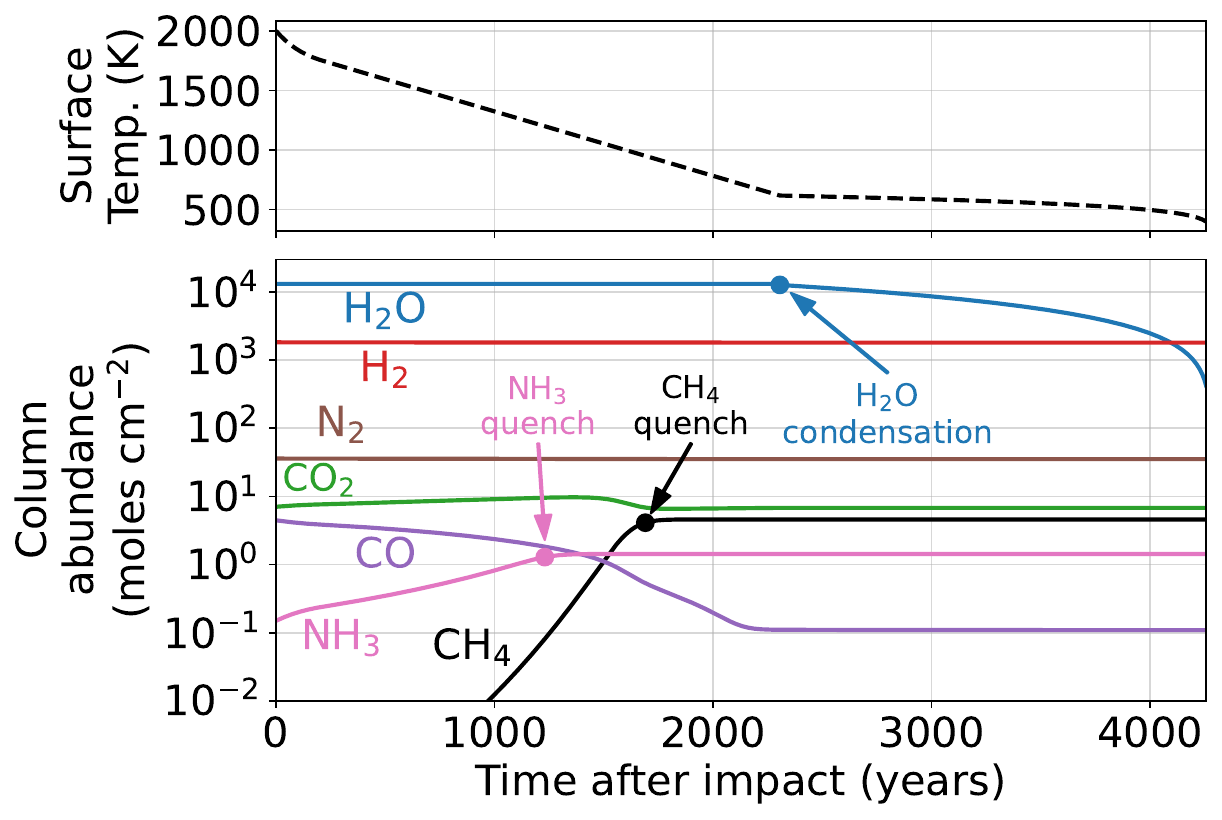}
  \caption{A kinetics-climate simulation of a cooling steam atmosphere caused by a $1.58 \times 10^{21}$ kg impactor. The model uses the Table \ref{tab:fiducial_parameters} nominal parameters. The top panel is surface temperature and the bottom panel shows atmospheric composition.}
  \label{fig:figure1}
\end{figure}

Figure \ref{fig:figure1} shows our model applied to a $1.58 \times 10^{21}$ kg ($\sim 900$ km diameter) impactor. As the steam cools, ammonia quenches when the atmosphere is $\sim 1200$ K, followed by CH$_4$ quenching at $\sim 950$ K. After quenching, nearly half of the total carbon in the atmosphere exists as CH$_4$. After 4200 years, the steam has largely rained out to form an ocean, leaving behind a H$_2$-dominated atmosphere containing CH$_4$ and NH$_3$. NH$_3$ is soluble in water, so a fraction should be removed from the atmosphere by dissolution in the newly formed ocean; however our simulations (e.g. Figure \ref{fig:figure1}) do not account for this effect. 

\begin{figure}
  \centering
  \includegraphics[width=0.55\textwidth]{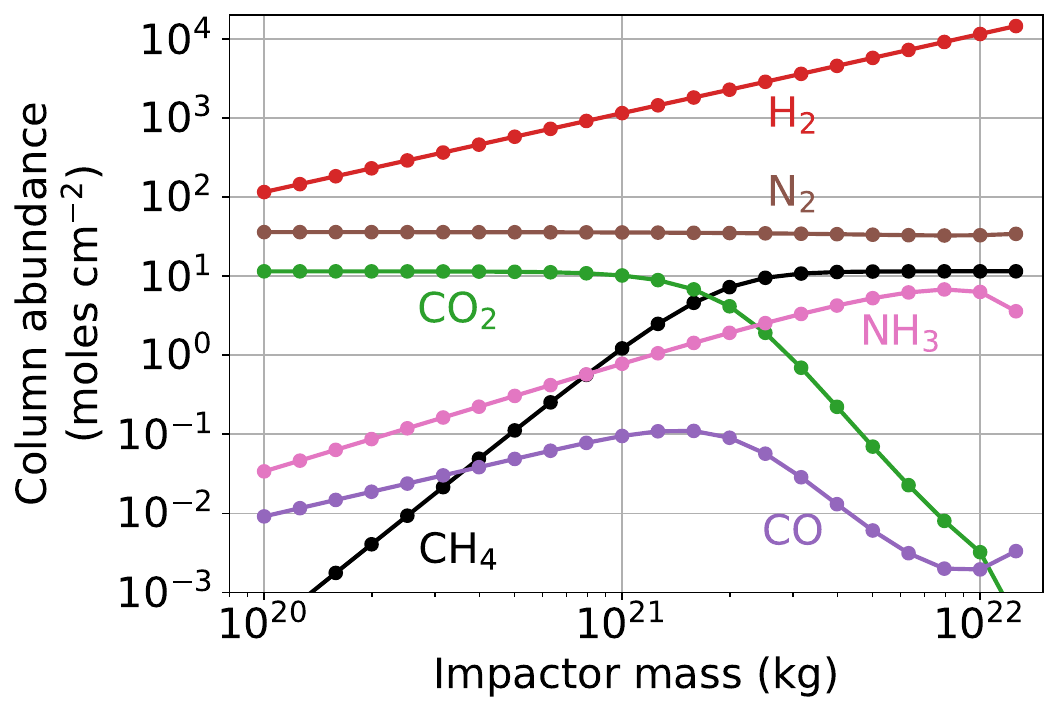}
  \caption{Predicted atmospheric composition as a function of impactor mass after steam has condensed to an ocean. We use our nominal modeling assumptions (Table \ref{tab:fiducial_parameters}), and also use gas-phase kinetics. Most CO$_2$ is converted to CH$_4$ for impactors larger than $1.6 \times 10^{21}$ kg.}
  \label{fig:figure2}
\end{figure}

Figure \ref{fig:figure2} shows predicted atmospheric composition at the end of the steam atmosphere (e.g. at 4200 years in Figure \ref{fig:figure1}) as a function of impactor mass. The calculations use gas-phase reactions, and our nominal model parameters (Table \ref{tab:fiducial_parameters}), including the assumption that 100\% of the iron delivered by the impactor reacts with the steam atmosphere to make H$_2$. For example, a $10^{20}$ kg impactor generates $1.2 \times 10^{2}$ H$_2$ moles cm$^{-2}$ which would have a partial pressure of 1.2 bars if the atmosphere did not contain water vapor. A $10^{22}$ kg impactor generates $1.1 \times 10^{4}$ H$_2$ moles cm$^{-2}$ which would have a ``dry'' partial pressure of 23.8 bars.\footnote{Partial pressures depend on the mean molecular weight of the atmosphere. The $10^{22}$ kg simulation in Figure \ref{fig:figure2} has 65.0 bars H$_2$ before ocean vapor condenses, and would have 23.8 bars H$_2$ if there was no water vapor in the atmosphere. Both scenarios have the same number of H$_2$ molecules in the atmosphere, but have different partial pressures because of dissimilar mean molecular weights. To avoid ambiguity, we occasionally report partial pressures in ``dry'' bars, which is the partial pressure of a gas if the atmosphere had no water vapor.}
We find that most of the CO$_2$ in the atmosphere is converted to CH$_4$ for impactors larger than $1.6 \times 10^{21}$ kg ($\sim 900$ km diameter), and that bigger impacts generate more NH$_3$, e.g., a $10^{22}$ kg impactor makes 0.013 ``dry'' bars of NH$_3$. Reduced species like CH$_4$ and NH$_3$ are thermodynamically preferred in the thick H$_2$ atmospheres generated by bigger impacts. Large impacts generate big amounts of hydrogen because they deliver more iron which more thoroughly reduces the atmosphere.

The Figure \ref{fig:figure2} calculations might underestimate the CH$_4$ produced in the post-impact atmosphere because they ignore reactions occurring on nickel surfaces that can catalyze CH$_4$ generation. If the impactors that struck the Earth during the Hadean resembled enstatite chondrite or carbonaceous chondrite composition then they would have contained 1\% - 2\% nickel \citep[Table 15]{Lewis_1992}. This nickel would have coexisted with the rock and iron vapor atmosphere that lasted months to years following a massive impact (Phase 1 in Figure \ref{fig:impact_diagram}). Metals along with silicates would have rained out as spherules covering the entire planet \citep{Genda_2017}. As the impact-generated steam cooled, chemical reactions catalyzing CH$_4$ production could have occurred on nickel surfaces in the bed of spherules \citep{Schmider_2021}. These surface reactions could lower the quench temperature of CH$_4$, causing more of the gas to be produced.

To estimate the effect of nickel catalysis on CH$_4$ production, we use our kinetics-climate box model (Appendix \ref{sec:kinetics_climate}) with the nickel-surface reaction network developed by \citet{Schmider_2021}. The network is based on quantum chemistry calculations and about a dozen experiments from the literature. Our micro-kinetics approach is distinct from the empirical one taken by, e.g. \citet{Kress_2004}, because our model tries to capture each elementary step of catalysis, rather than use a parameterization that is specific to certain experimental conditions. 

\begin{figure}
  \centering
  \includegraphics[width=0.7\textwidth]{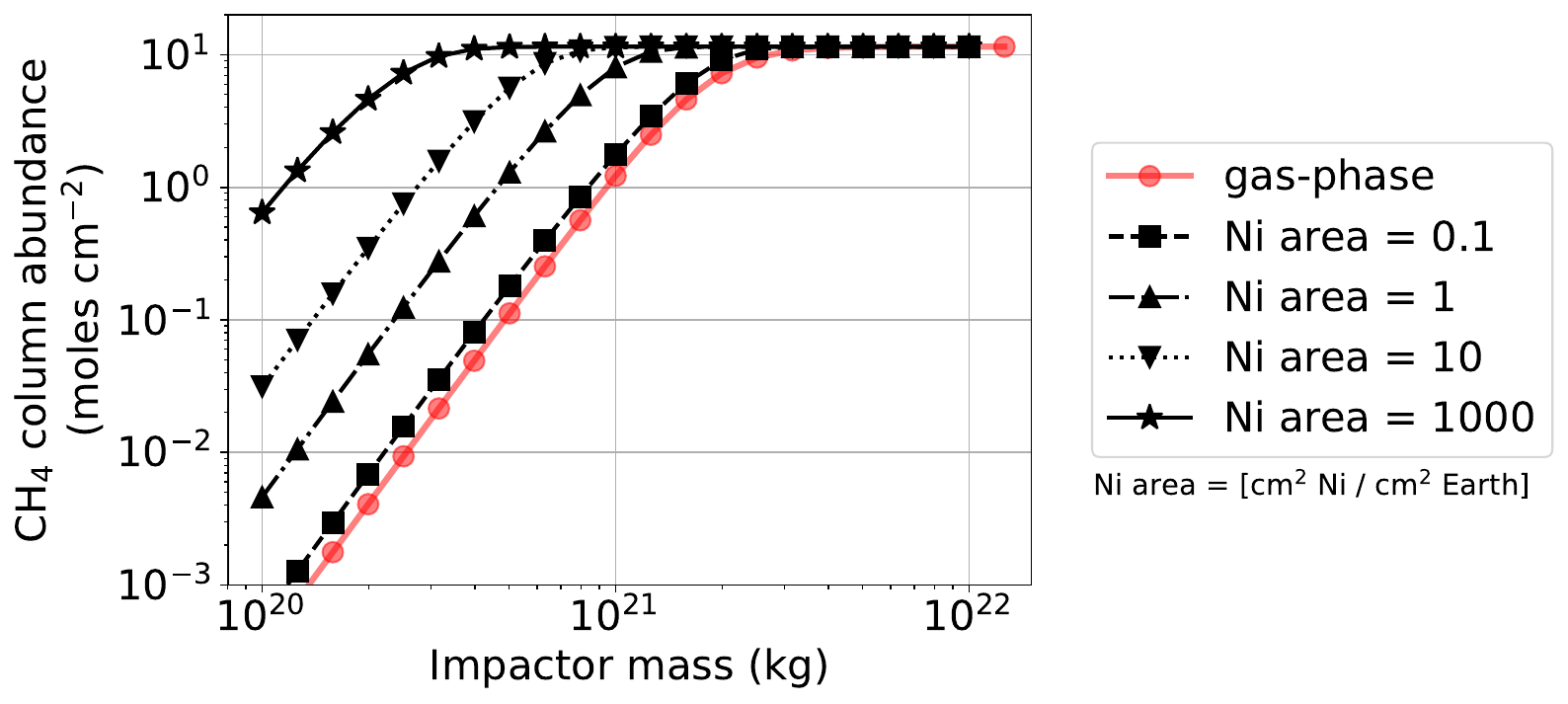}
  \caption{The effect of nickel catalysts on post-impact methane production. The calculations use the Table \ref{tab:fiducial_parameters} model parameters and the \citet{Schmider_2021} surface reaction network. Ni areas larger than 0.1 cm$^2$ nickel / cm$^2$ Earth generates more methane than our model that uses gas-phase reactions, e.g., Figure \ref{fig:figure2}.}
  \label{fig:figure3}
\end{figure}

Figure \ref{fig:figure3} shows the quenched methane abundance as a function of impactor mass predicted by our model that includes nickel catalysts. The amount of CH$_4$ generated depends strongly on the amount of available nickel surface area. Nickel areas bigger than 0.1 cm$^2$ nickel / cm$^2$ Earth permit more CH$_4$ production compared to our gas-phase only model. Assuming a nickel area of 1000 cm$^2$ nickel / cm$^2$ Earth, then a Vesta-size impactor ($2.6 \times 10^{20}$ kg, 500 km diameter) could convert most CO$_2$ in the pre-impact atmosphere to CH$_4$.

Unfortunately, a precise nickel surface area is hard to estimate. The correct value depends on how the rock, iron and nickel spherules mix and precipitate to the surface, and furthermore, how effectively the atmosphere can diffuse through and react on exposed nickel. We do not attempt to compute these effects here, and instead estimate possible upper bounds. Consider a $2.6 \times 10^{20}$ kg impactor (Vesta-sized) of enstatite chondrite composition, containing 2\% by mass Ni \citep{Lewis_1992}. If all this nickel is gathered into 1 mm spheres, a plausible droplet size according to \citet{Genda_2017}, then the total nickel surface area is $3.4 \times 10^3$ cm$^2$ nickel / cm$^2$ Earth. An impactor ten times more massive would deliver ten times more nickel resulting in an upper bound Ni area that is one order of magnitude larger. Significantly smaller nickel particles are conceivable. There is experimental support for the formation of ultra-fine $< 300$ nm particles in the wake of impacts colliding with an ocean \citep{Furukawa_2007}. For a Vesta-sized impactor, collecting all nickel into $100$ nm particles has a nickel area \hl{six} orders of magnitude large than the 1 mm case - $3.4 \times 10^{9}$ cm$^2$ nickel / cm$^2$ Earth. Overall, the larger nickel areas shown in Figure \ref{fig:figure3} may be within the realm of possibility. \hl{Alternatively, nickel might be buried by rock and iron when these materials condense out of the post-impact atmosphere, and that $< 0.1$ cm$^2$ nickel / cm$^2$ Earth is available for catalysis. In this case, gas-phase kinetics would determine the conversion of CO$_2$ to CH$_4$.}

Figures \ref{fig:figure2} and \ref{fig:figure3} optimistically assume that all iron delivered by the impactor reacts with steam to make H$_2$, however, this may not be the case (see Section \ref{sec:phase1}). Therefore, in Appendix Figures \ref{fig:figure2_citron} and \ref{fig:figure3_citron} we recalculate Figures \ref{fig:figure2} and \ref{fig:figure3}, but assume that only a fraction of the impactor's iron reduces the steam atmosphere by extrapolating SPH simulations of impacts (``Model 1B'' in Figure \ref{fig:melt_reaction_sup}). The resulting H$_2$, CH$_4$, and NH$_3$ production appear similar, except shifted by a factor of $\sim 5$ to larger impactors. The results are shifted by this amount because SPH simulations suggest approximately $\sim 1/5$ of impactor iron is delivered to the atmosphere, while the rest is either embedded in Earth, or ejected to space. We consider these supplementary calculations lower-bounds for impactor generated CH$_4$ and NH$_3$.

\subsection{Phase 3: Long-term photochemical-climate evolution} \label{sec:phase3}

Several thousand years after a massive impact, the steam-dominated atmosphere would condense to an ocean leaving behind a H$_2$-dominated atmosphere containing CH$_4$ and NH$_3$ (e.g. at 4200 years in Figure \ref{fig:figure1}). The reducing atmospheric state should persist for millions of years until hydrogen escapes \citep{Zahnle_2020}.

% Phases 1 of a post-impact atmosphere indicated in Figure \ref{fig:impact_diagram} occurs at $t = 0$ and is assumed to occur instantaneously. Phases 2 and 3 occur between 0 and 4200 years and between 4200 and $2 \times 10^6$ years.

\begin{figure}
  \centering
  \includegraphics[width=0.7\textwidth]{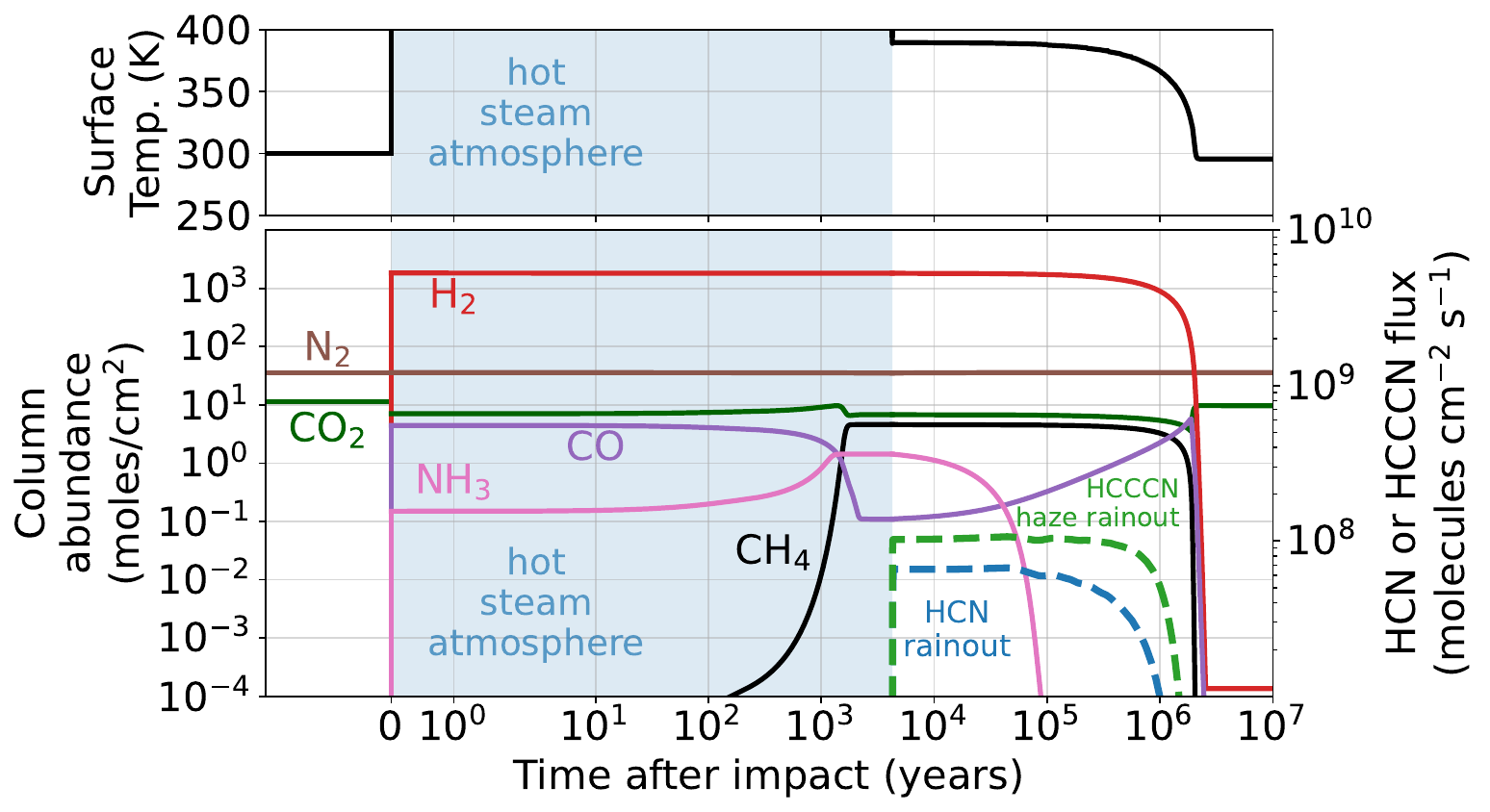}
  \caption{Simulated composition and climate of the Hadean atmosphere after a $1.58 \times 10^{21}$ kg impactor that produces 7.0 bars of H$_2$ once vaporized ocean water condenses. We use the Table \ref{tab:fiducial_parameters} model parameters. The blue shaded region labeled ``hot steam atmosphere'', also called Phase 2 in Figure \ref{fig:impact_diagram}, is simulated by the kinetics-climate model described in Appendix \ref{sec:kinetics_climate}. After this time-period, during Phase 3 of a post-impact atmosphere, we evolve the atmosphere with 1-D photochemical-climate model (Appendix \ref{sec:photochem}), which maintains 0.018 bar of CH$_4$ between $4 \times 10^3$ and $\sim 10^6$ years. Dashed lines are referenced to the right-hand axis. ``HCN rainout'' is HCN molecules raining out in droplets of water. ``HCCCN haze rainout'' is the rainout rate of HCCCN incorporated into particles formed from the reaction $\mathrm{C_4H} + \mathrm{HCCCN} \rightarrow \mathrm{polymer}$. CH$_4$ and N$_2$ photochemistry generates HCN and HCCCN for about one million years until H$_2$ escapes to space.}
  \label{fig:figure4}
\end{figure}

We simulate the long-term evolution of this hydrogen-rich atmosphere using a coupled one-dimensional photochemical-climate model (Appendix \ref{sec:photochem}). Figure \ref{fig:figure4} shows our model applied to the atmosphere following a $1.58 \times 10^{21}$ kg ($\sim 900$ km diameter) impactor. We assume a pre-impact atmosphere with 1 bar N$_2$ and 0.5 bars of CO$_2$, and simulate the cooling steam atmosphere with our kinetics-climate climate model (Section \ref{sec:phase2}). Next, we use the end of the steam atmosphere simulation as initial conditions for our 1-D photochemical-climate model. 

We find that N$_2$ and CH$_4$ photochemistry generates HCN in a hazy Titan-like atmosphere for about one million years until it is halted by hydrogen escape to space. In this model, the dominant channel producing HCN is $\mathrm{N} + \mathrm{^3CH_2} \rightarrow \mathrm{HCN} + \mathrm{H}$ where $\mathrm{^3CH_2}$ is ground (triplet) state of the methylene radical derived form methane photolysis. There are two other important paths. The first is $\mathrm{N} + \mathrm{CH} \rightarrow \mathrm{CN} + \mathrm{H}$ followed by $\mathrm{H_2} + \mathrm{CN} \rightarrow \mathrm{HCN} + \mathrm{H}$, and the second is $\mathrm{N} + \mathrm{CH_3} \rightarrow \mathrm{H} + \mathrm{H_2CN}$ and $\mathrm{H_2CN} + \mathrm{H} \rightarrow \mathrm{HCN} + \mathrm{H_2}$. In all pathways, hydrocarbon radicals (e.g., $\mathrm{^3CH_2}$ and $\mathrm{CH_3}$) are sourced from photolyzed CH$_4$ and atomic N is derived from photolyzed N$_2$, which both occur at high altitudes ($p < 10^{-5}$ bar, Appendix Figure \ref{fig:ch4_n_hcn}). The largest chemical loss of HCN is photolysis followed by $\mathrm{N} + \mathrm{CN} \rightarrow \mathrm{N_2} + \mathrm{C}$. Other significant losses are paths that form HCCCN \hl{haze} aerosols. HCN production and loss is our model is comparable to pathways discussed in similar studies \citep{Zahnle_1986,Tian_2011,Rimmer_2019}. \hl{We determined the chemical paths most important for producing and destroying HCN by studying column integrated reaction rates at 14,200 years in Figure \mbox{\ref{fig:figure4}}.}

In Figure \ref{fig:figure4}, HCN mixes to the surface and rains out in droplets of water at a rate of $\sim 10^7$ molecules cm$^{-2}$ s$^{-1}$. HCN also dissolves into the ocean at a similar rate, where we assume it is eventually destroyed by hydrolysis (not shown in Figure \ref{fig:figure4}). To emulate HCN dissolution and destruction in the ocean, we assume a $7 \times 10^{-3}$ cm s$^{-1}$ deposition velocity justified in Appendix \ref{sec:hcn_vdep}. Additionally, a relatively small amount of HCN polymerizes to haze particles in our model via $\mathrm{H_2CN} + \mathrm{HCN} \rightarrow \mathrm{polymer}$ following \citet{Lavvas_2008}, which falls and rains out in water droplets to the surface.

Our results differ from the simulations of \citet{Zahnle_2020}, which suggested that the duration of HCN production after an impact was limited by rapid photolysis of methane. The Figure \ref{fig:figure4} simulation finds that the CH$_4$ lifetime is 4.8 million years because, following photolysis, CH$_4$ efficiently recombines in a hydrogen rich atmosphere from the following reaction, which is well known in the atmospheres of the giant planets in out solar system (Appendix Figure \ref{fig:ch4_prod_loss}).

\begin{equation} \label{eq:ch4_recom}
  \mathrm{CH_3} + \mathrm{H} + \mathrm{M} \rightarrow \mathrm{CH_4} + \mathrm{M}  
\end{equation}
\citet{Zahnle_2020} did not account for Reaction \ref{eq:ch4_recom}. The lifetime of cyanide production is therefore instead determined by the timescale of hydrogen escape to space. Significant hydrogen escape permits the destruction of most atmospheric CH$_4$ because Reaction \ref{eq:ch4_recom} becomes inefficient, which in turn ceases CH$_4$-driven HCN production.

In Figure \ref{fig:figure4}, HCCCN is primarily destroyed by photolysis and produced by the following reaction from acetylene and the cyanide radical,

\begin{equation} \label{eq:hcccn}
  \mathrm{C_2H_2} + \mathrm{CN} \rightarrow \mathrm{HCCCN} + \mathrm{H}  
\end{equation}
A fraction of produced HCCCN reacts to form aerosols via $\mathrm{C_4H} + \mathrm{HCCCN} \rightarrow \mathrm{polymer}$ following \citet{Lavvas_2008}. These polymers fall and mix toward the surface where they rainout in droplets of water at a rate of $\sim 10^{8}$ molecules cm$^{-2}$ s$^{-1}$. Most gas-phase HCCCN is either destroyed by photolysis or incorporated into aerosols, causing vanishingly small surface HCCCN gas pressures ($< 10^{-{16}}$ bar).

\hl{Our model approximates haze formation with the following three reactions: $\mathrm{C_2H} + \mathrm{C_4H_2} \rightarrow \mathrm{polymer} + \mathrm{H}$, $\mathrm{H_2CN} + \mathrm{HCN} \rightarrow \mathrm{polymer}$, and $\mathrm{C_4H} + \mathrm{HCCCN} \rightarrow \mathrm{polymer}$. At 14,200 years in Figure \mbox{\ref{fig:figure4}}, the first pathway dominates, forming $\sim 1.8 \times 10^{13}$ g haze yr$^{-1}$. At this same point in time the second and third pathways produce $9 \times 10^{7}$ g yr$^{-1}$ and $2.8 \times 10^{12}$ g yr$^{-1}$, respectively. The total haze production rate ($2.1 \times 10^{13}$ g yr$^{-1}$) is comparable to values estimated by \mbox{\citet{Trainer_2006}} for the early Earth based on laboratory experiments. Haze particles fall and rainout to the surface where they can hydrolyze and participate in prebiotic chemistry \mbox{\citep{Neish_2010,Poch_2012}}.}

In Figure \ref{fig:figure4}, impact-generated ammonia persists for nearly $10^5$ years. NH$_3$ is primarily destroyed by photolysis, but then recombines from reactions with hydrogen:

\begin{gather} 
  \mathrm{NH} + \mathrm{H_2} + \mathrm{M} \rightarrow \mathrm{NH_3} + \mathrm{M} \label{eq:nh3_1} 
  \\
  \mathrm{NH_2} + \mathrm{H} + \mathrm{M} \rightarrow \mathrm{NH_3} + \mathrm{M} \label{eq:nh3_2}
\end{gather}
Reactions \ref{eq:nh3_1} and \ref{eq:nh3_2} are relatively efficient in a hydrogen-rich atmosphere. Ammonia \hl{photolysis} primarily occurs at the $10^{-3}$ bar altitude, while haze is largely produced above the $10^{-5}$ bar altitude. Therefore, haze particles partially shield ammonia from photolysis, extending the NH$_3$ lifetime \citep{Sagan_1997}. Our model assumes the haze particles are perfect spheres with optical properties governed by Mie theory. Observations of Titan's haze have revealed that hydrocarbon haze particles have a fractal structure which absorb and scatter UV more effectively than Mie spheres \citep{Wolf_2010}. Therefore, our model likely overestimates NH$_3$ photolysis in post-impact atmospheres. 

Figure \ref{fig:figure4} assumes that all NH$_3$ is in the atmosphere and that it does not rainout, but the gas is highly soluble in water and should dissolve in the ocean where it hydrolyzes to ammonium, NH$_4^+$. \hl{Later in Section \mbox{\ref{sec:climate_uncertainty}}, we show that for an atmosphere with 0.3 mol cm$^{-2}$ NH$_3$ and a 371 K ocean at $\text{pH} = 7$, $4\%$ of NH$_3$ would persist in the atmosphere, while the rest is dissolved in the ocean. For a hotter 505 K atmosphere with 6.8 mol cm$^{-2}$ NH$_3$, only 20\% of ammonia dissolves in the ocean because solubility decreases with increasing temperature (Section \mbox{\ref{sec:climate_uncertainty}}).} Ammonia dissolution in the ocean would protect it from photolysis perhaps lengthening the lifetime of ammonia in the atmosphere-ocean system. Overall, since our \hl{photochemical-climate} model neglects NH$_3$ ocean dissolution and likely overestimates NH$_3$ photolysis, then we probably underestimate the lifetime of NH$_3$ in Figure \ref{fig:figure4}.

While HCN and HCCCN are produced in Figure \ref{fig:figure4}, the surface temperature would be $\sim 390$ K primarily caused by H$_2$-H$_2$ collision-induced absorption (CIA), which has a significant greenhouse effect in thick H$_2$ atmospheres like this one of 8.5 bars total pressure. The atmosphere cools to $\sim 300$ K after H$_2$ escapes to space.

\begin{figure}
  \centering
  \includegraphics[width=1.0\textwidth]{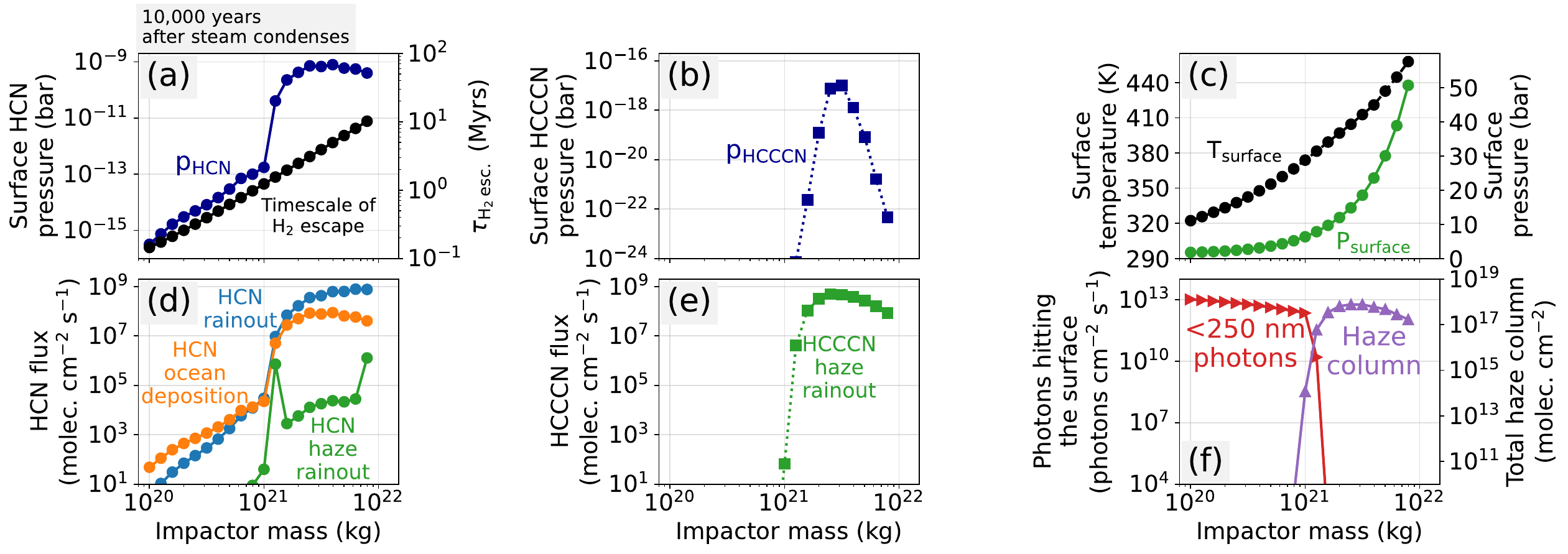}
  \caption{The state of the Hadean post-impact atmospheres 10,000 years after steam condenses to an ocean. This time period is within Phase 3 of a post-impact atmosphere indicated in Figure \ref{fig:impact_diagram}. The simulations assume the Table \ref{tab:fiducial_parameters} parameters and gas-phase reactions during the cooling steam atmosphere. (a) The surface HCN pressure and timescale of H$_2$ escape, which can be interpreted as the approximate duration of HCN and HCCCN production. (b) The HCCCN surface pressure. (c) The surface temperature and pressure. (d) The HCN deposition rate in the ocean and the rate HCN leaves the atmosphere in rain drops. ``HCN haze rainout'' is the rainout rate of a aerosol created via the reaction $\mathrm{H_2CN} + \mathrm{HCN} \rightarrow \mathrm{polymer}$. (e) The rainout rate of an aerosol formed from the reaction $\mathrm{C_4H} + \mathrm{HCCCN} \rightarrow \mathrm{polymer}$. (f) The $< 250$ nm photons hitting the surface, and the total hydrocarbon haze column abundance. Impactors larger than $10^{21}$ kg produce haze-rich atmospheres and a stepwise increase in HCN and HCCCN production.}
  \label{fig:figure5}
\end{figure}

Figure \ref{fig:figure5} applies our model to various impactor masses. The results show the Hadean atmosphere 10,000 years after the post-impact generated steam atmosphere has condensed to an ocean. We choose 10,000 years after ocean condensation because this is adequate time for the atmosphere to reach a quasi-photochemical steady-state \hl{that} does not change \hl{significantly} until hydrogen escapes (e.g. Figure \ref{fig:figure4}). Figure \ref{fig:figure5}d and \ref{fig:figure5}e show a sharp increase in the HCN and HCCCN production for impactors larger than $10^{21}$ kg ($\sim 780$ km). Such large impacts generate $\mathrm{CH_4}/\mathrm{CO_2} > 0.1$ (Figure \ref{fig:figure2}), which makes a thick Titan-like haze \citep{Trainer_2006}. Haze shielding causes CH$_4$ photolysis to be higher in the atmosphere and closer to N$_2$ photolysis, therefore the photolysis products of both species can more efficiently combine to make cyanides (Appendix Figure \ref{fig:ch4_n_hcn}). Additionally, HCCCN production requires acetylene (Reaction \ref{eq:hcccn}), which is a haze precursor that accumulates when $\mathrm{CH_4}/\mathrm{CO_2} > 0.1$. These Titan-like atmospheres have $\sim 10^{-9}$ bar surface HCN, and HCN ocean deposition and rainout rates between $10^7$ and $10^9$ HCN molecules cm$^{-2}$ s$^{-1}$ persisting on hydrogen escape timescales ($> 1$ million years). HCCCN is incorporated into aerosols before raining out to the surface at a rate of up to $10^9$ HCCCN molecules cm$^{-2}$ s$^{-1}$.

In addition to photochemistry, lightning should also generate HCN \citep{Chameides_1981,Stribling_1987}. Appendix Figure \ref{fig:figure5_lightning} shows HCN production from lighting for the same time period as the Figure \ref{fig:figure5} simulation using methods described in \citet{Chameides_1981}. Assuming the same lightning dissipation rate as modern Earth's, we find that lightning produces up to $\sim 10^{4}$ HCN molecules cm$^{-2}$ s$^{-1}$. This value is small compared to the $10^{7}$ - $10^{9}$ HCN molecules cm$^{-2}$ s$^{-1}$ produced from photochemistry after $> 10^{21}$ kg impacts.

Larger impacts generate a thicker H$_2$ atmosphere which make the atmosphere warmer (Figure \ref{fig:figure5}c). For impactors $> 10^{21}$ kg, which generate substantial HCN and HCCCN, the surface temperature is $> 380$ K. Figure \ref{fig:figure5}f shows that impactors that produce substantial haze shield the surface from $< 250$ nm photons, which means that prebiotic schemes that require high energy UV light \citep[e.g.,][]{Patel_2015} would need to rely on stockpiling of the nitriles for later use. 

The Hadean Earth CO$_2$ concentration is uncertain. Models of the Hadean geologic carbon cycle argue for CO$_2$ levels between $\sim 10^{-5}$ and $1$ bar at 4 Ga with a median value of $\sim 0.5$ bar and a 95\% uncertainty spanning $10^{-5}$ to $1$ bar \citep{Kadoya_2020}. However, these values might be unrealistically small because a large impact would warm surface rocks possibly causing carbonates to degass thereby increasing the atmospheric CO$_2$ reservoir. Up to $\sim 80$ bars of CO$_2$ may potentially be liberated from surface carbonates \citep{Krissansen_2021}.

\begin{figure}
  \centering
  \includegraphics[width=1.0\textwidth]{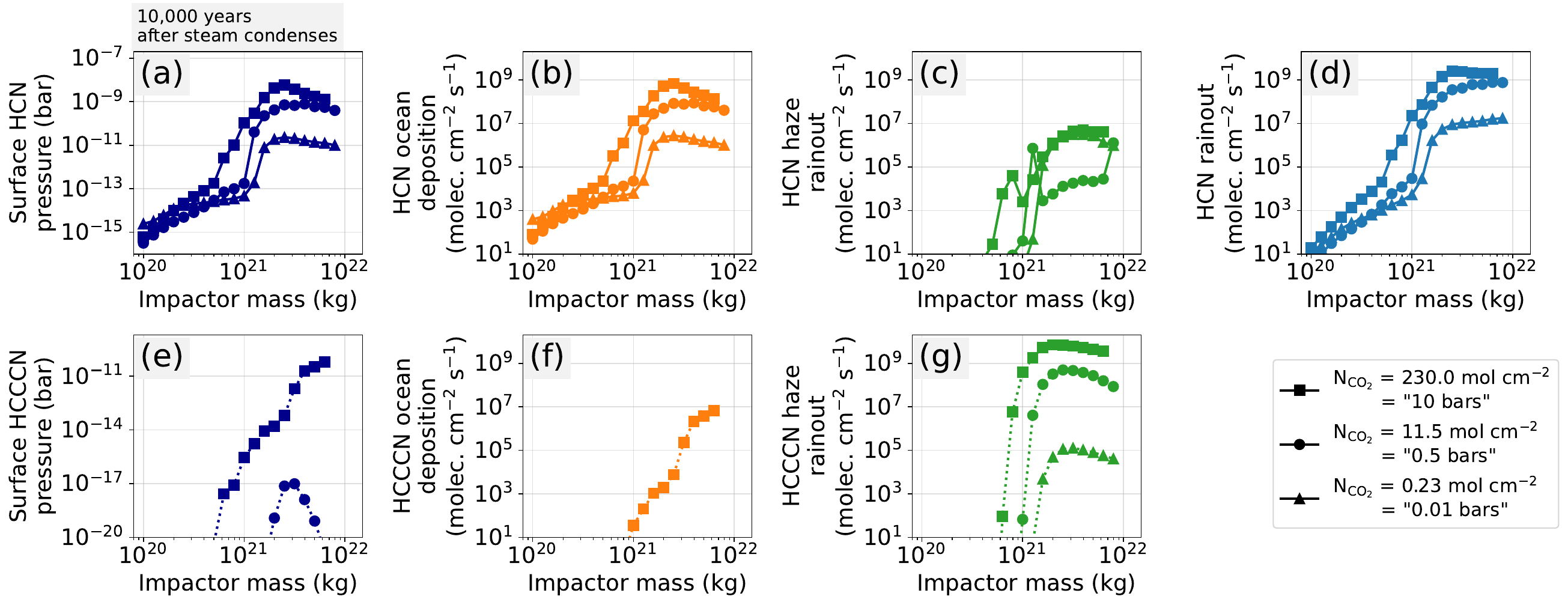}
  \caption{The effect of the pre-impact CO$_2$ abundance on HCN and HCCCN production in post-impact atmospheres. All values are for the atmosphere 10,000 years after the steam condenses to an ocean, which is within Phase 3 of a post-impact atmosphere (Figure \ref{fig:impact_diagram}). The simulations assume the Table 1 parameters, except vary the pre-impact CO$_2$ inventory between 0.01 bar (triangles), 0.5 bar (circles), and 10 bars (squares). The calculations use gas-phase chemistry during the cooling steam atmosphere. Panels (a) - (d) show the surface HCN abundance and fluxes, while (e) - (g) show HCCCN production. Our model assumes that HCCCN is not soluble in water and does not rainout, therefore we omit a panel showing HCCCN rainout. Prebiotic nitrile production is directly correlated with the pre-impact CO$_2$ inventory.}
  \label{fig:figure6}
\end{figure}

Figure \ref{fig:figure6} explores the effect of different pre-impact CO$_2$ abundances on HCN and HCCCN production in post-impact atmospheres. The simulations are snapshots of the atmosphere 10,000 years after the impact-vaporized steam has condensed to an ocean. Larger pre-impact CO$_2$ causes larger HCN and HCCCN production because it allows more CH$_4$ to form in the cooling steam atmosphere. As discussed previously, CH$_4$ is closely tied to photochemical cyanide generation. Regardless of the pre-impact CO$_2$ concentrations, HCN and HCCCN production sharply increases for impactors larger than $\sim 10^{21}$ kg due to more efficient haze production (Figure \ref{fig:figure5}, and corresponding text).

\begin{figure}
  \centering
  \includegraphics[width=1.0\textwidth]{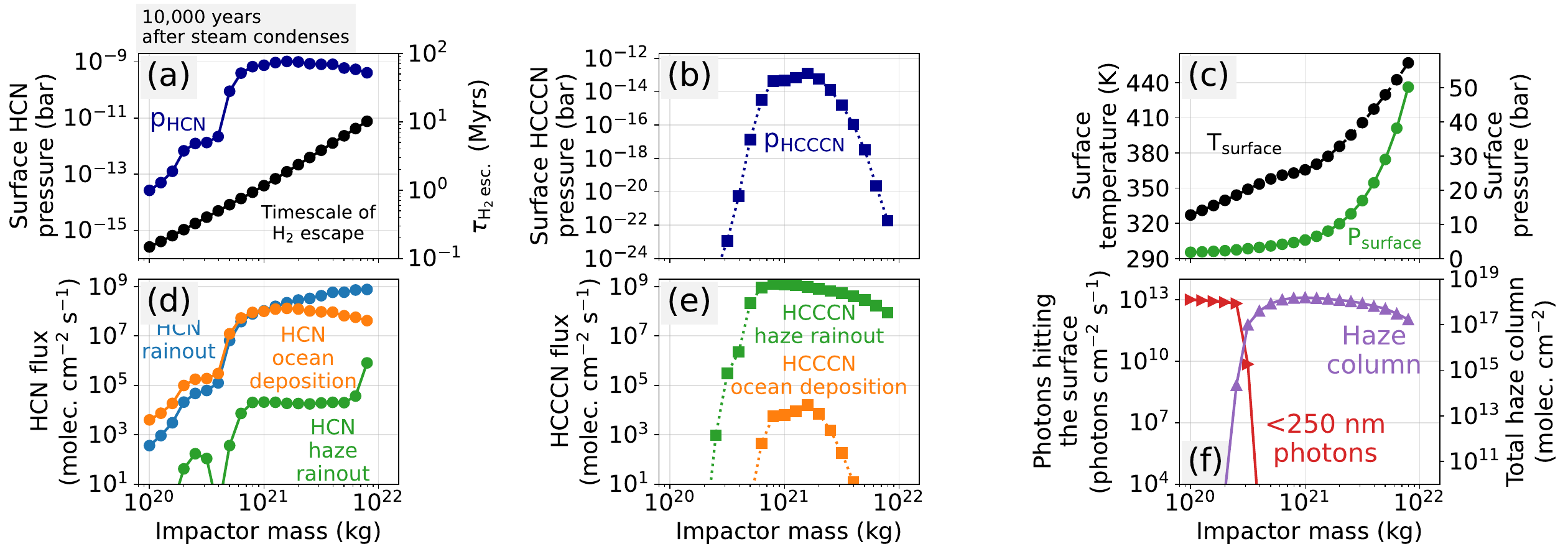}
  \caption{Identical to Figure \ref{fig:figure5}, except simulations account for nickel-surface reactions which catalyze methane production as the steam atmosphere cools \citep{Schmider_2021}. We assume a nickel surface area of 10 cm$^2$ nickel / cm$^2$ Earth (for context, see Figure \ref{fig:figure3}). Nickel catalysts cause more efficient CH$_4$ generation, permitting bigger HCN and HCCCN production for smaller impactors compared to the gas-phase only scenario (Figure \ref{fig:figure5}).}
  \label{fig:figure7}
\end{figure}

Figure \ref{fig:figure7} shows the state of the atmosphere after impacts of various size assuming 10 cm$^2$ nickel / cm$^2$ Earth is present in the steam atmosphere to catalyze methane production. The nickel causes more efficient conversion of CO$_2$ to CH$_4$ compared to the gas-phase only scenario (Figure \ref{fig:figure5}) permitting \hl{greater production of} HCN and HCCCN for smaller impactors. For example, a $5 \times 10^{20}$ kg ($\sim 610$ km) impactor which accounts for nickel catalysts (Figure \ref{fig:figure7}) has comparable HCN and HCCCN production to a $1.6 \times 10^{21}$ kg ($\sim 900$ km diameter) impactor if no nickel catalysts are assumed in the cooling steam atmosphere (Figure \ref{fig:figure5}).

A critical assumption in this section is that 100\% of the iron delivered by impactors reacts with steam to generate H$_2$. As discussed in Section \ref{sec:phase1}, it is possible that the post-impact atmosphere is less thoroughly reduced by impactor iron. Appendix Figures \ref{fig:figure5_citron} and \ref{fig:figure7_citron} recalculate main text Figures \ref{fig:figure5} and \ref{fig:figure7}, assuming that a fraction (approximately 15\% to 30\%) of impactor iron reduces the steam atmosphere based on SPH simulations (``Model 1B'' in Figure \ref{fig:melt_reaction_sup}). This alternative assumption requires that impactors $\sim 5$ times more massive are required to generate a haze-rich post-impact atmosphere with copious HCN and HCCCN production. For example, in Figure \ref{fig:figure5}, recall that there is a sharp increase in cyanide production for impactors larger than $10^{21}$ kg ($\sim 780$ km). Appendix Figure \ref{fig:figure5_citron}, which instead assumes a fraction of iron reduces the steam atmosphere, finds that the sharp increase in cyanide production occurs for impacts larger than $5 \times 10^{21}$ kg ($\sim 1330$ km). However, the presence of nickel catalysts may permit large prebiotic nitrile production for smaller impactors, even under pessimistic post-impact H$_2$ generation (Appendix Figure \ref{fig:figure7_citron}).

\section{Discussion}

\subsection{Comparison to previous work}

Recently, \citet{Zahnle_2020} performed calculations of post-impact atmospheres using simpler models than the ones used in this article. Our results differ in several important ways. First, we find that our purely gas-phase model of the post-impact steam atmosphere (Section \ref{sec:phase2}) predicts less CH$_4$ generation than the model used in \citet{Zahnle_2020}. For example, Figure \ref{fig:figure2} predicts that most CO$_2$ is converted to CH$_4$ for impactors larger than $1.6 \times 10^{21}$ kg. Figure 2 (top panel) in \citet{Zahnle_2020}, which is a comparable scenario, suggests a $5 \times 10^{20}$ kg impactor is required to convert most of the atmospheric CO$_2$ to CH$_4$. The difference is likely caused by different approaches to computing CH$_4$ quenching, or freeze-out, as the atmosphere cools. Our kinetics-climate model automatically computes CH$_4$ quenching by tracking the elementary reactions producing and destroying CH$_4$ along with many other atmospheric species. In most of our simulations of cooling post-impact atmospheres, CH$_4$ quenches when the temperature is between $900$ and $1000$ K. \citet{Zahnle_2020} instead used equilibrium chemistry modeling with a parameterization for CH$_4$ quenching derived from kinetics calculations of H$_2$-dominated brown dwarf atmospheres \citep{Zahnle_2014}. This parameterization predicts $\sim 800$ K CH$_4$ quenching temperatures. The different quenching temperatures between our model and the \citet{Zahnle_2020} model suggests that the \citet{Zahnle_2020} kinetics parameterization is likely not suitable for a cooling steam-rich atmosphere.

The new photochemical model predicts longer post-impact CH$_4$ lifetimes than the \citet{Zahnle_2020} model. As mentioned previously, \citet{Zahnle_2020} included CH$_4$ photolysis, but neglected Reaction \ref{eq:ch4_recom}, which efficiently recombine photolysis products in hydrogen-rich atmospheres. In our model, these recombination reactions allow CH$_4$ to persist in most post-impact atmospheres until hydrogen escapes to space ($\sim $ millions of years). \citet{Zahnle_2020} instead finds that CH$_4$ is eradicated from the atmosphere before hydrogen escape. 

Finally, nitrile production and rainout in our new model depend strongly on the presence of haze and the $\mathrm{CH_4}/\mathrm{CO_2}$ ratio, which was not the case in \citet{Zahnle_2020}. Our model finds that up to $\sim 10^9$ molecules cm$^{-2}$ s$^{-1}$ HCN and HCCCN is rained out in hazy post-impact atmospheres with $\mathrm{CH_4}/\mathrm{CO_2} > 0.1$ (Figure \ref{fig:figure5}). When $\mathrm{CH_4}/\mathrm{CO_2} < 0.1$, there is little haze, and HCN production is $< \sim 10^5$ molecules cm$^{-2}$ s$^{-1}$ and HCCCN production is negligible. Haze causes CH$_4$ and N$_2$ photolysis products to be close in altitude so that they efficiently react to make cyanides (Appendix Figure \ref{fig:ch4_n_hcn}). Additionally, HCCCN generation requires C$_2$H$_2$ in our model (Reaction \ref{eq:hcccn}), which is only abundant in hazy atmospheres. In contrast, \citet{Zahnle_2020} finds that cyanide production rate in post-impact atmospheres is $10^8$ to $10^{10}$ molecules cm$^{-2}$ s$^{-1}$ regardless of the presence of haze and the $\mathrm{CH_4}/\mathrm{CO_2}$ ratio. Our results differ largely because our model is 1-D (has vertical transport), while the \citet{Zahnle_2020} is a zero dimensional box model. HCN production depends on the proximity of CH$_4$ and N$_2$ photolysis, but a box model cannot account for this 1-D effect. Furthermore, \citet{Zahnle_2020} does not distinguish between different prebiotic nitriles (e.g. HCN and HCCCN), or determine their surface concentrations and rainout rates. Also, \citet{Zahnle_2020} does not have a coupled climate model.

Cometary and lightning sources of HCN are relatively small compared to our estimated photochemical production rates in haze-rich post-impact atmospheres. \citet{Todd_2020} calculated that comets could deliver $\sim 1.8 \times 10^{5}$ HCN molecules cm$^{-2}$ s$^{-1}$ to the Hadean Earth, a value $\sim 4$ orders of magnitude smaller than HCN from photochemistry in our most optimistic models. As discussed in Section \ref{sec:phase3}, we find that HCN production from lightning in post-impact atmospheres to be at most $\sim 10^{4}$ HCN molecules cm$^{-2}$ s$^{-1}$ which is also small compared to UV photochemistry in a CH$_4$ rich atmosphere. This result agrees with \citet{Pearce_2022}, who also finds that lightning-produced HCN is relatively insignificant.

\citet{Rimmer_2019b} suggested that localized ultra-reducing magma rich in carbon and nitrogen might outgas HCN and HCCCN. They imagine this gas interacting with subsurface water causing high concentrations of dissolved prebiotic molecules, and therefore a setting for origin of life chemistry. While this idea may have merit, their calculations do not account for graphite saturation in magma, which may inhibit outgassing of reduced carbon-bearing species, like HCN \citep{Hirschmann_2008,Wogan_2020,Thompson_2022}. Additionally, \citet{Rimmer_2019b} did not self-consistently account for the solubility of gases in magma, which has been been hypothesized to prevent the outgassing of H-bearing gases, like CH$_4$ or HCN \citep{Wogan_2020}. Therefore, we argue that a hypothesized volcanic source of HCN and HCCCN requires further modeling and experiments before it can be compared to a photochemical source, but, in general, seems challenging.

\hl{\mbox{\citet{Cerrillo_2022}} recently used a climate model to predict the surface temperature of post-impact atmospheres with compositions predicted by \mbox{\citet{Zahnle_2020}}. They find surface temperatures $> 600$ K in some cases. However, the \mbox{\citet{Cerrillo_2022}} calculations do not include the effects of water vapor on the lapse rate in the troposphere. Latent heat from water condensation alters convection in the troposphere, greatly reducing the lapse rate when compared to a dry lapse rate. The result is a much cooler surface. Our climate calculations include the effects of water vapor on the lapse rate, which is why we predict surface temperatures $\lesssim 500$ K, even in the wake of a $7.9 \times 10^{21}$ kg impact in Figure \mbox{\ref{fig:figure5}}. All our climate simulations of post-impact atmospheres allow surface liquid water.}

\hl{If a post-impact atmosphere of 2000 mol cm$^{-2}$ H$_2$ is all lost to space ($\sim 4$ bar pure H$_2$ atmosphere, or $\sim 13\%$ of the H$_2$ in an ocean), that would shift D/H of the ocean heavier by 1.4\% by hydrodynamic escape and Rayleigh fractionation (following Equation (16) in \mbox{\citet{Zahnle_2019}} using an escape fractionation factor $\sim 0.9$ appropriate for an atmosphere where H$_2$ dominates over CO$_2$). This may be an underestimate of the D/H shift because the immediate post-impact oxidation of iron by steam probably produce H$_2$ with a lower concentration of D than the steam that condense into an ocean. Experiments show isotopic fractionation in reaction of iron powder with steam at low temperatures \mbox{\citep{Smith_1957}}, but we are unaware of high temperature experiments corresponding to post-impact conditions. In any case, cumulative big impacts during the Hadean that created highly reducing atmospheres would be expected to increase oceanic D/H additively, raising the ocean D/H from starting values that were ten \mbox{\citep{Piani_2020}} to tens of percent \mbox{\citep{Alexander_2012}} lighter than the modern ocean. Such an evolution with intermittent hydrogen escape in the Hadean is consistent both with D/H constraints and with the xenon isotope record \mbox{\citep{Avice_2018}}, in which ionic xenon is dragged out to space by early hydrogen escape and the distribution of xenon isotopes becomes heavier \mbox{\citep{Zahnle_2019}}.}

\subsection{Origin of life setting and stockpiling of cyanides}

The Hadean Earth may have had less land but was likely speckled with hot-spot volcanic islands similar to modern-day Hawaii \citep{Bada_2018}, and possibly had continental land \citep{Korenaga_2021} where nitriles could accumulate. The majority of HCCCN and HCN produced in post-impact atmospheres would dissolve or rainout into the ocean where it would be diluted and gradually removed by hydrolysis reactions \citep{Miyakawa_2002} or complexation with dissolved ferrous iron \citep{Keefe_1996}. However, some of the nitriles would be deposited in lakes or ponds on land. We consider, first, equilibrium with atmospheric $p_\mathrm{HCN}$ and, second, time-integrated deposition.

Nitrile concentrations in waterbodies on land in equilibrium with the atmosphere according to Henry's law would be too small to participate in prebiotic schemes that form ribonucleotides. Our models predict HCN surface pressures up to $10^{-9}$ bar (Figure \ref{fig:figure5}). For a warm 373 K pond, Henry's law predicts the dissolved HCN concentration is $4 \times 10^{-11}$ mol L$^{-1}$. Yet, $\sim 0.01$ mol L$^{-1}$ HCN is required for polymerization \citep{Sanchez_1967} and published prebiotic schemes can use 1 mol L$^{-1}$ HCN \citep{Patel_2015}.

Additionally, while nitriles are produced in post-impact atmospheres, waterbodies on land would likely be too warm for prebiotic chemistry. In the Figure \ref{fig:figure4} simulation, substantial HCN and HCCCN production occurs in the aftermath of big impacts when the surface temperature is $\sim 390$ K caused by a H$_2$-H$_2$ CIA greenhouse. Nickel catalysts permit big HCN and HCCCN production for surface temperatures as small as $\sim 360$ K (Figure \ref{fig:figure7}). Nucleotide building blocks are fragile at such hot temperatures and conditions may not be conducive to an RNA world \citep{Bada_2002}. 

We propose that cyanides produced in hot post-impact atmospheres may instead be preserved, stockpiled, and concentrated, and used in prebiotic schemes at a later time when the climate is colder. Cyanide rainout and stockpiling could occur for millions of years until HCN production is halted by H$_2$ escape to space (Figure \ref{fig:figure4}). \hl{For example, if HCN rains out at $10^{9}$ molecules cm$^{-2}$ s$^{-1}$ over one million years (Figure \mbox{\ref{fig:figure5}}), then $\sim 1.4$ g cm$^{-2}$ HCN could be stockpiled assuming all molecules are preserved.} Once H$_2$ escapes, the surface temperature would drop to $\sim 300$ K (Figure \ref{fig:figure4}), and over longer timescales the carbonate-silicate cycle might settle on even colder climates because impact ejecta promotes CO$_2$ sequestration \citep{Kadoya_2020}. In this cold climate, cyanide stockpiled into salts could be released as HCN or CN$^-$ into water bodies on land because of rehydration, volcanic or impact heating \citep{Patel_2015,Sasselov_2020}, or UV exposure \citep{Todd_2022}. Liberation of cyanide could enable the prebiotic schemes that make RNA.

\citet{Toner_2019} investigated a mechanism for stockpiling cyanides. Their thermodynamic calculations show that HCN can be preserved as ferrocyanide salts in evaporating carbonate-rich lakes. However, the \citet{Toner_2019} numerical experiments were at 273 K and 298 K, which are far colder environments than the $> 360$ K surface temperatures that coincide with large HCN production in post-impact atmospheres (Figure \ref{fig:figure7}). Although, \citet{Toner_2019} did not address stockpiling of HCCCN, cyanoacetylene can be captured by 4,5-dicyanoimidazole (DCI), a byproduct of adenine synthesis, to make crystals of 4,5-dicyanoimidazole (CV-DCI) \citep{Ritson_2022} and it is possible that other capture mechanisms are yet to be discovered. Overall, the feasibility of stockpiling prebiotic nitriles in post-impact conditions requires further geochemical modeling and experiments.

\subsection{Impactor size and the likelihood of the origin of life}

\begin{figure}
  \centering
  \includegraphics[width=0.4\textwidth]{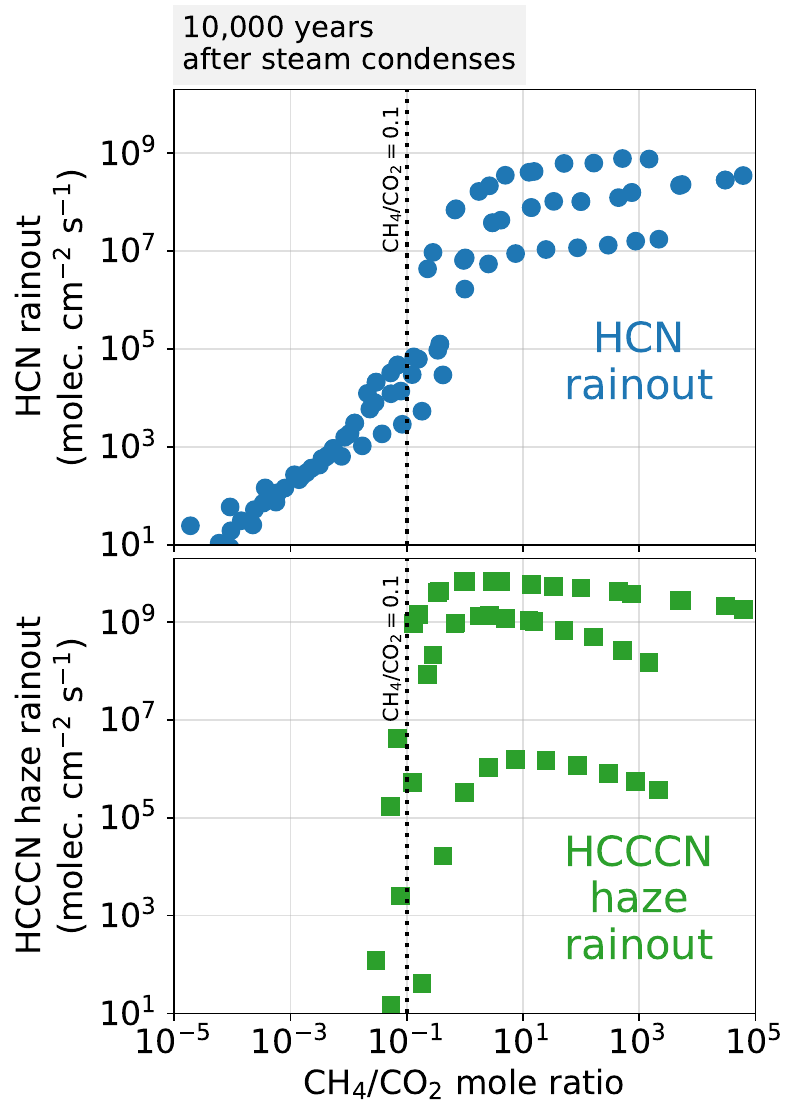}
  \caption{HCN rainout and HCCCN haze rainout as a function of the $\mathrm{CH_4}/\mathrm{CO_2}$ mole ratio in all simulated post-impact atmospheres. The figure considers simulations shown in Figure \ref{fig:figure5}, \ref{fig:figure6}, \ref{fig:figure7}, \ref{fig:figure5_citron} and \ref{fig:figure7_citron}. All values are for the atmosphere 10,000 years after steam has condensed to an ocean. The HCN and HCCCN haze production is significantly larger for atmospheres with $\mathrm{CH_4}/\mathrm{CO_2} \gtrsim 0.1$.}
  \label{fig:figure7_5}
\end{figure}

We hypothesize that $\mathrm{CH_4}/\mathrm{CO_2} > 0.1$ might be an important threshold required for a post-impact atmosphere to produce useful concentrations of nitriles for origin of life chemistry. Figure \ref{fig:figure7_5} shows HCN and HCCCN haze rainout as a function of the atmospheric $\mathrm{CH_4}/\mathrm{CO_2}$ mole ratio for every post-impact simulation in this article. When $\mathrm{CH_4}/\mathrm{CO_2} > 0.1$, the atmosphere is hazy, and HCN and HCCCN are delivered to the surface at a rate of up to $\sim 10^9$ molecules cm$^{-2}$ s$^{-1}$. In contrast, atmospheres with $\mathrm{CH_4}/\mathrm{CO_2} < 0.1$ rainout less than $10^5$ HCN molecules cm$^{-2}$ s$^{-1}$ and have surface HCN concentrations less than $10^{-13}$ bar (Figure \ref{fig:figure5}). Such small HCN concentrations may be challenging to stockpile as ferrocyanides \citep{Toner_2019}. Additionally, modeled atmospheres with $\mathrm{CH_4}/\mathrm{CO_2} < 0.1$ produce negligible HCCCN, yet the molecule is required in prebiotic schemes to synthesize pyrimidine (cytosine and uracil) nucleobase precursors to RNA \citep{Powner_2009,Okamura_2019,Becker_2019}.

The impactor mass required to generate an atmosphere with $\mathrm{CH_4}/\mathrm{CO_2} > 0.1$ is uncertain. Our optimistic model, which considers the effect of nickel-catalyzed methane production, requires a $> 4 \times 10^{20}$ kg ($> 570$ km) impactor (Figure \ref{fig:figure7}). The lunar cratering record and abundance of highly siderophile elements Earth's mantle imply that between 4 and 7 such impacts occurred during the Hadean \citep{Marchi_2014, Zahnle_2020}. Our least optimistic model needs a $> 5 \times 10^{21}$ kg ($> 1330$ km) impact to create a post-impact atmosphere with $\mathrm{CH_4}/\mathrm{CO_2} > 0.1$ because it assumes only a fraction of iron delivered to Earth reacts with the ocean to create atmospheric H$_2$ (Appendix Figure \ref{fig:figure5_citron}). The Hadean only experienced 0 to 2 collisions this large \citep{Zahnle_2020}. The precise minimum impactor mass to make an atmosphere with $\mathrm{CH_4}/\mathrm{CO_2} > 0.1$ depends on the importance of atmospheric equilibration with a melt pond (Section \ref{sec:phase1}, and \citet{Itcovitz_2022}), the fraction of impactor iron that reduces the atmosphere, and the effect of Nickel and other surface catalysts on CH$_4$ kinetics.

An additional consideration is that any progress toward the origin of life caused by an impact could be erased by a subsequent impact that sterilizes the planet. For example, suppose a $> 500$ km impact that vaporizes the ocean sterilizes the globe \citep{Citron_2022}. With our most pessimistic calculations for post-impact CH$_4$ generation a $> 1330$ km ($> 5 \times 10^{21}$ kg) impact is required to create an atmosphere that generates significant HCN and HCCCN. In this scenario, the last $> 1330$ km impact favorable for prebiotic chemistry would likely be followed by a $500$ to $1330$ km impact that would destroy any primitive life without rekindling it. Alternatively, our optimistic model for post-impact CH$_4$ generation only requires a $> 4 \times 10^{20}$ kg ($> 570$ km) impact to create an atmosphere with $\mathrm{CH_4}/\mathrm{CO_2} > 0.1$. In this case, the final $> 570$ km impact that might kickstart the origin of life is unlikely to be followed by a slightly smaller $500$ km to $570$ km sterilizing impact.

A caveat to the reasoning in the previous paragraph is that ocean-vaporization may not have sterilized the planet because microbes could have possibly survived in the deep subsurface \citep{Sleep_1989,Grimm_2018}. 

In summary, we suggest that $\mathrm{CH_4}/\mathrm{CO_2} > 0.1$ may be an important threshold for post-impact atmospheres to be conducive to the origin of life because they generate $> 4$ orders of magnitude larger surface HCN concentrations, and are the only modeled atmospheres capable of generating HCCCN. We find that the minimum impactor mass required to create a post-impact atmosphere with $\mathrm{CH_4}/\mathrm{CO_2} > 0.1$ is between $4 \times 10^{20}$ and $5 \times 10^{21}$ kg  ($570$ to $1330$ km). The value is uncertain because we do not know how effectively iron delivered by an impact reduces the atmosphere (Section \ref{sec:phase1}), the importance of atmospheric equilibration with a melt pond (Section \ref{sec:phase1}), and because it is hard to estimate a realistic surface area of nickel catalysts available during the cooling steam atmosphere (Section \ref{sec:phase2}).

\subsection{Model caveats and uncertainties}

\subsubsection{Hydrogen from crust-atmosphere reactions}

Perhaps the most significant caveat to the modeling effort described above is that we did not consider H$_2$ \hl{production} from reactions between a hot post-impact atmosphere and solid, non-melted crust. Section \ref{sec:phase1} explores impact H$_2$ made by two mechanisms: (1) reduction of the atmosphere by impact-derived iron and (2) atmospheric equilibration with a melt pond made by the impact. However, it is also conceivable that while the atmosphere is hot and  steam-rich in the $\sim 10^3$ years following an impact (i.e. Phase 2), water vapor could permeate through and react with the solid crust to produce H$_2$ by a process like serpentinization. Specifically, H$_2$O reduction by FeO in the solid crust could make H$_2$: 

\begin{equation} \label{eq:serp}
  \mathrm{H_2O} + 3 \mathrm{FeO} \rightarrow \mathrm{H_2} + \mathrm{Fe_3O_4}
\end{equation}

In our nominal model (Figure \ref{fig:figure2} and \ref{fig:figure5}), we require a post-impact atmosphere has $> 2 \times 10^3$ H$_2$ mol cm$^{-2}$ (i.e. the equivalent of converting 13\% of Earth's ocean to H$_2$) in order to reach a $\mathrm{CH_4}/\mathrm{CO_2} > 0.1$ and big nitrile production rates. Assuming a crustal FeO content of 8 wt\% \citep{Takahashi_1986}, then $2 \times 10^3$ H$_2$ mol cm$^{-2}$ could be produced by reacting water with FeO in the top $\sim 16$ km of Earth's lithosphere. The feasibility of extensive water-rock H$_2$ generation depends on the permeability of the crust and the pressure gradients driving subsurface fluid circulation. For example, low permeability rocks with slow water circulation may not permit serpentinization of the upper crust within $\sim 10^3$ years while the atmosphere is hot and steam-rich. A comprehensive model is out of the scope of this article, but if attainable, significant water-rock reactions might produce a thick H$_2$ atmosphere after relatively small impacts (e.g. $10^{20}$ kg) which favors a $\mathrm{CH_4}/\mathrm{CO_2} > 0.1$ and significant nitrile generation. 

Another possibility, which we do not investigate in detail, is that atmosphere-crust reactions occur in the immediate aftermath of a giant impact (i.e. Phase 1), rather than over $\sim 10^3$ as previously discussed. A large impact could produce a global ejecta blanket several kilometers thick of mixed hot water and rock. As water was vaporized to form a steam atmosphere, the water and rock slurry could chemically equilibrate, producing H$_2$.

\citet{Zahnle_2020} attempted to account for atmosphere-crust interaction by equilibrating the post-impact steam atmosphere (Phase 2) to a mineral redox buffer. For example, their Figure 5 assumes the atmosphere has a fixed oxygen fugacity set by the FMQ buffer at an assumed 650 K methane quench temperature. The calculation predicts most CO$_2$ is converted to CH$_4$ for impacts as small as $\sim 5 \times 10^{19}$ kg, but \citet{Zahnle_2020} did not determine whether such significant atmosphere-crust interaction is physically plausible.

\subsubsection{Climate} \label{sec:climate_uncertainty}

\hl{A shortcoming of this work} is that our climate model is relatively simple. Throughout the Results section, our climate code assumes an isothermal 200 K stratosphere, a saturated adiabatic troposphere (i.e. relative humidity, $\phi = 1.0$), and ignores clouds. However, many of our simulated post-impact atmospheres contain a hydrocarbon haze which should absorb sunlight and warm the stratosphere \citep{Arney_2016}. Also, in a hydrogen-dominated atmosphere, water vapor has a larger molecular weight compared to the background gas which could inhibit convection \citep{Leconte_2017} and perhaps cause low relative humidities. Furthermore, low-altitude clouds reflect sunlight and should cool a planet while high clouds have a greenhouse warming effect \citep{Goldblatt_2011}.

\begin{figure}
  \centering
  \includegraphics[width=0.4\textwidth]{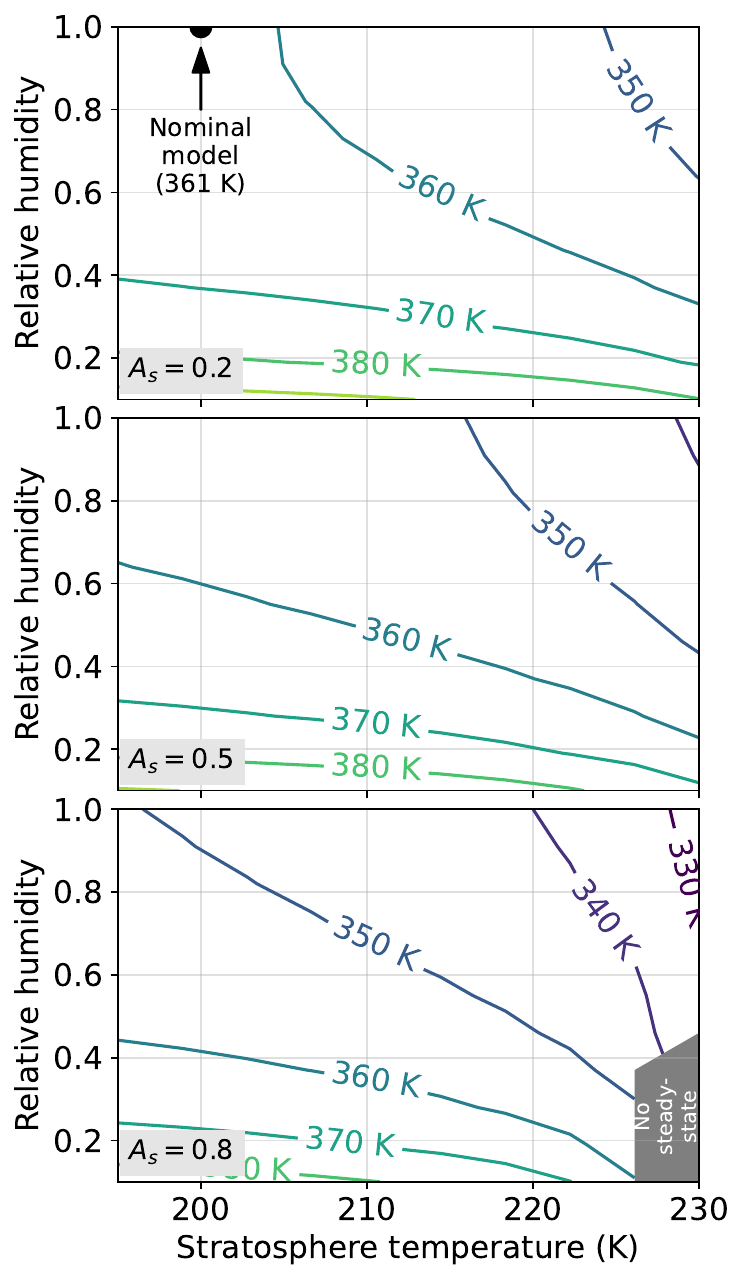}
  \caption{Surface temperature of a post-impact atmosphere as a function of stratosphere temperature, relative humidity, and surface clouds which we crudely approximate with the surface albedo ($A_b$). The atmosphere has 5.9 mol cm$^{-2}$ CO$_2$, 5.6 mol cm$^{-2}$ CH$_4$, 35.8 mol cm$^{-2}$ N$_2$, 556 mol cm$^{-2}$ H$_2$, 0.01 mol cm$^{-2}$ CO, 0.3 mol cm$^{-2}$ NH$_3$, and a liquid water ocean at the surface. This is the same composition of the atmosphere after a $5 \times 10^{20}$ kg impact in Figure \ref{fig:figure7} once steam has condensed to an ocean. The gray shaded region labeled ``no steady-state'' has no steady-state climate solutions that balance incoming shortwave and outgoing longwave energy. Uncertainties in our assumed stratosphere temperature, relative humidity and the effects of low-altitude clouds predict surface temperatures from $\sim 330$ K to $\sim 390$ K with a nominal value of 361 K.}
  \label{fig:figure8}
\end{figure}

Figure \ref{fig:figure8} attempts to show the uncertainty in our climate calculations as a function of three free parameters: stratosphere temperature, relative humidity, and low-altitude clouds which we crudely approximate by varying the surface albedo. The calculation uses the composition of the atmosphere after a $5 \times 10^{20}$ kg impact in Figure \ref{fig:figure7} immediately after the steam atmosphere has condensed to an ocean. Our nominal climate parameters ($T_\mathrm{strat} = 200$ K, $\phi = 1$, $A_s = 0.2$) predict a 361 K surface temperature. A warm stratosphere caused by hydrocarbon UV absorption and high albedo \hl{low altitude} clouds might cause the surface to be $\sim 30$ K colder than our nominal model, assuming water vapor is saturated. On the other hand, low relative humidities, which might be favored in convection-inhibited H$_2$ dominated atmospheres, \hl{increase the} troposphere \hl{lapse rate} which warms the surface \citep{Leconte_2017}. While Figure \ref{fig:figure8} gives a sense for the possible uncertainty in our climate calculations, it does not self-consistently simulate haze, relative humidity and clouds feedbacks. A more comprehensive model is required to resolve these nuances.

\hl{A further caveat is that our climate calculations ignore greenhouse warming from NH$_3$ (Table \mbox{\ref{tab:used_opacities}}). We choose to disregard the influence of NH$_3$ because a substantial fraction of the gas should dissolve in the ocean \mbox{\citep{Zahnle_2020}}, a process that our coupled photochemical-climate model cannot self-consistently account for. However, our climate model (Appendix \mbox{\ref{sec:clima}}), when uncoupled to photochemistry, can partition gases between the atmosphere and ocean according to gas solubility and ocean chemistry. Below, we use this stand-alone climate model to determine the climate affects of NH$_3$ in a post-impact atmosphere. NH$_3$ should dissolve into an ocean by henry's law, then hydrolyze to NH$_4^+$:}

\begin{gather}
  \mathrm{NH_3(g)} \leftrightarrow \mathrm{NH_3(aq)} \\
  \mathrm{NH_3(aq)} + \mathrm{H_2O} \leftrightarrow \mathrm{NH_4^+} + \mathrm{OH^-} \label{eq:nh3_hydro} \\
  \mathrm{H_2O} \leftrightarrow \mathrm{OH^-} + \mathrm{H^+} \label{eq:h2o_oh_h}
\end{gather}
\hl{Therefore, the concentration of aqueous NH$_3$ (in mol kg$^{-1}$) is given by $m_\mathrm{NH_3} = p_\mathrm{NH_3} \alpha_\mathrm{NH_3}$, where $p_\mathrm{NH_3}$ is the surface partial pressure of NH$_3$ in bars and $\alpha_\mathrm{NH_3}$ is the Henry's law constant (mol kg$^{-1}$ bar$^{-1}$). Reactions \mbox{\ref{eq:nh3_hydro}} and \mbox{\ref{eq:h2o_oh_h}} give the ammonium concentration to be $m_\mathrm{NH_4^+} = (K_9/K_{10}) m_\mathrm{NH_3} m_\mathrm{H^+}$, where $K_9$ and $K_{10}$ are equilibrium constants for each reaction. The Henry's law constant for NH$_3$ (in mol kg$^{-1}$ bar$^{-1}$) is $\alpha_\mathrm{NH_3} = 61 \exp(4200 (\frac{1}{T} - \frac{1}{298.15}))$ \mbox{\citep{Linstrom_1998}}. The equilibrium constants for Reactions \mbox{\ref{eq:nh3_hydro}} and \mbox{\ref{eq:h2o_oh_h}} are approximately $\log_{10}(K_9) = - 84.63 \exp(- 0.0161 T) - 4.05$ and $\log_{10}(K_{10}) = - 39.96 \exp(- 0.00639 T) - 8.06$. We derived both of these parameterizations using the SUPCRT thermodynamic database \mbox{\citep{Johnson_1992}}.}

\hl{We implement this ocean chemistry into our stand-alone climate model (Appendix \mbox{\ref{sec:clima}}), and compute the surface temperature after the $5 \times 10^{20}$ kg impact in Figure \mbox{\ref{fig:figure7}} once the steam atmosphere has condensed to an ocean. The mol cm$^{-2}$ of each gas are given in the Figure \mbox{\ref{fig:figure8}} caption. We use our nominal climate parameters ($T_\mathrm{strat} = 200$ K, $\phi = 1$, $A_s = 0.2$), assume the total H$_2$O reservoir is 1 modern ocean (15,000 mol cm$^{-2}$) with pH = 7, and account for the radiative affects of NH$_3$ in addition to the Table \mbox{\ref{tab:used_opacities}} opacities. The model predicts a 371 K surface temperature, which is 10 K warmer than calculations that do not include NH$_3$ opacities or ocean dissolution. 96\% of the ammonia reservoir is dissolved in the ocean.}

\hl{Ammonia has a more substantial effect on climate after larger impactors. Consider the atmosphere after a $7.9 \times 10^{21}$ kg impact in Figure \mbox{\ref{fig:figure3}} once steam has condensed to an ocean (0.008 mol cm$^{-2}$ CO$_2$, 11.5 mol cm$^{-2}$ CH$_4$, 32.6 mol cm$^{-2}$ N$_2$, 9122 mol cm$^{-2}$ H$_2$, 0.002 mol cm$^{-2}$ CO, and 6.8 mol cm$^{-2}$ NH$_3$). Our climate model, which includes NH$_3$ opacities and ocean dissolution, predicts a 505 K surface temperature with only 20\% of the NH$_3$ dissolved in the ocean because solubility decreases with increased temperature. This is 42 K hotter than our model that ignores NH$_3$ greenhouse contributions (Figure \mbox{\ref{fig:figure5}}). Overall, our climate calculations throughout most of this article perhaps underestimate the greenhouse warming by $10$ to $\sim 40$ K by ignoring NH$_3$ opacities, but instead may overestimate surface temperature because do not account for the cooling effects of haze and low altitude clouds (Figure \mbox{\ref{fig:figure8}}). Additional warming from NH$_3$ would only be relevant for a fraction of the post-impact atmosphere, before ammonia is destroyed by photolysis (Figure \mbox{\ref{fig:figure4}}).}

\subsubsection{Unknown chemical reactions and the effect of ions}

\hl{While our chemical scheme for HCCCN successfully reproduces the HCCCN abundances in Titan's atmosphere (Appendix Figure \mbox{\ref{fig:earth_titan_valid}}), it may lack many reactions relevant to post-impact atmospheres. Our sparse HCCCN network is necessary because currently few kinetic measurements are published in the literature.}

Our photochemical model does not include ion chemistry, which is likely a reasonable simplification because ions are not important for HCN or HCCCN formation on Titan \citep{Loison_2015}. Only some heavy hydrocarbons, like benzene (C$_6$H$_6$), rely on coupled neutral-ion chemistry to explain their observed abundances in Titan's atmosphere \citep{Horst_2017}.

\section{Conclusions}

We use atmospheric models to investigate the production of prebiotic feedstock molecules in impact-generated reducing atmospheres on the Hadean Earth, updating simpler calculations made by \citet{Zahnle_2020}. We find that massive asteroid impacts can generate temporary H$_2$-, CH$_4$- and NH$_3$-rich atmospheres, which photochemically generate HCN and HCCCN for the duration of hydrogen escape to space ($10^5$ to $10^7$ years). The production of nitriles increases dramatically for haze-rich atmospheres that have mole ratios of $\mathrm{CH_4}/\mathrm{CO_2} > 0.1$. In these cases, HCN can rain out onto land surfaces at a rate of $\sim 10^9$ molecules cm$^{-2}$ s$^{-1}$, and HCCCN incorporated in haze rains out at a similar rate. Atmospheres with $\mathrm{CH_4}/\mathrm{CO_2} < 0.1$ produce 3 to 4 orders of magnitude less HCN, and generate negligible HCCCN. The impactor mass required to create an atmosphere with $\mathrm{CH_4}/\mathrm{CO_2} > 0.1$ is uncertain and depends on how efficiently atmosphere-iron, atmosphere-melt and atmosphere-crust reactions generate H$_2$ and the surface area of nickel catalysts exposed to the cooling steam atmosphere. In an optimistic modeling scenario a $> 4 \times 10^{20}$ kg ($> 570$ km) impactor is sufficient, while in our least optimistic scenario a $> 5 \times 10^{21}$ kg ($> 1330$ km) impactor is required. 

We find that post-impact atmospheres that generate significant prebiotic molecules have $> 360$ K surface temperatures caused by a H$_2$-H$_2$ greenhouse which may be too hot for prebiotic chemistry, although the temperature may be cooler if reflective clouds occur. An alternative is that HCN and HCCCN generated in post-impact atmosphere are stockpiled. Cyanide can plausibly be stockpiled and concentrated in ferrocyanide salts and cyanoacetylene could be captured by byproducts of adenine synthesis into imidazole-based crystals \citep{Ritson_2022}. HCN and HCCCN can be used to create nucleotide precursors to RNA millions of years after the impact, once the H$_2$ has escaped to space, and the atmosphere has cooled to a more temperate state.

Nominally, the Hadean Earth appears to have experienced several impacts that would have produced an atmosphere that made significant prebiotic feedstock molecules. Like Earth, all rocky exoplanets accreted from impacts. Consequently, impact-induced reducing atmospheres may be a common planetary processes that provides windows of opportunity for the origin of exoplanet life.

\section*{Acknowledgements}

We thank Joshua Krissansen-Totton for numerous conversations that have improved the atmospheric models used in this article. Conversations with Maggie Thompson, Sandra Bastelberger, and Shawn Domagal-Goldman also helped us create the \emph{Photochem} and \emph{Clima} models. We also thank Eric Wolf for advice on computing reliable k-distributions for climate modeling. \hl{Additionally, we thank Paul Molli\`ere for conversations that helped us build \emph{Clima}. Finally, we thank our two anonymous reviewers for constructive feedback that improved this article.} N.F.W. and D.C.C. were supported by the Simon's Collaboration on Origin of Life Grant 511570 (to D.C.C.). Also, N.F.W., D.C.C., and K.J.Z. were supported by NASA Astrobiology Program Grant 80NSSC18K0829 and benefited from participation in the NASA Nexus for Exoplanet Systems Science research coordination network. N.F.W. and D.C.C. also acknowledge support from Sloan Foundation Grant G-2021-14194. R.L. and K.J.Z. were supported by NASA Exobiology Grant 80NSSC18K1082. \hl{R.L. was additionally supported by NASA XRP 80NSSC22K0953.}

\appendix
\renewcommand{\thefigure}{A\arabic{figure}}
\renewcommand{\thetable}{A\arabic{table}}

\setcounter{figure}{0}
\setcounter{table}{0}

\begin{figure}
  \centering
  \includegraphics[width=0.8\textwidth]{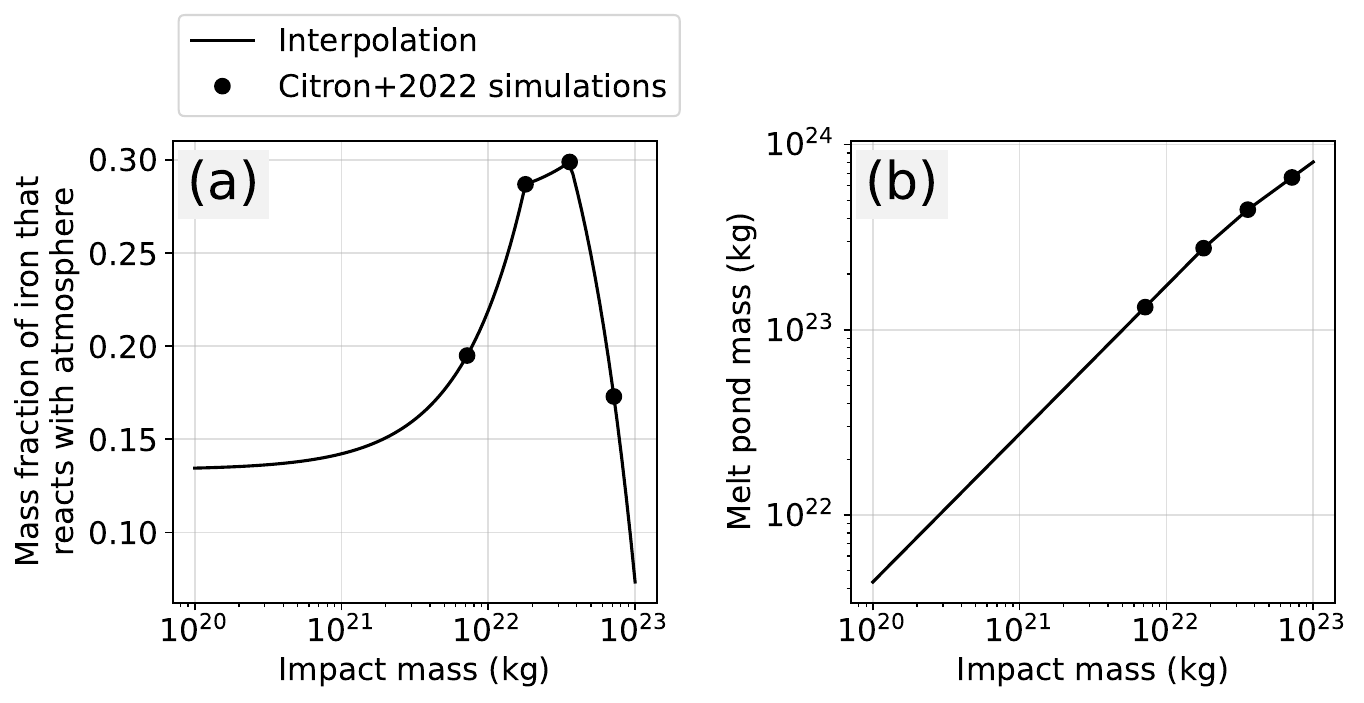}
  \caption{Interpolation and extrapolation of the \citet{Citron_2022} SPH simulations of impacts that collide with Earth at a $45^\circ$ angle and with a velocity of twice Earth's escape velocity. (a) is the mass fraction of iron that reacts with the atmosphere and (b) is the mass of the melt pool produced by an impact.  These interpolations are relevant to Figure \ref{fig:melt_reaction_sup} in the main text, and Appendix Figures \ref{fig:figure2_citron}, \ref{fig:figure3_citron}, \ref{fig:figure5_citron}, and \ref{fig:figure7_citron}.}
  \label{fig:citron_interpolations}
\end{figure}

\begin{figure}
  \centering
  \includegraphics[width=0.55\textwidth]{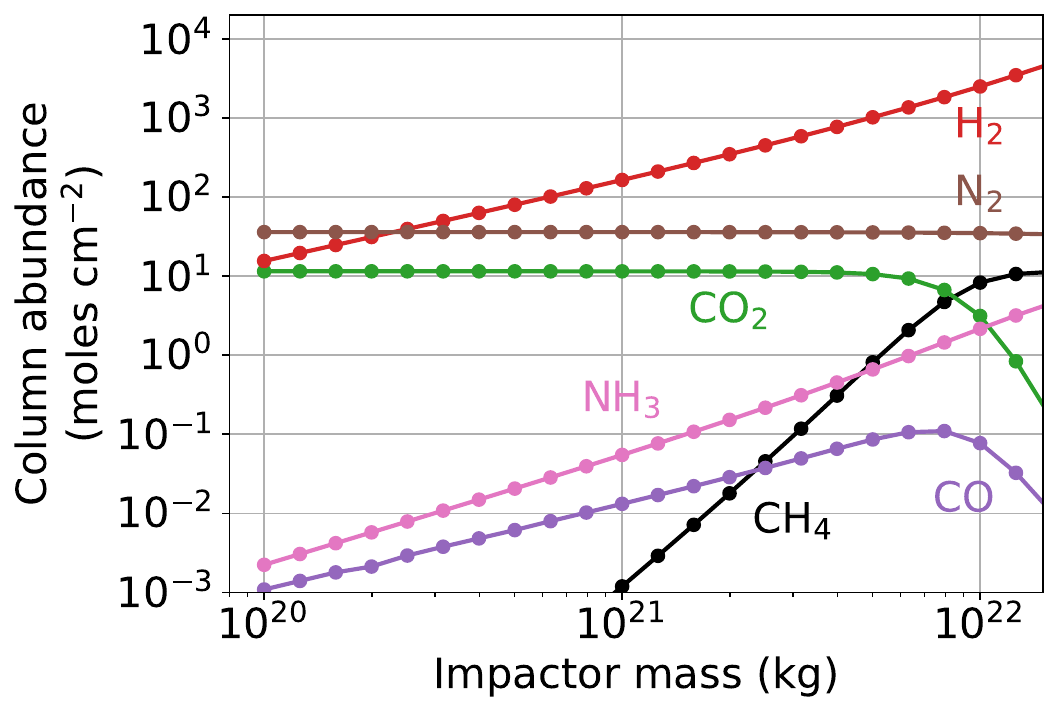}
  \caption{Identical to Figure \ref{fig:figure2} in the main text, but instead assumes post-impact H$_2$ generation if governed by ``Model 1B'' described in Figure \ref{fig:melt_reaction_sup}.}
  \label{fig:figure2_citron}
\end{figure}

\begin{figure}
  \centering
  \includegraphics[width=0.7\textwidth]{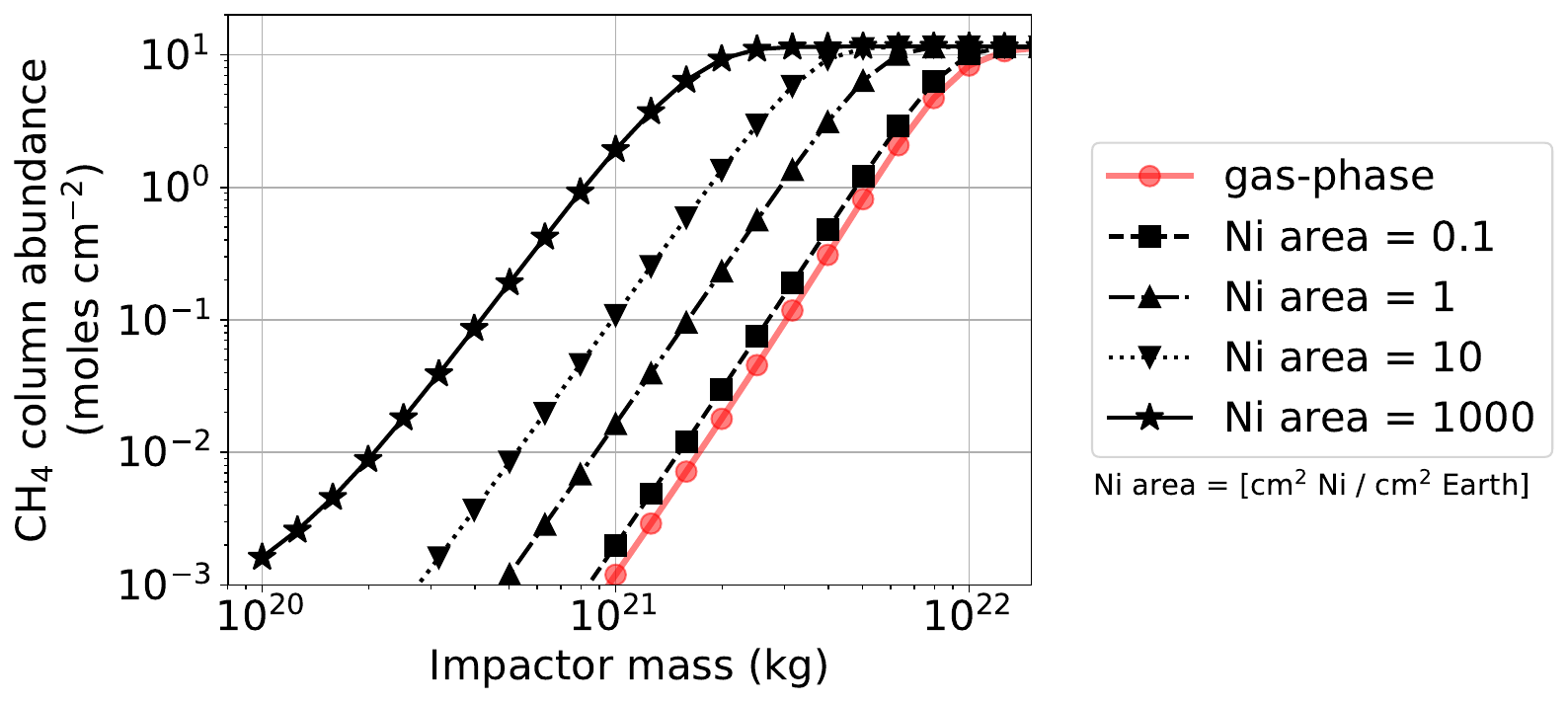}
  \caption{Identical to Figure \ref{fig:figure3} in the main text, but instead assumes post-impact H$_2$ generation if governed by ``Model 1B'' described in Figure \ref{fig:melt_reaction_sup}.}
  \label{fig:figure3_citron}
\end{figure}

\begin{figure}
  \centering
  \includegraphics[width=0.7\textwidth]{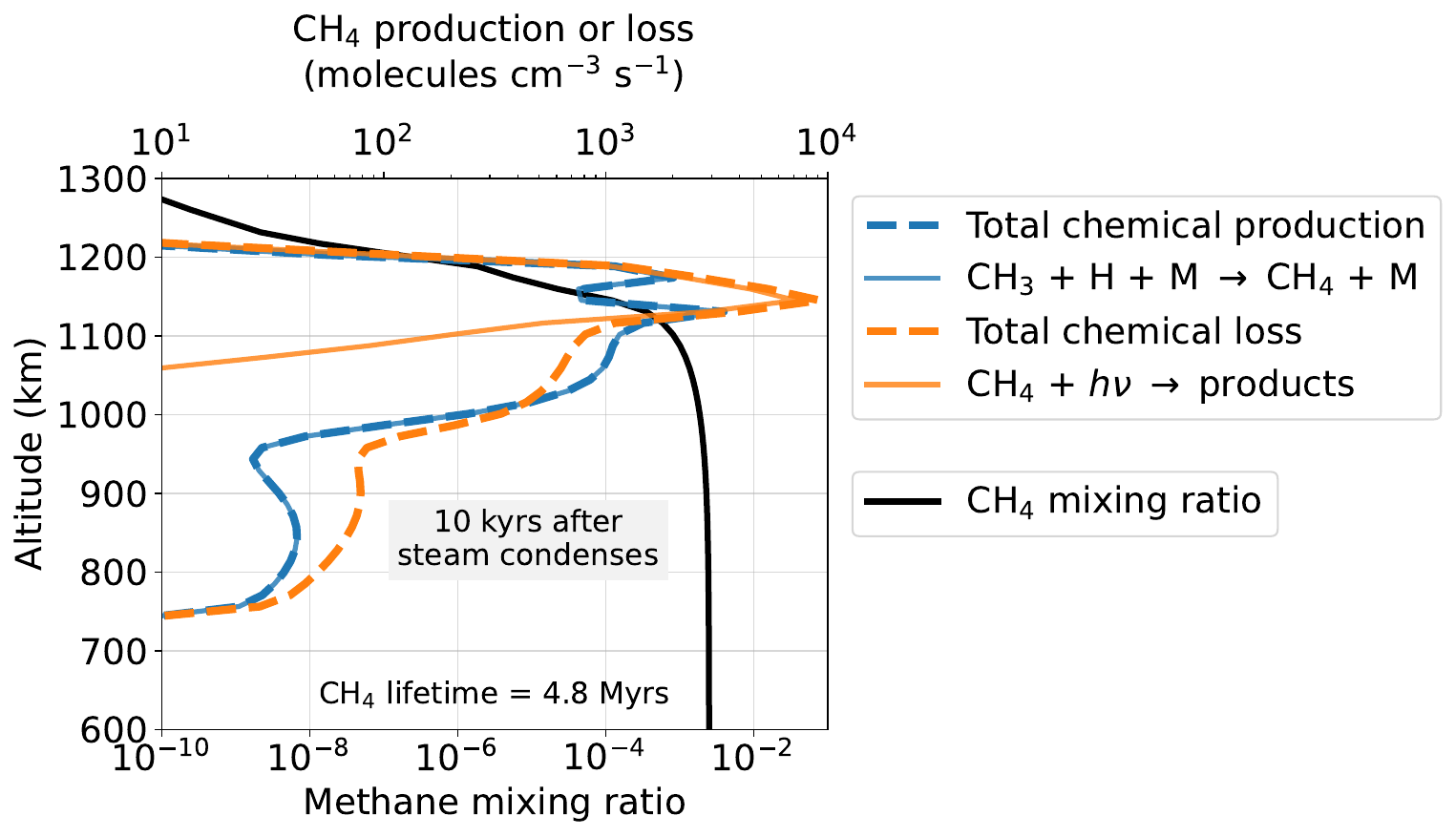}
  \caption{The methane photochemical lifetime in post-impact atmospheres. The plot shows the CH$_4$ mixing ratio and production and loss as a function of altitude 10,000 years after the steam atmosphere has condensed to an ocean following the $1.58 \times 10^{21}$ kg impact described in Figure \ref{fig:figure4}. CH$_4$ is primarily destroyed by photolysis, but reforms efficiently in the H$_2$ rich atmosphere from $\mathrm{CH_3} + \mathrm{H} + \mathrm{M} \rightarrow \mathrm{CH_4} + \mathrm{M}$. The result is a 4.8 million year CH$_4$ photochemical lifetime. CH$_4$ only persists in the atmosphere for about one million years because H$_2$ escapes to space in this amount of time which inhibits CH$_4$ recombination.}
  \label{fig:ch4_prod_loss}
\end{figure}

\begin{figure}
  \centering
  \includegraphics[width=1.0\textwidth]{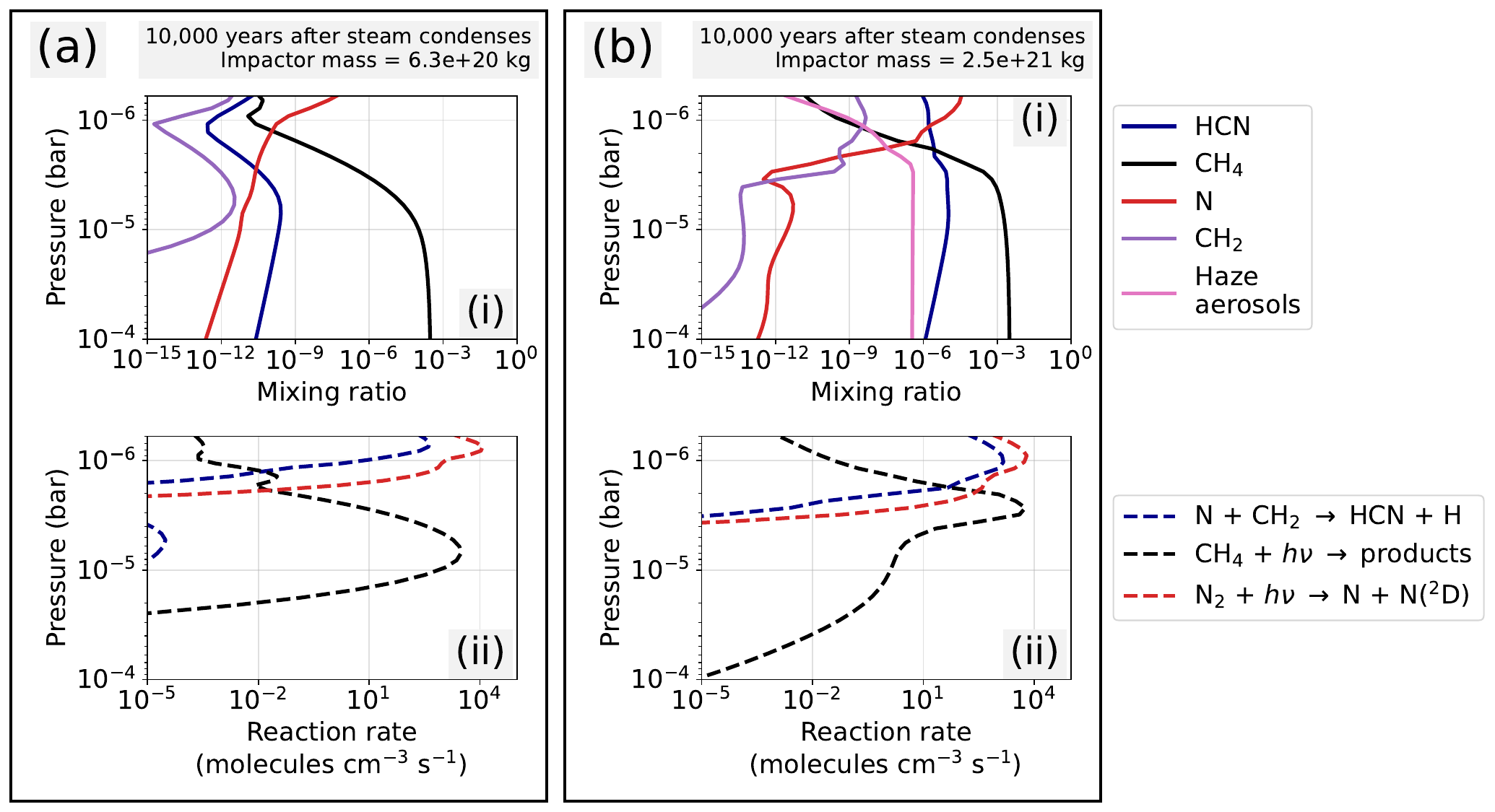}
  \caption{HCN production and precursor species after a (a) $6.3 \times 10^{20}$ kg and (b) $2.5 \times 10^{21}$ kg impactor. In both (a) and (b) panels (i) shows mixing ratios, while (ii) shows photolysis and reaction rates. Panel (b) contains haze aerosols which cause CH$_4$ photolysis to be higher in the atmosphere compared to panel (a). High altitude CH$_4$ photolysis, closer to N$_2$ photolysis, promotes HCN production because photolysis produced can more readily combine to make cyanides.}
  \label{fig:ch4_n_hcn}
\end{figure}

\begin{figure}
  \centering
  \includegraphics[width=1.0\textwidth]{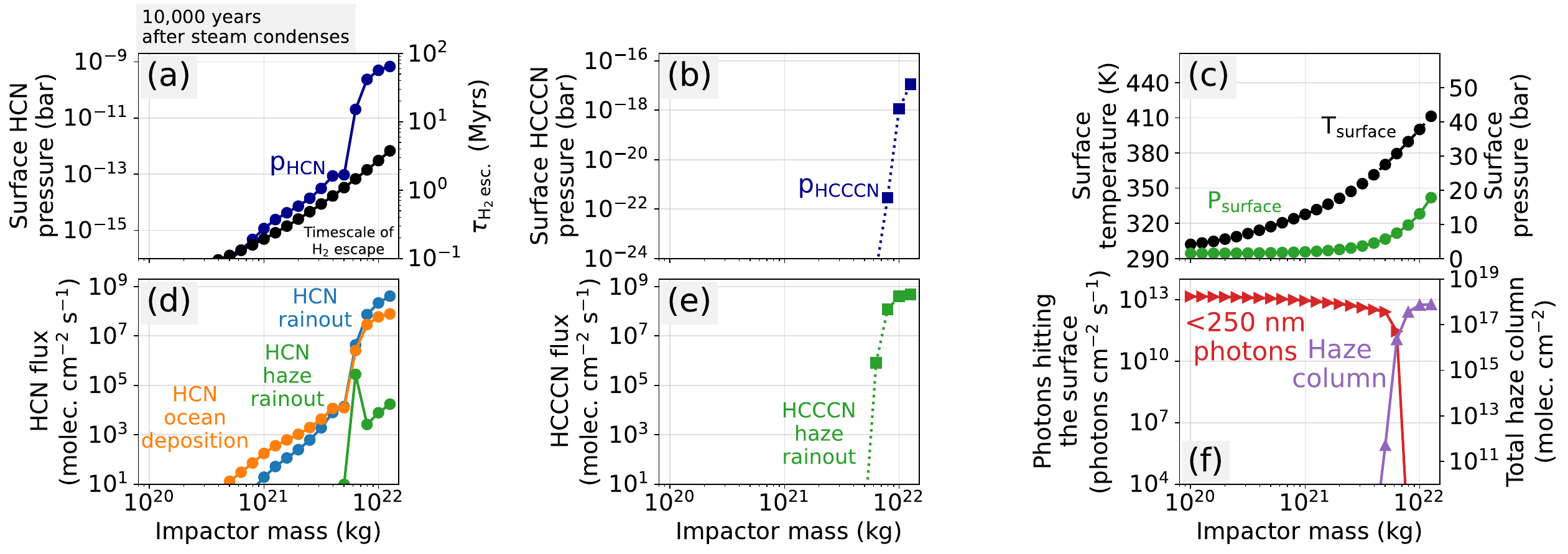}
  \caption{Identical to Figure \ref{fig:figure5} in the main text, but instead assumes post-impact H$_2$ generation if governed by ``Model 1B'' described in Figure \ref{fig:melt_reaction_sup}.}
  \label{fig:figure5_citron}
\end{figure}

\begin{figure}
  \centering
  \includegraphics[width=1.0\textwidth]{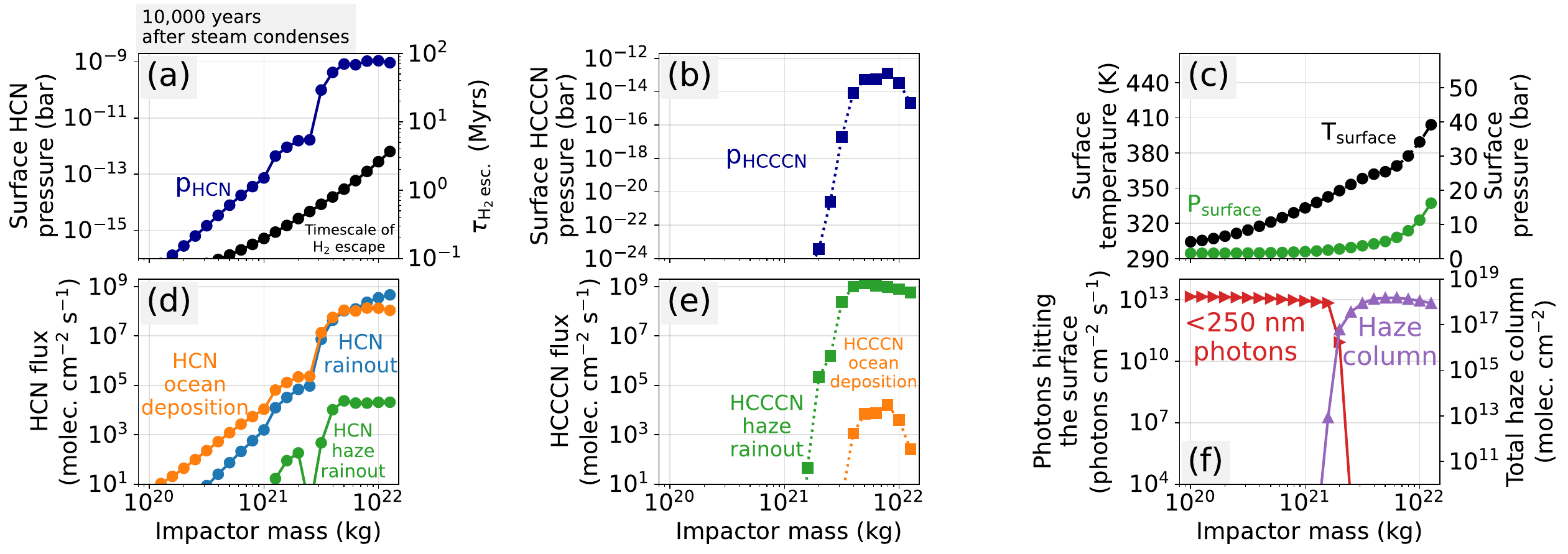}
  \caption{Identical to Figure \ref{fig:figure7} in the main text, but instead assumes post-impact H$_2$ generation if governed by ``Model 1B'' described in Figure \ref{fig:melt_reaction_sup}.}
  \label{fig:figure7_citron}
\end{figure}

\begin{figure}
  \centering
  \includegraphics[width=0.4\textwidth]{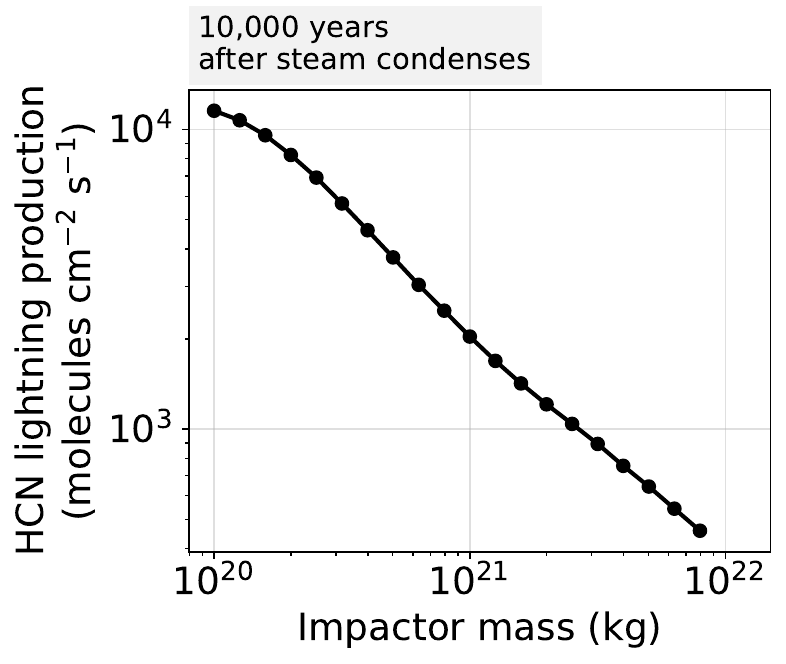}
  \caption{HCN production from lighting for the same simulations and time-period shown in Figure \ref{fig:figure5}. The calculations use methods described in \citet{Chameides_1981} assuming modern Earth's lightning dissipation rate ($9.8 \times 10^{-9}$ J cm$^{-2}$ s$^{-1}$), and a 2250 K HCN freeze-out temperature. HCN production from lightning is small compared to what is achievable with photochemistry.}
  \label{fig:figure5_lightning}
\end{figure}

\section{H$_2$ generation from iron and molten crust} \label{sec:phase1_appendix}

Here, we describe our model for atmospheric H$_2$ generation in the days to months following a massive asteroid impact (Phase 1 in Figure \ref{fig:impact_diagram}). All our simulations assume a pre-impact atmosphere containing CO$_2$, N$_2$, and ocean water. First, we assume that half of the impactor's kinetic energy heats the atmosphere and ocean water to $\sim 2000$ K. We assume the atmosphere is heated to $\sim 2000$ K because this is roughly the evaporation temperature of silicates. For our assumed impact velocity of $20.7$ km s$^{-1}$, all impactor masses that we consider in the main text ($10^{20}$ to $10^{22}$ kg) have kinetic energies $> 2 \times 10^{28}$ joules delivering $> 10^{28}$ joules to the atmosphere which is larger than the $5 \times 10^{27}$ joules required to vaporize an ocean \citep{Sleep_1989}. 

% Recently, \citet{Citron_2022} used a SPH impact code to find that an impact with $> 4 \times 10^{28}$ joules is required to vaporize an ocean for a 45$^\circ$ impact angle. Therefore our model assumption that an impact $> 2 \times 10^{28}$ joules vaporizes an ocean is only a factor of two different than what was found by \citet{Citron_2022}.

Next, our model assumes each mole of iron delivered reacts with the atmosphere and removes one mole of oxygen. The moles cm$^{-2}$ of iron delivered to the atmosphere is

\begin{equation}
  N_\mathrm{Fe,atmos} = \frac{X_\mathrm{Fe,atmos} X_\mathrm{Fe,imp} M_\mathrm{imp}}{\mu_\mathrm{Fe} A_\oplus}
\end{equation}
Here, $M_\mathrm{imp}$ is the mass of the impact in grams, $X_\mathrm{Fe,imp}$ is the iron mass fraction of the impact, $X_\mathrm{Fe,atmos}$ is the fraction of the impactor iron that reacts with the atmosphere, $\mu_\mathrm{Fe}$ is the molar weight of iron, and $A_\oplus$ is the area of Earth in cm$^2$. Following \citet{Zahnle_2020} we take $X_\mathrm{Fe,imp} = 0.33$. In main text, we assume $X_\mathrm{Fe,atmos} = 1$ (e.g. ``Model 1A'' in Figure \ref{fig:melt_reaction_sup}), while the Appendix contains calculations with $X_\mathrm{Fe,atmos} = 0.15$ to $0.3$ based on extrapolations of the \citet{Citron_2022} SPH impact simulations for $45^{\circ}$ impactors traveling at 20.7 km s$^{-1}$ (e.g. ``Model 1B'' in Figure \ref{fig:melt_reaction_sup}). To approximate equilibration between the delivered iron and the atmosphere, we simply remove $N_\mathrm{Fe,atmos}$ of \hl{oxygen atoms} from the atmosphere.

Our model also optionally considers reactions between the atmosphere and a melt pond generated by the impact. Our approach is similar to the one described in \citet{Itcovitz_2022}. We estimate the total mass of the melt pond ($M_\mathrm{melt}$) by interpolating SPH impact simulations from \citet{Citron_2022} for a $45^{\circ}$ impact angle. The smallest impact they consider is $7 \times 10^{21}$ kg, so we extrapolate their results down to $10^{20}$ kg. We additionally take the melted crust to be basaltic in composition except with variable initial amounts of ferric and ferrous iron. Effectively, this means that the initial oxygen fugacity of the melted crust is a free parameter because iron redox state is related to oxygen fugacity through the equilibrium reaction,

\begin{equation} 
  \label{eq:kress_carmichael_1}
  0.5 \mathrm{O_2} + 2 \mathrm{FeO} \leftrightarrow \mathrm{Fe_2O_3}
\end{equation}
We assume the oxygen atoms can flow from the atmosphere into the melt (or vice-versa) in order to bring Reaction \ref{eq:kress_carmichael_1} to an equilibrium state defined by \citet{Kress_1991} thermodynamic data. Our model also considers H$_2$O gas dissolution in the melt using the Equation (19) solubility relation in \citet{Itcovitz_2022}.

Finally, given a heated post-impact atmosphere that has been reduced by impactor iron and, optionally, in contact with an melt pool, we compute thermodynamic equilibrium of the atmosphere-melt system at $1900$ K. We choose 1900 K because any impact-produced silicate vapors should have condensed and rained out of the atmosphere, and the melt pool should have not yet solidified \citep{Itcovitz_2022}. To find an equilibrium state, we first compute an equilibrium composition for the atmosphere alone using the equilibrium solver in the Cantera chemical engineering package \citep{cantera} with our thermodynamic data (Appendix \ref{sec:photochem_reactions}). Next, to equilibrate the atmosphere-melt system, we perform a zero-dimensional kinetics integration for 1000 years at constant temperature and pressure with our reaction network (Appendix \ref{sec:photochem_reactions}). All reactions in our network are reversible thermodynamically, therefore integrating the kinetics forward in time should ultimately reach a state of thermodynamic equilibrium. Our integration includes additional reactions representing Reaction \ref{eq:kress_carmichael_1} and H$_2$O dissolution in the melt. We arbitrarily choose forward reaction rates of $10^{-10}$ s$^{-1}$ for both reactions, then reverse the rates using the \citet{Kress_1991} equilibrium constant, and the Equation (19) solubility relation in \citet{Itcovitz_2022}. Overall, our approach finds a chemical equilibrium state between the atmosphere and the melt pond, and therefore an estimation of the amount of H$_2$ generated from atmosphere-iron and atmosphere-melt reactions.

Our code for solving melt-atmosphere equilibrium is available at the following Zenodo link: https://doi.org/10.5281/zenodo.7802966.

\section{Kinetics model of a cooling steam atmosphere} \label{sec:kinetics_climate}

We simulate the chemistry of a cooling post-impact atmosphere using a zero-dimensional kinetics-climate model. We assume the atmosphere's composition, pressure, and temperature are homogeneous in all directions, and has a vertical extent of one atmospheric scale height ($H_a$). For these assumptions, the following system of ordinary differential equations govern our model:

\begin{equation} \label{eq:prod_loss}
  \frac{\partial N_i}{\partial t} = \frac{H_a}{N_a} (P_i - L_i) + \frac{A_c}{N_a} (P_{i,\text{surf}} - L_{i,\text{surf}})
\end{equation}
\begin{equation} \label{eq:temp}
  \frac{\partial T_s}{\partial t} = -\frac{1}{\rho c_p} \left(\frac{F_\text{net}}{H_a}\right) - \frac{1}{\rho c_p}\left(\frac{d M_\mathrm{H_2O}}{dt} \frac{l_\mathrm{H_2O}}{A_\oplus H_a}\right) 
\end{equation}

All variables and units are in Table \ref{tab:variables}. In Equation \eqref{eq:prod_loss}, $N_i$ is the column abundance of species $i$ in mole cm$^{-2}$, which changes because of gas-phase chemical reactions (production rate $P_i$ and loss rate $L_i$) and reactions occurring on surfaces ($P_{i,\text{surf}}$ and $L_{i,\text{surf}}$). In Equation \eqref{eq:temp}, $T$ is surface temperature, which changes because of energy radiated to space ($F_\text{net}$), and because of latent heat from H$_2$O condensation ($\frac{d M_\mathrm{H_2O}}{dt}$), where $M_\mathrm{H_2O}$ is the mass of H$_2$O in the atmosphere. We approximate the energy radiated to space in ergs cm$^{-2}$ s$^{-1}$ from a steam-dominated atmosphere with the following parameterization:

\begin{equation}
  F_\text{net} = 8.3 \times 10^4 + 1000 \max(T_s - 1750,0)
\end{equation}
This parameterization fits calculations from our radiative transfer model (see Appendix \ref{sec:clima}), which uses the a solar spectrum at 4.0 Ga derived from methods described in \citet{Claire_2012}. 

We can rewrite Equation \eqref{eq:temp}, replacing $\rho H_a$ using the ideal gas law and the definition of atmospheric scale height,

\begin{equation}
  \rho H_a = \frac{p \bar\mu}{N_a k T} \frac{N_a k T}{\bar\mu g} = \frac{p}{g}
\end{equation}
Here, $p$ is the total atmospheric pressure in dynes cm$^{-2}$, $g$ is gravitational acceleration in cm s$^{-2}$, $k$ is the Boltzmann constant, $\bar\mu$ is the mean molecular weight in g mol$^{-1}$, and $N_a$ is Avogadro's number. Therefore,

\begin{equation} \label{eq:temp0}
  \frac{\partial T_s}{\partial t} = -\frac{g}{p c_p} F_\text{net} - \frac{g}{p c_p}\left(\frac{d M_\mathrm{H_2O}}{dt} \frac{l_\mathrm{H_2O}}{A_\oplus}\right) 
\end{equation}

Next, we must derive an expression for the steam condensation rate ($d M_\mathrm{H_2O}/dt$) in terms of known variables. Working in CGS units, the total pressure of the atmosphere is given by its gravitational force divided by Earth's surface area ($5.1 \times 10^{18}$ cm$^2$):

\begin{equation}
 p = \frac{Mg}{A_\oplus}
\end{equation}
Here, $M$ is the mass of the atmosphere in grams. We are considering steam-dominated atmospheres, therefore, the mass and pressure in the above relation is approximately equal to the mass of atmospheric H$_2$O and the H$_2$O partial pressure.

\begin{align}
  p_\mathrm{H_2O} &\approx \frac{M_\mathrm{H_2O}g}{A_\oplus} \\
  M_\mathrm{H_2O} &\approx \frac{p_\mathrm{H_2O}A_\oplus}{g} \label{eq:atmos_pressure}
\end{align}
Taking a time derivative of Equation \eqref{eq:atmos_pressure} yields

\begin{equation}
  \frac{d M_\mathrm{H_2O}}{dt} \approx \frac{A_\oplus}{g} \frac{d p_\mathrm{H_2O}}{dt} \label{eq:conden}
\end{equation}
We assume that the only processes changing the H$_2$O mass in the atmosphere is condensation, which occurs in our model when steam becomes saturated. We further assume that the H$_2$O partial pressure is fixed at saturation once steam condensation begins. We approximate the saturation vapor pressure of H$_2$O, $p_\mathrm{H_2O}^\text{sat}$, using the Clausius-Clapeyron equation, assuming a temperature-independent latent heat, $l_\mathrm{H_2O}$,

\begin{equation}
  p_\mathrm{H_2O}^\text{sat} = p_0 \exp\left(\frac{l_\mathrm{H_2O} \mu_\mathrm{H_2O}}{R}\left( \frac{1}{T_0} - \frac{1}{T}  \right)\right) \label{eq:sat_press}
\end{equation}
$p_0$ and $T_0$ are reference pressures and temperatures, respectively. Taking a time derivative of Equation \eqref{eq:sat_press} yields

\begin{equation}
  \frac{dp_\mathrm{H_2O}^\text{sat}}{dt} = \left(\frac{l_\mathrm{H_2O} \mu_\mathrm{H_2O}}{RT^2}\right) \frac{dT_s}{dt} p_\mathrm{H_2O}^\text{sat} \label{eq:sat_press_deriv}
\end{equation}
Substituting Equation \eqref{eq:sat_press_deriv} into Equation \eqref{eq:conden} gives

\begin{equation}
  \frac{d M_\mathrm{H_2O}}{dt} = \frac{A_\oplus}{g} \left(\frac{l_\mathrm{H_2O} \mu_\mathrm{H_2O}}{RT^2}\right) \frac{dT_s}{dt} p_\mathrm{H_2O}^\text{sat} \label{eq:conden2}
\end{equation}
Finally, we can substitute Equation \eqref{eq:conden2} into Equation \eqref{eq:temp0} and rearrange to solve for $dT/dt$. The result below gives the rate of change of temperature when the steam is too hot to condense ($p_\mathrm{H_2O} > p_\mathrm{H_2O}^\text{sat}$), and when the steam is condensing ($p_\mathrm{H_2O} = p_\mathrm{H_2O}^\text{sat}$).
\begin{equation} \label{eq:temp1}
  \frac{d T_s}{d t} = 
  \begin{cases} 
    -\frac{g}{p c_p} F_\text{net} & p_\mathrm{H_2O} > p_\mathrm{H_2O}^\text{sat} \\
    -\frac{g}{p c_p} F_\text{net}\left( 1 + \frac{l_\mathrm{H_2O}^2 \mu_\mathrm{H_2O} p_\mathrm{H_2O}^\text{sat}}{p c_p R T^2} \right)^{-1} & p_\mathrm{H_2O} = p_\mathrm{H_2O}^\text{sat}
 \end{cases}
\end{equation} 

Equations \eqref{eq:prod_loss} and \eqref{eq:temp1} are a system of ordinary differential equations, which we approximately solve over time using the CVODE BDF method developed by Sundials Computing \citep{Hindmarsh_2005}. Additionally, for either gas-phase or surface reactions, we make use of the Cantera software library \citep{cantera} to compute chemical production and destruction rates. Our code for solving the equations derived in this section is available at the following Zenodo link: https://doi.org/10.5281/zenodo.7802966.

\begin{table}
  \caption{Variables}
  \label{tab:variables}
  \begin{center}
  \begin{tabularx}{\linewidth}{p{0.15\linewidth} | p{0.55\linewidth} | p{0.3\linewidth}}
  \hline \hline
  Variable & Definition & Units \\
  \hline
  $f_{i}$ & Mixing ratio of species $i$ & dimensionless 
  \\
  $n_{i}$ & Number density of species $i$ & molecules cm$^{-3}$ 
  \\
  $n$ & Total number density & molecules cm$^{-3}$ 
  \\
  $N_{i}$ & Column abundance of species $i$ & mol cm$^{-2}$ 
  \\
  $N$ & Total column abundance & mol cm$^{-2}$ 
  \\
  $\rho$ & Density of the atmosphere & g cm$^{-3}$ 
  \\
  $c_p$ & Specific heat capacity of the atmosphere & erg g$^{-1}$ K$^{-1}$
  \\
  $F_\text{net}$ & Net radiative energy leaving the atmosphere & erg cm$^{-2}$ s$^{-1}$
  \\
  $z$ & Altitude & cm 
  \\
  $t$ & Time & seconds 
  \\
  $P_{i}$ & Total chemical production of species $i$ & molecules cm$^{-3}$ s$^{-1}$ 
  \\
  $L_{i}$ & Total chemical loss of species $i$ & molecules cm$^{-3}$ s$^{-1}$ 
  \\
  $P_{i,\text{surf}}$ & Total chemical production of species $i$ from surface reactions & molecules
  cm$^{-2}$ s$^{-1}$ 
  \\
  $L_{i,\text{surf}}$ & Total chemical loss of species $i$ from surface reactions & molecules
  cm$^{-2}$ s$^{-1}$ 
  \\
  $R_{\text{i, rainout}}$ & Production and loss of species $i$ from rainout & molecules cm$^{-3}$ s$^{-1}$
  \\
  $Q_{i\text{, cond}}$ & Production and loss of species $i$ from condensation and evaporation & molecules cm$^{-3}$ s$^{-1}$
  \\
  $\Phi_{i}$ & Vertical flux of species $i$ & molecules cm$^{-2}$ s$^{-1}$ 
  \\
  $K_{zz}$ & Eddy diffusion coefficient & cm$^{-2}$ s$^{-1}$ 
  \\
  $D_{i}$ & Molecular diffusion coefficient & cm$^{-2}$ s$^{-1}$ 
  \\
  $H_{i}$ & $= N_{a}\text{kT}\text{/}\mu_{i}g$, The scale heights of species $i$ & cm 
  \\
  $H_{a}$ & $= N_{a}\text{kT}\text{/}\overline{\mu}g$, The average scale height. & cm 
  \\
  $N_{a}$ & Avogadro's number & molecules mol$^{-1}$ 
  \\
  $k$ & Boltzmann's constant & erg K$^{-1}$ 
  \\
  $R$ & Gas constant & erg mol$^{-1}$ K$^{-1}$ 
  \\
  $\mu$ & Molar mass. $\overline{\mu}$ is mean molar mass of the atmosphere, and $\mu_{i}$ is the molar mass of species $i$ & g mol$^{-1}$ 
  \\
  $A_c$ & Catalyst surface area per atmospheric column & cm$^2$ catalyst / cm$^2$ Earth
  \\
  $l_\mathrm{H_2O}$ & Latent heat of H$_2$O condensation & erg g$^{-1}$ 
  \\
  $A_\oplus$ & Area of Earth's surface & cm$^2$
  \\
  $\phi$ & Relative humidity & dimensionless
  \\
  $A_s$ & Optical surface albedo & dimensionless
  \\
  $p$ & Atmospheric pressure. $p_\mathrm{i}$ is the partial pressure of species $i$. & dynes cm$^{-2}$ 
  \\
  $M$ & Mass of the atmosphere. $M_\mathrm{i}$ is the mass of species $i$. $M_\mathrm{imp}$ is the mass of an impactor. & g
  \\
  $g$ & Gravitational acceleration & cm s$^{-2}$ 
  \\
  $\alpha_{\text{Ti}}$ & Thermal diffusion coefficient of species $i$. We neglect this term ($\alpha_{\text{Ti}} = 0$) & dimensionless
  \\
  \hl{$w_i$} & Fall velocity of a particle & cm s$^{-1}$
  \\
  $T$ & Temperature. $T_s$ is the surface temperature. & K 
  \\
  \end{tabularx}
  \end{center}
\end{table}

\section{The \emph{Photochem} model} \label{sec:photochem}

To simulate the photochemistry of post-impact reducing atmospheres, we developed a photochemical model called \emph{Photochem}. The model is a re-written and vastly updated version of \emph{PhotochemPy} \citep{Wogan_2022}. \emph{Photochem} is written in modern Fortran and C, with a Python interface made possible by Cython \citep{Behnel_2010}. This article uses \emph{Photochem} version v0.3.14 archived in the following Zenodo repository: https://doi.org/10.5281/zenodo.7802921.

The following sections briefly describe the fundamental model equations solved by \emph{Photochem}, our chemical network, and validates the model against observations of Earth and Titan.

\subsection{Model equations}
We begin our derivation of the equations governing \emph{Photochem} with modified versions of Equations B.1, B.2 and B.29 in \citet{Catling_2017}:

\begin{gather}
  \frac{\partial n_{i}}{\partial t} = - \frac{\partial}{\partial z}\Phi_{i} + P_{i} - L_{i} - R_{i\text{, rainout}} + Q_{i\text{, cond}} \label{eq:continuity} 
  \\
  \Phi_{i\text{,gas}} = - K_{zz} n\frac{\partial}{\partial z}\left( \frac{n_{i}}{n} \right) - n_i D_{i} \left( \frac{1}{n_i} \frac{\partial n_i}{\partial z} + \frac{1}{H_i} + \frac{1}{T} \frac{\partial T}{\partial z}  + \frac{\alpha_{Ti}}{T} \frac{\partial T}{\partial z} \right) \label{eq:flux_gas} 
  \\
  \Phi_{i\text{,particle}} = - K_{zz} n\frac{\partial}{\partial z}\left( \frac{n_{i}}{n} \right) - w_i n_i \label{eq:flux_part}
\end{gather}
Table \ref{tab:variables} explains the variables and their units. Equation \eqref{eq:continuity} states that molecule concentration ($n_i$ in molecules cm$^{-3}$) changes over time at a point in space because of vertical movement of particles ($\frac{\partial}{\partial z}\Phi_{i}$), and chemical reactions, rainout or condensation/evaporation ($P_{i}$, $L_{i}$, $R_{i\text{, rainout}}$, and $C_{i\text{, cond}}$). The equation is 1-D, because it only considers vertical gas transport and differs from Equation B.1 in \citet{Catling_2017} because we explicitly include rainout and condensation. Equation \eqref{eq:flux_gas} states that the flux of gases ($\Phi_{i\text{,gas}}$) is determined by eddy and molecular diffusion, and Equation \eqref{eq:flux_part} assumes that the flux of particles ($\Phi_{i\text{,particle}}$) is given by eddy diffusion and the rate particles fall through the atmosphere.

Many 1-D photochemical models further simplify Equation \eqref{eq:continuity} by assuming that total number density does not change over time ($\partial n / \partial t \approx 0$). Using this assumption, Equation \eqref{eq:continuity} is recast in terms of evolving mixing ratios ($f_i$) rather than number densities (see Appendix B.1 in \citet{Catling_2017} for a derivation). Such models assume a time-constant temperature profile. The surface pressure is also prescribed, and pressures above the surface are computed with the hydrostatic equation. In order to guarantee that all mixing ratios in the atmosphere sum to 1, models assume a background filler gas with a mixing ratio $f_\mathrm{background} = 1 - \sum_i f_i$. N$_2$, CO$_2$ or H$_2$ are common choices for the background gas, depending on the atmosphere under investigation. By definition, the background gas is not conserved. This approach is valid for steady-state photochemical calculations, and is also reasonable for atmospheric transitions which maintain approximately constant surface pressure and atmospheric temperature. The \emph{Photochem} code contains an implementation of this traditional approach to photochemical modeling.

Unfortunately, solving a simplified version of Equation \eqref{eq:continuity} in terms of mixing ratios does not work well for post-impact atmospheric modeling. For example, a post-impact atmosphere can contain 10 bars of H$_2$ which escapes to space over millions of years, lowering the surface pressure to a 1 bar N$_2$ dominated atmosphere (e.g. Figure \ref{fig:figure4}). Traditional photochemical models fail to simulate this scenario because it is not reasonable to assume a single background gas and time-constant surface pressure. Additionally, most models fix atmospheric temperature during any single model integration, but surface temperature should change significantly as impact-generated H$_2$ escapes to space.

Therefore, \emph{Photochem} implements a code that solves
Equation \eqref{eq:continuity} in terms of number densities ($n_i$) without the assumption of fixed surface pressure or a background gas. This approach requires slight modifications to Equation \eqref{eq:flux_gas} and \eqref{eq:flux_part} which we describe below. Consider the hydrostatic equation and ideal gas law

\begin{gather}
  \frac{\partial p}{\partial z} = \frac{-g p \bar \mu}{N_a k T} \\
  p = n k T
\end{gather}
Substituting the ideal gas law in  the hydrostatic equation yields

\begin{gather}
  \frac{\partial}{\partial z} (n T) = \frac{-g n \bar \mu}{N_a k} \\
  n \frac{\partial T}{\partial z} + T \frac{\partial n}{\partial z} = \frac{-g n \bar \mu}{N_a k}
\end{gather}
After rearrangement and substituting the definition of scale height,

\begin{equation}
  \frac{1}{n} \frac{\partial n}{\partial z} = - \frac{1}{H_a} - \frac{1}{T} \frac{\partial T}{\partial z} \label{eq:hydrostatic_ideal_gas}
\end{equation}
Now consider the following expansion using the quotient rule

\begin{equation}
  \frac{\partial }{\partial z} \left(\frac{n_i}{n}\right) = \frac{1}{n} \frac{\partial n_i}{\partial z} - \frac{n_i}{n^2} \frac{\partial n}{\partial z} \label{eq:quotient_expansion}
\end{equation}
Substituting Equation \eqref{eq:hydrostatic_ideal_gas} into Equation \eqref{eq:quotient_expansion} and rearrangement gives

\begin{equation}
  n \frac{\partial }{\partial z} \left(\frac{n_i}{n}\right) = \frac{\partial n_i}{\partial z} + \frac{n_i}{H_a} + \frac{n_i}{T} \frac{\partial T}{\partial z} \label{eq:densub}
\end{equation}
Finally, we can substitute Equation \eqref{eq:densub} into Equations \eqref{eq:flux_gas} and \eqref{eq:flux_part} to derive new equations for the flux of gases and particles

\begin{gather}
  \Phi_{i\text{,gas}} = - K_{zz} n_i \left(\frac{1}{n_i} \frac{\partial n_i}{\partial z} + \frac{1}{H_a} + \frac{1}{T} \frac{\partial T}{\partial z}\right)
  - n_i D_{i} \left( \frac{1}{n_i} \frac{\partial n_i}{\partial z} + \frac{1}{H_i} + \frac{1}{T} \frac{\partial T}{\partial z}  + \frac{\alpha_{Ti}}{T} \frac{\partial T}{\partial z} \right) \label{eq:flux_gas1} 
  \\
  \Phi_{i\text{,particle}} = - K_{zz} n_i \left(\frac{1}{n_i} \frac{\partial n_i}{\partial z} + \frac{1}{H_a} + \frac{1}{T} \frac{\partial T}{\partial z}\right)
  - w_i n_i \label{eq:flux_part1}
\end{gather}

We then apply a finite-volume approximation to the Equation \eqref{eq:continuity} system of particle differential equations using fluxes for gases and particles given by Equations \eqref{eq:flux_gas1} and \eqref{eq:flux_part1}, which results in a system of ordinary differential equations. We use a second-order centered scheme for all spatial derivatives except falling particles, which use a first-order upwind scheme for stability. \emph{Photochem} evolves the finite volume approximate forward in time using the CVODE BDF method developed by Sundials Computing \citep{Hindmarsh_2005}. The model assumes no background gas, and surface pressure can evolve over time as, for example, gases escape to space. Additionally, our model computes a self-consistent temperature structure within each time step using the \emph{Clima} radiative transfer code (Appendix \ref{sec:clima}) assuming a pseudo-moist adiabatic troposphere connected to an isothermal upper atmosphere.

An additional challenge of post-impact atmospheres is that the scale height changes by a factor of $\sim 10$ or more when H$_2$ escapes leaving behind a N$_2$ or CO$_2$ dominated atmosphere (Figure \ref{fig:figure4}). Most relevant photochemistry occurs at pressures $> 10^{-7}$ bar, and so we choose a model domain which starts at the surface and extends to an altitude that is approximately this pressure. However, suppose we choose a model domain extending to $\sim 1000$ km (i.e. the $10^{-7}$ bar level) appropriate for an H$_2$ dominated atmosphere. After H$_2$ escapes to space, all relevant photochemistry would occur below $~100$ km, in the bottom several grid cells of the model. Therefore, the important photochemistry would be poorly resolved and inaccurate, and the extremely small pressures at the top of the model domain would likely cause numerical instability. Our solution is to adaptively adjust the model domain so it is always appropriate for atmospheres scale height. We use the root finding functionally in CVODE BDF to halt integration whenever the pressure at the top of the atmosphere falls below $10^{-7}$ bar and lower the top of the model domain before continuing integration. This procedure is done automatically tens to hundreds of times during each post-impact integration.

\subsection{Chemical network, photolysis cross sections and thermodynamic data} \label{sec:photochem_reactions}

Our chemical reactions, photolysis cross sections, and thermodynamic data used for all gas-phase kinetics are archived in the following Zenodo repository: https://doi.org/10.5281/zenodo.7802962. Chemical reactions and thermodynamic data are in the file ``reaction\_mechanisms/zahnle\_earth.yaml'', and photolysis cross sections are in the folder ``xsections/''. All thermodynamic data is from the NIST Chemistry WebBook \hl{\mbox{\citep{Linstrom_1998}}}. The chemical and photolysis reactions are an updated version of rates presented in \citet{Zahnle_2016}.

\hl{In this article, our model simulates rainout in droplets of water for the following species: particles, OH, CN, HCN, C$_2$H$_4$, NO, HO$_2$, N$_2$O, H$_2$O$_2$, O$_3$, NO$_2$, NO$_3$, HNO$_2$, HNO$_3$, C$_2$H$_6$, CH$_3$OH, CH$_3$CHO, C$_3$H$_6$, CH$_3$CN.}

\subsection{Model validation}

\begin{figure}
  \centering
  \includegraphics[width=0.8\textwidth]{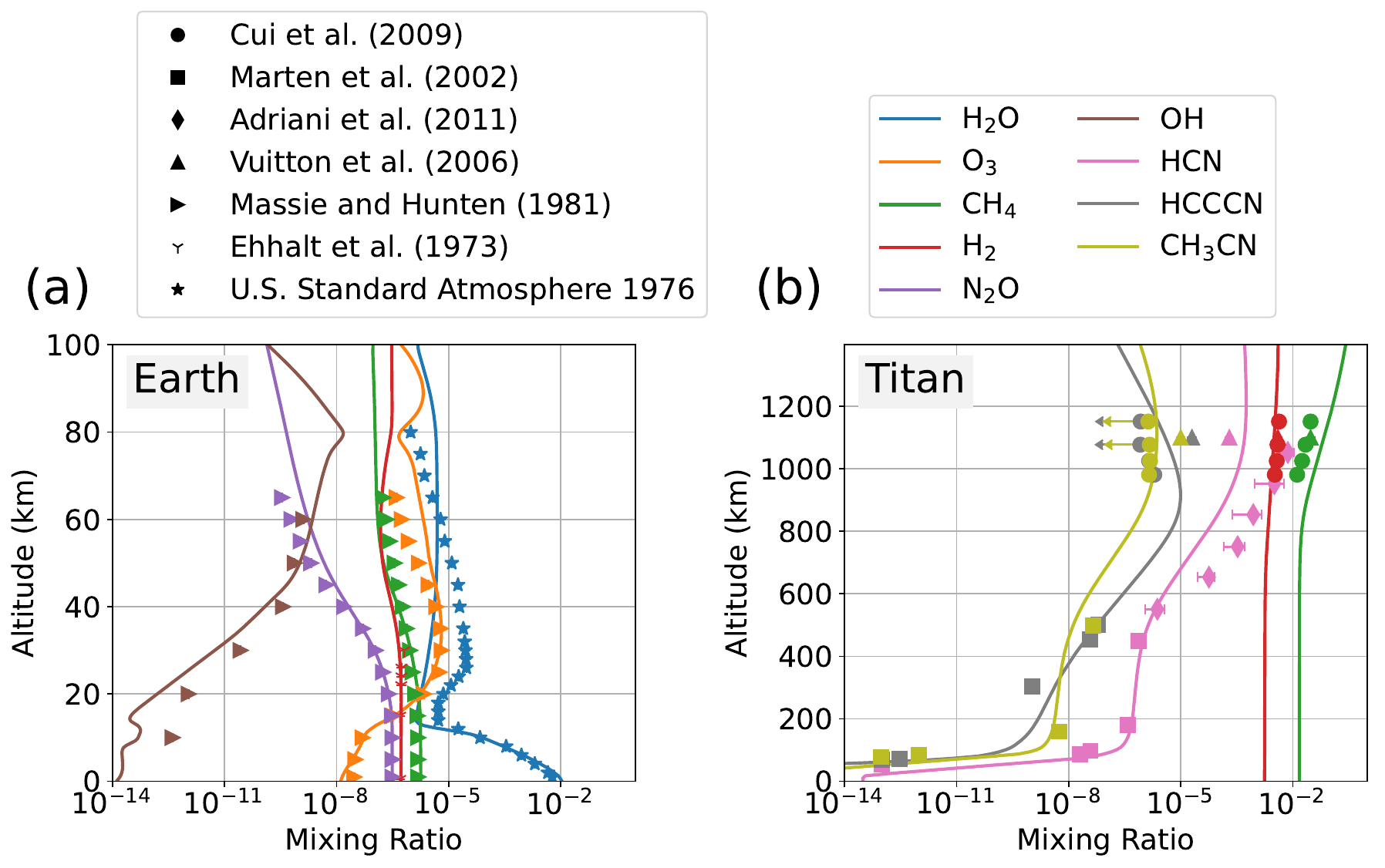}
  \caption{Earth and Titan photochemical model validation. (a) and (b) shows the \emph{Photochem} model applied to Earth and Titan, respectively, compared to data from the literature \citep{Cui_2009,Marten_2002,Adriani_2011,Vuitton_2006,Massie_1981,Ehhalt_1975}.}
  \label{fig:earth_titan_valid}
\end{figure}

Figure \ref{fig:earth_titan_valid} shows \emph{Photochem} applied to Earth and Titan compared to observations gathered from the literature. All boundary conditions and settings for each model are archived in the ``ModernEarth'' and ``Titan'' templates in the following Zenodo repository: https://doi.org/10.5281/zenodo.7802921. Our model of Titan fixes the surface CH$_4$ mixing ratio to $0.015$ volume mixing ratio, permits H$_2$ escape at the diffusion-limited rate, and allows aerosols to fall to Titan's surface, but otherwise has zero-flux boundary conditions. We ignore the effects of galactic cosmic rays, which causes our model to under-predict the nitrile haze production in the lower atmosphere \citep{Lavvas_2008b}. Additionally, we neglect ion chemistry which is argued to be important for the formation of large hydrocarbons (e.g., $\mathrm{C_6H_6}$), but inconsequential for smaller molecular weight species. Despite these omissions, \emph{Photochem} broadly reproduces the main cyanide chemistry on Titan.

\subsection{Deposition velocity of HCN} \label{sec:hcn_vdep}

Our photochemical-climate simulations of post-impact atmosphere assume a HCN surface deposition velocity of $7 \times 10^{-3}$ cm s$^{-1}$. Here, we describe a simple model of HCN hydrolysis in the ocean which justifies this value.

Motivated by Appendix 3 in \citet{Kharecha_2005}, we imagine a two-box ocean model with a surface ocean of depth $\sim 100$ m and a deep ocean ($\sim 4$ km). We assume HCN transport into the ocean is governed by a stagnant boundary layer model (see Figure 3 in \citet{Kharecha_2005}), where it is destroyed by hydrolysis reactions. HCN is mixed between the surface and deep ocean reservoirs by a turnover velocity, $v_\text{over}$, which we nominally take to be $1.2 \times 10^{-5}$ cm s$^{-1}$ which is appropriate for modern Earth. Under these circumstances, the following system of ordinary differential equations governs the concentration of HCN in the surface and deep ocean.

\begin{gather}
  \frac{d m_\mathrm{HCN,s}}{dt} = \frac{\Phi_\mathrm{HCN}}{C z_s} - k_\mathrm{tot} m_\mathrm{HCN,s} - \left(\frac{v_\mathrm{over}}{z_s}\right)(m_\mathrm{HCN,s} - m_\mathrm{HCN,d}) \label{eq:hydro1}
  \\
  \frac{d m_\mathrm{HCN,d}}{dt} = - k_\mathrm{tot} m_\mathrm{HCN,s} + \left(\frac{v_\mathrm{over}}{z_d}\right)(m_\mathrm{HCN,s} - m_\mathrm{HCN,d}) \label{eq:hydro2}
\end{gather}

Here, $m_\mathrm{HCN,s}$ and $m_\mathrm{HCN,d}$ are the concentration of HCN in the surface and deep ocean, respectively, in mol L$^{-1}$, $C$ is a constant equal to $6.022 \times 10^{20}$ molecules mol$^{-1}$ L cm$^{-3}$, $z_s$ is the depth of the surface ocean, and $z_d$ is the depth of the deep ocean. We compute the temperature and pH dependent hydrolysis rate coefficient, $k_\mathrm{tot}$, following \citet{Miyakawa_2002}. $\Phi_\mathrm{HCN}$ is the HCN flux into the ocean in molecules cm$^{-2}$ s$^{-1}$, which is determined by a stagnant boundary layer model:

\begin{equation} \label{eq:flux_lid}
  \Phi_\mathrm{HCN} = v_\mathrm{p,HCN} (\alpha_\mathrm{HCN} 10^{-6} p_\mathrm{HCN} - m_\mathrm{HCN,s}) C
\end{equation}
We assume the piston velocity of HCN is $5 \times 10^{-3}$ cm s$^{-1}$, which is the same as the piston velocity of CO \citep[Table 1]{Kharecha_2005}. Also, $\alpha_\mathrm{HCN}$ is the henry's law coefficient for HCN. The flux of a gas can also be parameterized with a deposition velocity ($v_\mathrm{d,HCN}$):

\begin{align}
\begin{split}
  \Phi_\mathrm{HCN} &= n_\mathrm{HCN} v_\mathrm{d,HCN} \\
  &= \frac{p_\mathrm{HCN}}{k T} v_\mathrm{d,HCN} \label{eq:dep_vel}
\end{split}
\end{align}

Assuming a steady state ($d m_\mathrm{HCN,s}/dt = d m_\mathrm{HCN,d}/dt = 0$) and solving for $v_\mathrm{d,HCN}$ in Equations \eqref{eq:hydro1} - \eqref{eq:dep_vel} yields

\begin{equation}
  v_\mathrm{d,HCN} = 10^{-6} k T \alpha_\mathrm{HCN} C k_\mathrm{tot} v_\mathrm{p,HCN} \frac{k_\mathrm{tot} z_d z_s + v_\mathrm{over} (z_d + z_s)}{k_\mathrm{tot} z_d (v_\mathrm{p,HCN} + k_\mathrm{tot} z_s) + v_\mathrm{over}(v_\mathrm{p,HCN} + k_\mathrm{tot} (z_d + z_s))} \label{eq:HCN_vdep}
\end{equation}
Here, we assume that the temperature and pH of the ocean is uniform, and that the temperature of the surface air is the same as the temperature of the ocean. 

\begin{figure}
  \centering
  \includegraphics[width=0.5\textwidth]{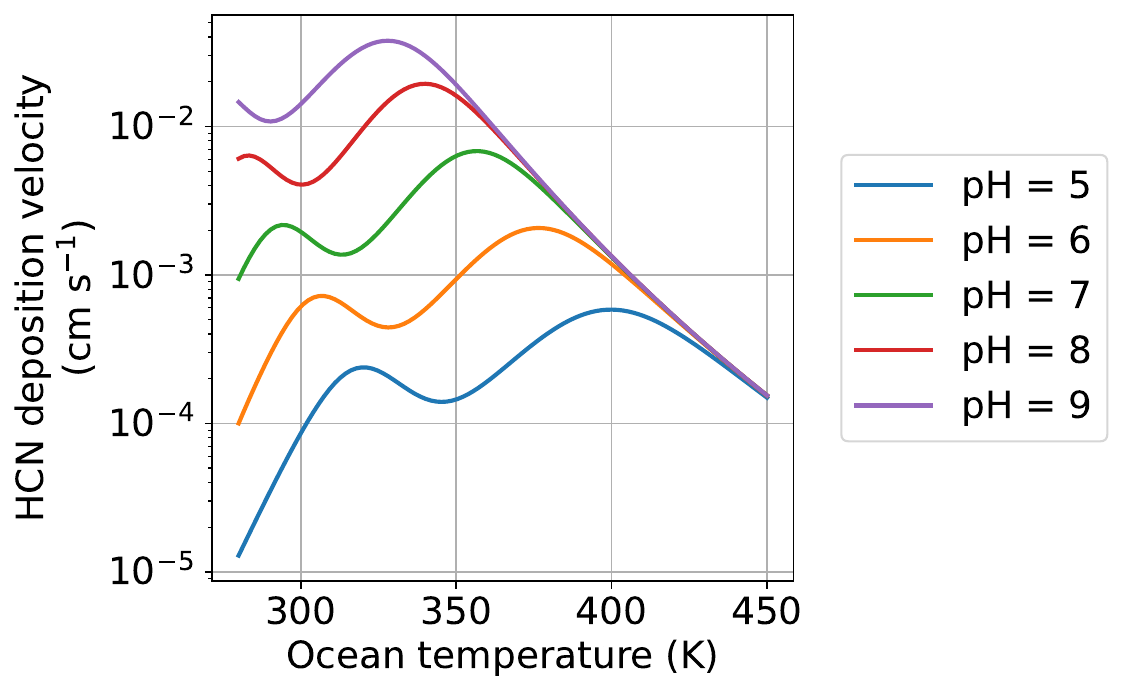}
  \caption{The deposition velocity of HCN caused by hydrolysis in the ocean. Calculations use Equation \eqref{eq:HCN_vdep} assuming $v_\mathrm{over} = 1.2 \times 10^{-5}$ cm s$^{-1}$, $z_s = 10^{4}$ cm, $z_d = 4 \times 10^{5}$ cm.}
  \label{fig:vdep_hcn}
\end{figure}

Figure \ref{fig:vdep_hcn} computes the deposition velocity of HCN using Equation \eqref{eq:HCN_vdep} over a wide range of ocean temperatures and pH. \citet{Kadoya_2020} used a model of the geologic carbon cycle to argue that the Hadean ocean was moderately alkaline (pH $\approx 8$). Therefore, we choose a HCN deposition velocity of $7 \times 10^{-3}$ cm s$^{-1}$ for our nominal model because it a reasonable approximation of the pH $= 8$ case over a wide range of temperatures. Additionally, we assume that HCCCN has the same deposition velocity as HCN, also caused by hydrolysis reactions in the ocean.

We have re-run our photochemical-climate simulations of post-impact atmospheres with order-of-magnitude larger and smaller HCN deposition velocities. The results are qualitatively unchanged. For example, assuming $v_\mathrm{d,HCN} = 7 \times 10^{-4}$ cm s$^{-1}$ for a $2 \times 10^{21}$ kg impactor in Figure \ref{fig:figure5} causes one order of magnitude smaller HCN ocean deposition. However, the HCN rainout and HCN surface pressure is unchanged because HCN rainout dominates over the HCN ocean deposition.

\section{The \emph{Clima} radiative transfer and climate model} \label{sec:clima}

To simulate the climate of post-impact atmospheres, we developed a new radiative transfer and climate code called \emph{Clima}. We approximately solve the radiative transfer equation using standard two-stream methods \citep{Toon_1989}. The code includes opacities representing photolysis, Rayleigh scattering, collision-induced absorption, and approximates line absorption with k-distributions. All available opacities and citations, except photolysis cross sections, are listed in Table \ref{tab:climate_opacities}. In this article, to account for line absorption of multiple species, we use the ``random overlap with resorting and rebinning'' method described in \citet{Amundsen_2017}. 

\begin{figure}
  \centering
  \includegraphics[width=0.5\textwidth]{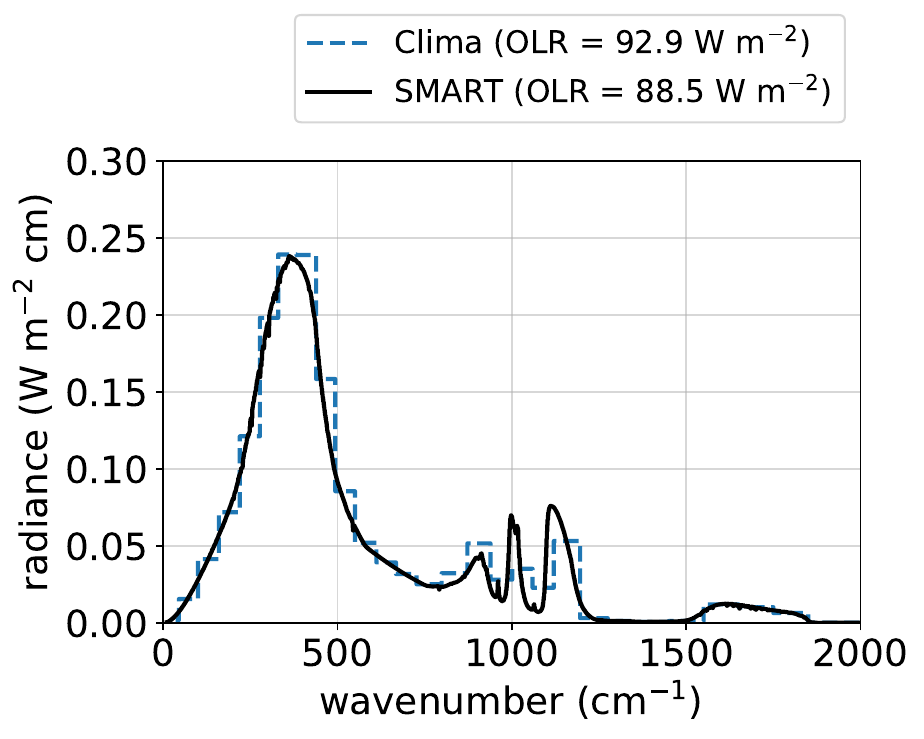}
  \caption{Outgoing longwave radiation of a 2 bar CO$_2$ atmosphere on Mars with a 250 K surface temperature, computed with \emph{Clima} (this work) and SMART \citep{Meadows_1996}. The agreement between the two codes validates \emph{Clima}.}
  \label{fig:early_mars_validation}
\end{figure}

Figure \ref{fig:early_mars_validation} shows a thermal emission spectra computed with \emph{Clima} for a two bar pure CO$_2$ atmosphere on Mars with a 250 K surface temperature. This same benchmark has been computed by several other radiative transfer codes: SOCRATES \citep[Figure 2]{Wolf_2022}, ExoRT \citep[Figure 2]{Wolf_2022}, SMART (Figure \ref{fig:early_mars_validation}), and the radiative transfer code used in \citet{Kopparapu_2013} (their Figure 1). All codes estimate the total outgoing thermal energy to be between 86 and 94 W m$^{-2}$, which is comparable to the value computed by \emph{Clima} (92.9 W m$^{-2}$).

The \emph{Clima} code also includes an adiabatic climate model \hl{which we use in Section \mbox{\ref{sec:climate_uncertainty}}}. Given partial pressures of gases at the surface, the code draws a \hl{pseudo-adiabat} temperature profile upward using Equation (1) in \citet{Graham_2021} until the temperature reaches an assume isothermal stratosphere. The code is general and can consider any number of condensing species, but H$_2$O is the only relevant condensible for post-impact atmospheres. Finally, to find an equilibrium climate, we solve a nonlinear equation for the surface temperature that balances incoming solar and outgoing longwave radiation. Each iteration of the nonlinear solve involves drawing an adiabat upward then computing the solar and infrared radiative fluxes.

\hl{We have validated the stand-alone climate model in \emph{Clima} by reproducing the calculations in \mbox{\citet{Wordsworth_2017}} of early Mars with CO$_2$ and H$_2$ atmospheres (left panel of Figure 2 in \mbox{\citet{Wordsworth_2017}}). Furthermore, we predict the runaway greenhouse limit to be 291 W m$^{-2}$, which is in acceptable agreement with the literature \mbox{\citep[e.g.,][]{Kopparapu_2013}}. Finally, we have also confirmed that our code for drawing pseudo-moist adiabats reproduces the code used in \mbox{\citet{Graham_2021}}.}

% Furthermore, we have used \emph{Clima} to estimate the habitable zone to be between 0.97 and 1.60 AU for a Sun-like star. For comparison, \mbox{\citet{Kopparapu_2013}} predicts the habitable zone is between 0.99 and 1.70 AU. The discrepancy at the outer edge is partially accounted for by different CO$_2$-CO$_2$ CIA opacities. 

The version of \emph{Clima} used in this article (\hl{v0.3.7}) is archived on Zenodo (https://doi.org/10.5281/zenodo.8060772), while the most up-to-date version can be found on GitHub (https://github.com/Nicholaswogan/clima).

\begin{table}
  \caption{Opacities used in the \emph{Clima} radiative transfer code}
  \label{tab:climate_opacities}
  \begin{center}
  \begin{tabularx}{\linewidth}{p{0.13\linewidth} | p{0.1\linewidth} | p{0.38\linewidth} | p{0.3\linewidth}}
    \hline \hline
    Opacity type & Opacity & Notes & Citation \\
    \hline 

    k-distributions$^\text{a}$ & H$_2$O & HITEMP2010 for 0 to 30,000 cm$^{-1}$ and HITRAN2016 for 30,000 to 42,000 $^{-1}$. Voigt line shape with 25 cm$^{-1}$ cutoff. Assumes Earth air broadening coefficients. Plinth or base is removed because this opacity is combined with MT\_CKD H$_2$O continuum. & \citet{Rothman_2010,Gordon_2017} \\

     & CO$_2$ & HITEMP2010. Sub-Lorentzian line shape with 500 cm$^{-1}$ cutoff. Assumes self-broadening coefficients. & \citet{Rothman_2010} \\

     & CH$_4$ & HITEMP2020. Voigt line shape with 25 cm$^{-1}$ cutoff. Assumes Earth air broadening coefficients. & \citet{Hargreaves_2020} \\

     & CO & HITEMP2019. Voigt line shape with 25 cm$^{-1}$ cutoff. Assumes self-broadening coefficients. & \citet{Li_2015} \\

     & O$_2$ & HITRAN2016. Voigt line shape with 25 cm$^{-1}$ cutoff. Assumes Earth air broadening coefficients. & \citet{Gordon_2017} \\

     & O$_3$ & HITRAN2016. Voigt line shape with 25 cm$^{-1}$ cutoff. Assumes Earth air broadening coefficients. & \citet{Gordon_2017} \\

     & \hl{NH$_3$} & HITRAN2016. Voigt line shape with 25 cm$^{-1}$ cutoff. Assumes Earth air broadening coefficients. & \citet{Gordon_2017} \\

    \hline

    CIA & H$_2$-H$_2$ & - & \citet{Molliere_2019} \\
    & H$_2$-He & - & \citet{Molliere_2019} \\
    & N$_2$-N$_2$ & - & \citet{Molliere_2019} \\
    & CH$_4$-CH$_4$ & - & \citet{Karman_2019} \\
    & N$_2$-O$_2$ & - & \citet{Karman_2019} \\
    & O$_2$-O$_2$ & - & \citet{Karman_2019} \\
    & H$_2$-CH$_4$ & - & \citet{Karman_2019} \\
    & CO$_2$-CO$_2$ & - & \citet{Karman_2019} \\
    & CO$_2$-CH$_4$ & - & \citet{Karman_2019} \\
    & CO$_2$-N$_2$ & - & \citet{Karman_2019} \\
    & N$_2$-H$_2$ & - & \citet{Karman_2019} \\

    \hline

    Rayleigh scattering$^\text{b}$ & N$_2$ & - & \citet{Keady_2002,Penndorf_1957} \\
    & CO$_2$ & - & \citet{Keady_2002,Shemansky_1972} \\
    & O$_2$ & - & \citet{Keady_2002,Penndorf_1957} \\
    & H$_2$O & - & \citet{Keady_2002,Ranjan_2017,Murphy_1977} \\
    & H$_2$ & - & \citet{Keady_2002} \\

    \hline

  \multicolumn{4}{p{1.0\linewidth}}{
    $^\text{a}$ All k-distributions are computed using HELIOS-K \citep{Grimm_2021}.

    $^\text{b}$ Rayleigh scattering opacities are computed using a parameterization from \citet{Vardavas_1984}.
  }

  \end{tabularx}
  \end{center}
\end{table}

\bibliography{bib}

\begin{thebibliography}{}
\expandafter\ifx\csname natexlab\endcsname\relax\def\natexlab#1{#1}\fi
\providecommand{\url}[1]{\href{#1}{#1}}
\providecommand{\dodoi}[1]{doi:~\href{http://doi.org/#1}{\nolinkurl{#1}}}
\providecommand{\doeprint}[1]{\href{http://ascl.net/#1}{\nolinkurl{http://ascl.net/#1}}}
\providecommand{\doarXiv}[1]{\href{https://arxiv.org/abs/#1}{\nolinkurl{https://arxiv.org/abs/#1}}}

\bibitem[{Adriani {et~al.}(2011)Adriani, Dinelli, Lopez-Puertas, Garcia-Comas,
  Moriconi, \& Funke}]{Adriani_2011}
Adriani, A., Dinelli, B., Lopez-Puertas, M., {et~al.} 2011, Icarus, 214, 584

\bibitem[{Alexander {et~al.}(2012)Alexander, Bowden, Fogel, Howard, Herd, \&
  Nittler}]{Alexander_2012}
Alexander, C.~O., Bowden, R., Fogel, M., {et~al.} 2012, Science, 337, 721

\bibitem[{Amundsen {et~al.}(2017)Amundsen, Tremblin, Manners, Baraffe, \&
  Mayne}]{Amundsen_2017}
Amundsen, D.~S., Tremblin, P., Manners, J., Baraffe, I., \& Mayne, N.~J. 2017,
  Astronomy \& Astrophysics, 598, A97

\bibitem[{Arney {et~al.}(2016)Arney, Domagal-Goldman, Meadows, Wolf,
  Schwieterman, Charnay, Claire, H{\'e}brard, \& Trainer}]{Arney_2016}
Arney, G., Domagal-Goldman, S.~D., Meadows, V.~S., {et~al.} 2016, Astrobiology,
  16, 873

\bibitem[{Aulbach \& Stagno(2016)}]{Aulbach_2016}
Aulbach, S., \& Stagno, V. 2016, Geology, 44, 751

\bibitem[{Avice {et~al.}(2018)Avice, Marty, Burgess, Hofmann, Philippot,
  Zahnle, \& Zakharov}]{Avice_2018}
Avice, G., Marty, B., Burgess, R., {et~al.} 2018, Geochimica et Cosmochimica
  Acta, 232, 82

\bibitem[{Bada \& Korenaga(2018)}]{Bada_2018}
Bada, J.~L., \& Korenaga, J. 2018, Life, 8, 55

\bibitem[{Bada \& Lazcano(2002)}]{Bada_2002}
Bada, J.~L., \& Lazcano, A. 2002, Science, 296, 1982

\bibitem[{Becker {et~al.}(2019)Becker, Feldmann, Wiedemann, Okamura, Schneider,
  Iwan, Crisp, Rossa, Amatov, \& Carell}]{Becker_2019}
Becker, S., Feldmann, J., Wiedemann, S., {et~al.} 2019, Science, 366, 76

\bibitem[{Behnel {et~al.}(2010)Behnel, Bradshaw, Citro, Dalcin, Seljebotn, \&
  Smith}]{Behnel_2010}
Behnel, S., Bradshaw, R., Citro, C., {et~al.} 2010, Computing in Science \&
  Engineering, 13, 31

\bibitem[{Benner {et~al.}(2020)Benner, Bell, Biondi, Brasser, Carell, Kim,
  Mojzsis, Omran, Pasek, \& Trail}]{Benner_2020}
Benner, S.~A., Bell, E.~A., Biondi, E., {et~al.} 2020, ChemSystemsChem, 2,
  e1900035

\bibitem[{Catling \& Kasting(2017)}]{Catling_2017}
Catling, D.~C., \& Kasting, J.~F. 2017, Atmospheric Evolution on Inhabited and
  Lifeless Worlds (Cambridge University Press)

\bibitem[{Catling \& Zahnle(2020)}]{Catling_2020}
Catling, D.~C., \& Zahnle, K.~J. 2020, Science advances, 6, eaax1420

\bibitem[{Cech(2012)}]{Cech_2012}
Cech, T.~R. 2012, Cold Spring Harbor perspectives in biology, 4, a006742

\bibitem[{Cerrillo(2022)}]{Cerrillo_2022}
Cerrillo, K. 2022, PhD thesis

\bibitem[{Chameides \& Walker(1981)}]{Chameides_1981}
Chameides, W., \& Walker, J.~C. 1981, Origins of life, 11, 291

\bibitem[{Citron \& Stewart(2022)}]{Citron_2022}
Citron, R.~I., \& Stewart, S.~T. 2022, The Planetary Science Journal, 3, 116

\bibitem[{Claire {et~al.}(2012)Claire, Sheets, Cohen, Ribas, Meadows, \&
  Catling}]{Claire_2012}
Claire, M.~W., Sheets, J., Cohen, M., {et~al.} 2012, The Astrophysical Journal,
  757, 95

\bibitem[{Cui {et~al.}(2009)Cui, Yelle, Vuitton, Waite~Jr, Kasprzak, Gell,
  Niemann, M{\"u}ller-Wodarg, Borggren, Fletcher, {et~al.}}]{Cui_2009}
Cui, J., Yelle, R., Vuitton, V., {et~al.} 2009, Icarus, 200, 581

\bibitem[{Di~Giulio(1997)}]{Di_1997}
Di~Giulio, M. 1997, Journal of molecular evolution, 45, 571

\bibitem[{Ehhalt {et~al.}(1975)Ehhalt, Heidt, Lueb, \& Pollock}]{Ehhalt_1975}
Ehhalt, D., Heidt, L., Lueb, R., \& Pollock, W. 1975, Pure and Applied
  Geophysics, 113, 389

\bibitem[{Furukawa {et~al.}(2007)Furukawa, Nakazawa, Sekine, \&
  Kakegawa}]{Furukawa_2007}
Furukawa, Y., Nakazawa, H., Sekine, T., \& Kakegawa, T. 2007, Earth and
  Planetary Science Letters, 258, 543

\bibitem[{Genda {et~al.}(2017)Genda, Brasser, \& Mojzsis}]{Genda_2017}
Genda, H., Brasser, R., \& Mojzsis, S. 2017, Earth and Planetary Science
  Letters, 480, 25

\bibitem[{Gilbert(1986)}]{Gilbert_1986}
Gilbert, W. 1986, nature, 319, 618

\bibitem[{Goldblatt \& Zahnle(2011)}]{Goldblatt_2011}
Goldblatt, C., \& Zahnle, K.~J. 2011, Climate of the Past, 7, 203

\bibitem[{Goldman \& Kacar(2021)}]{Goldman_2021}
Goldman, A.~D., \& Kacar, B. 2021, Journal of Molecular Evolution, 89, 127

\bibitem[{Goodwin {et~al.}(2022)Goodwin, Moffat, Schoegl, Speth, \&
  Weber}]{cantera}
Goodwin, D.~G., Moffat, H.~K., Schoegl, I., Speth, R.~L., \& Weber, B.~W. 2022,
  \dodoi{10.5281/zenodo.6387882}

\bibitem[{Gordon {et~al.}(2017)Gordon, Rothman, Hill, Kochanov, Tan, Bernath,
  Birk, Boudon, Campargue, Chance, {et~al.}}]{Gordon_2017}
Gordon, I.~E., Rothman, L.~S., Hill, C., {et~al.} 2017, Journal of Quantitative
  Spectroscopy and Radiative Transfer, 203, 3

\bibitem[{Graham {et~al.}(2021)Graham, Lichtenberg, Boukrouche, \&
  Pierrehumbert}]{Graham_2021}
Graham, R., Lichtenberg, T., Boukrouche, R., \& Pierrehumbert, R.~T. 2021, The
  Planetary Science Journal, 2, 207

\bibitem[{Grimm \& Marchi(2018)}]{Grimm_2018}
Grimm, R.~E., \& Marchi, S. 2018, Earth and Planetary Science Letters, 485, 1,
  \dodoi{10.1016/j.epsl.2017.12.043}

\bibitem[{Grimm {et~al.}(2021)Grimm, Malik, Kitzmann, Guzm{\'a}n-Mesa,
  Hoeijmakers, Fisher, Mendon{\c{c}}a, Yurchenko, Tennyson, Alesina,
  {et~al.}}]{Grimm_2021}
Grimm, S.~L., Malik, M., Kitzmann, D., {et~al.} 2021, The Astrophysical Journal
  Supplement Series, 253, 30

\bibitem[{Hargreaves {et~al.}(2020)Hargreaves, Gordon, Rey, Nikitin, Tyuterev,
  Kochanov, \& Rothman}]{Hargreaves_2020}
Hargreaves, R.~J., Gordon, I.~E., Rey, M., {et~al.} 2020, The Astrophysical
  Journal Supplement Series, 247, 55

\bibitem[{Hindmarsh {et~al.}(2005)Hindmarsh, Brown, Grant, Lee, Serban,
  Shumaker, \& Woodward}]{Hindmarsh_2005}
Hindmarsh, A.~C., Brown, P.~N., Grant, K.~E., {et~al.} 2005, ACM Transactions
  on Mathematical Software (TOMS), 31, 363

\bibitem[{Hirschmann \& Withers(2008)}]{Hirschmann_2008}
Hirschmann, M.~M., \& Withers, A.~C. 2008, Earth and Planetary Science Letters,
  270, 147

\bibitem[{Holland(1984)}]{Holland_1984}
Holland, H.~D. 1984, The Chemical Evolution of the Atmosphere and Oceans
  (Princeton University Press)

\bibitem[{H{\"o}rst(2017)}]{Horst_2017}
H{\"o}rst, S.~M. 2017, Journal of Geophysical Research: Planets, 122, 432

\bibitem[{Itcovitz {et~al.}(2022)Itcovitz, Rae, Citron, Stewart, Sinclair,
  Rimmer, \& Shorttle}]{Itcovitz_2022}
Itcovitz, J.~P., Rae, A.~S., Citron, R.~I., {et~al.} 2022, The Planetary
  Science Journal, 3, 115

\bibitem[{Johnson \& Wing(2020)}]{Johnson_2020}
Johnson, B.~W., \& Wing, B.~A. 2020, Nature Geoscience, 13, 243

\bibitem[{Johnson {et~al.}(1992)Johnson, Oelkers, \& Helgeson}]{Johnson_1992}
Johnson, J.~W., Oelkers, E.~H., \& Helgeson, H.~C. 1992, Computers \&
  Geosciences, 18, 899

\bibitem[{Kadoya {et~al.}(2020)Kadoya, Krissansen-Totton, \&
  Catling}]{Kadoya_2020}
Kadoya, S., Krissansen-Totton, J., \& Catling, D.~C. 2020, Geochemistry,
  Geophysics, Geosystems, 21, e2019GC008734

\bibitem[{Karman {et~al.}(2019)Karman, Gordon, van Der~Avoird, Baranov, Boulet,
  Drouin, Groenenboom, Gustafsson, Hartmann, Kurucz, {et~al.}}]{Karman_2019}
Karman, T., Gordon, I.~E., van Der~Avoird, A., {et~al.} 2019, Icarus, 328, 160

\bibitem[{Keady \& Kilcrease(2002)}]{Keady_2002}
Keady, J., \& Kilcrease, D. 2002, in Allen’s Astrophysical Quantities
  (Springer), 95--120

\bibitem[{Keefe \& Miller(1996)}]{Keefe_1996}
Keefe, A.~D., \& Miller, S.~L. 1996, Origins of Life and Evolution of the
  Biosphere, 26, 111

\bibitem[{Kharecha {et~al.}(2005)Kharecha, Kasting, \& Siefert}]{Kharecha_2005}
Kharecha, P., Kasting, J., \& Siefert, J. 2005, Geobiology, 3, 53

\bibitem[{Kopparapu {et~al.}(2013)Kopparapu, Ramirez, Kasting, Eymet, Robinson,
  Mahadevan, Terrien, Domagal-Goldman, Meadows, \& Deshpande}]{Kopparapu_2013}
Kopparapu, R.~K., Ramirez, R., Kasting, J.~F., {et~al.} 2013, The Astrophysical
  Journal, 765, 131

\bibitem[{Korenaga(2021)}]{Korenaga_2021}
Korenaga, J. 2021, Life, 11, 1142

\bibitem[{Kress \& McKay(2004)}]{Kress_2004}
Kress, M.~E., \& McKay, C.~P. 2004, Icarus, 168, 475

\bibitem[{Kress \& Carmichael(1991)}]{Kress_1991}
Kress, V.~C., \& Carmichael, I.~S. 1991, Contributions to Mineralogy and
  Petrology, 108, 82

\bibitem[{Krissansen-Totton {et~al.}(2021)Krissansen-Totton, Kipp, \&
  Catling}]{Krissansen_2021}
Krissansen-Totton, J., Kipp, M.~A., \& Catling, D.~C. 2021, Geobiology, 19, 342

\bibitem[{Lavvas {et~al.}(2008{\natexlab{a}})Lavvas, Coustenis, \&
  Vardavas}]{Lavvas_2008}
Lavvas, P., Coustenis, A., \& Vardavas, I. 2008{\natexlab{a}}, Planetary and
  Space Science, 56, 27

\bibitem[{Lavvas {et~al.}(2008{\natexlab{b}})Lavvas, Coustenis, \&
  Vardavas}]{Lavvas_2008b}
---. 2008{\natexlab{b}}, Planetary and Space Science, 56, 67

\bibitem[{Leconte {et~al.}(2017)Leconte, Selsis, Hersant, \&
  Guillot}]{Leconte_2017}
Leconte, J., Selsis, F., Hersant, F., \& Guillot, T. 2017, Astronomy \&
  Astrophysics, 598, A98

\bibitem[{L{\'e}cuyer {et~al.}(1998)L{\'e}cuyer, Gillet, \&
  Robert}]{Lecuyer_1998}
L{\'e}cuyer, C., Gillet, P., \& Robert, F. 1998, Chemical Geology, 145, 249

\bibitem[{Lewis(1992)}]{Lewis_1992}
Lewis, J.~S. 1992, Space resources materials: National Aeronautics and Space
  Administration Special Publication, 509, 59

\bibitem[{Li {et~al.}(2015)Li, Gordon, Rothman, Tan, Hu, Kassi, Campargue, \&
  Medvedev}]{Li_2015}
Li, G., Gordon, I.~E., Rothman, L.~S., {et~al.} 2015, The Astrophysical Journal
  Supplement Series, 216, 15

\bibitem[{Linstrom \& Mallard(1998)}]{Linstrom_1998}
Linstrom, P., \& Mallard, W. 1998, NIST Chemistry WebBook, NIST Standard
  Reference Database Number 69.
\newblock \url{https://doi.org/10.18434/T4D303}

\bibitem[{Loison {et~al.}(2015)Loison, H{\'e}brard, Dobrijevic, Hickson,
  Caralp, Hue, Gronoff, Venot, \& B{\'e}nilan}]{Loison_2015}
Loison, J., H{\'e}brard, E., Dobrijevic, M., {et~al.} 2015, Icarus, 247, 218

\bibitem[{Marchi {et~al.}(2014)Marchi, Bottke, Elkins-Tanton, Bierhaus,
  Wuennemann, Morbidelli, \& Kring}]{Marchi_2014}
Marchi, S., Bottke, W., Elkins-Tanton, L., {et~al.} 2014, Nature, 511, 578

\bibitem[{Marten {et~al.}(2002)Marten, Hidayat, Biraud, \&
  Moreno}]{Marten_2002}
Marten, A., Hidayat, T., Biraud, Y., \& Moreno, R. 2002, Icarus, 158, 532

\bibitem[{Massie \& Hunten(1981)}]{Massie_1981}
Massie, S., \& Hunten, D. 1981, Journal of Geophysical Research: Oceans, 86,
  9859

\bibitem[{Meadows \& Crisp(1996)}]{Meadows_1996}
Meadows, V.~S., \& Crisp, D. 1996, Journal of Geophysical Research: Planets,
  101, 4595

\bibitem[{Miyakawa {et~al.}(2002)Miyakawa, James~Cleaves, \&
  Miller}]{Miyakawa_2002}
Miyakawa, S., James~Cleaves, H., \& Miller, S.~L. 2002, Origins of Life and
  Evolution of the Biosphere, 32, 195

\bibitem[{Miyazaki \& Korenaga(2022)}]{Miyazaki_2022}
Miyazaki, Y., \& Korenaga, J. 2022, Nature, 603, 86

\bibitem[{Molli{\`e}re {et~al.}(2019)Molli{\`e}re, Wardenier, Van~Boekel,
  Henning, Molaverdikhani, \& Snellen}]{Molliere_2019}
Molli{\`e}re, P., Wardenier, J., Van~Boekel, R., {et~al.} 2019, Astronomy \&
  Astrophysics, 627, A67

\bibitem[{Muller {et~al.}(2022)Muller, Escobar, Xu, Wegrzyn, Nainyte, Amatov,
  Chan, Pichler, \& Carell}]{Muller_2022}
Muller, F., Escobar, L., Xu, F., {et~al.} 2022, Nature, 605, 279

\bibitem[{Murphy(1977)}]{Murphy_1977}
Murphy, W.~F. 1977, The Journal of Chemical Physics, 67, 5877

\bibitem[{Neish {et~al.}(2010)Neish, Somogyi, \& Smith}]{Neish_2010}
Neish, C.~D., Somogyi, A., \& Smith, M.~A. 2010, Astrobiology, 10, 337

\bibitem[{Nicklas(2019)}]{Nicklas_2019}
Nicklas, R.~W. 2019, PhD thesis, University of Maryland, College Park

\bibitem[{Okamura {et~al.}(2019)Okamura, Crisp, H{\"u}bner, Becker, Rov{\'o},
  \& Carell}]{Okamura_2019}
Okamura, H., Crisp, A., H{\"u}bner, S., {et~al.} 2019, Angewandte Chemie
  International Edition, 58, 18691

\bibitem[{Patel {et~al.}(2015)Patel, Percivalle, Ritson, Duffy, \&
  Sutherland}]{Patel_2015}
Patel, B.~H., Percivalle, C., Ritson, D.~J., Duffy, C.~D., \& Sutherland, J.~D.
  2015, Nature Chemistry, 7, 301

\bibitem[{Pearce {et~al.}(2022)Pearce, Molaverdikhani, Pudritz, Henning, \&
  Cerrillo}]{Pearce_2022}
Pearce, B.~K., Molaverdikhani, K., Pudritz, R.~E., Henning, T., \& Cerrillo,
  K.~E. 2022, The Astrophysical Journal, 932, 9

\bibitem[{Penndorf(1957)}]{Penndorf_1957}
Penndorf, R. 1957, Journal of the Optical Society of America, 47, 176

\bibitem[{Piani {et~al.}(2020)Piani, Marrocchi, Rigaudier, Vacher, Thomassin,
  \& Marty}]{Piani_2020}
Piani, L., Marrocchi, Y., Rigaudier, T., {et~al.} 2020, Science, 369, 1110

\bibitem[{Poch {et~al.}(2012)Poch, Coll, Buch, Ram{\'\i}rez, \&
  Raulin}]{Poch_2012}
Poch, O., Coll, P., Buch, A., Ram{\'\i}rez, S., \& Raulin, F. 2012, Planetary
  and Space Science, 61, 114

\bibitem[{Powner {et~al.}(2009)Powner, Gerland, \& Sutherland}]{Powner_2009}
Powner, M.~W., Gerland, B., \& Sutherland, J.~D. 2009, Nature, 459, 239

\bibitem[{Ranjan \& Sasselov(2017)}]{Ranjan_2017}
Ranjan, S., \& Sasselov, D.~D. 2017, Astrobiology, 17, 169

\bibitem[{Rimmer \& Rugheimer(2019)}]{Rimmer_2019}
Rimmer, P.~B., \& Rugheimer, S. 2019, Icarus, 329, 124

\bibitem[{Rimmer \& Shorttle(2019)}]{Rimmer_2019b}
Rimmer, P.~B., \& Shorttle, O. 2019, Life, 9, 12

\bibitem[{Ritson {et~al.}(2022)Ritson, Poplawski, Bond, \&
  Sutherland}]{Ritson_2022}
Ritson, D.~J., Poplawski, M.~W., Bond, A.~D., \& Sutherland, J.~D. 2022,
  Journal of the American Chemical Society, 144, 19447

\bibitem[{Rothman {et~al.}(2010)Rothman, Gordon, Barber, Dothe, Gamache,
  Goldman, Perevalov, Tashkun, \& Tennyson}]{Rothman_2010}
Rothman, L.~S., Gordon, I., Barber, R., {et~al.} 2010, Journal of Quantitative
  Spectroscopy and Radiative Transfer, 111, 2139

\bibitem[{Sagan \& Chyba(1997)}]{Sagan_1997}
Sagan, C., \& Chyba, C. 1997, Science, 276, 1217

\bibitem[{Sanchez {et~al.}(1967)Sanchez, Ferbis, \& Orgel}]{Sanchez_1967}
Sanchez, R.~A., Ferbis, J.~P., \& Orgel, L.~E. 1967, Journal of Molecular
  Biology, 30, 223

\bibitem[{Sasselov {et~al.}(2020)Sasselov, Grotzinger, \&
  Sutherland}]{Sasselov_2020}
Sasselov, D.~D., Grotzinger, J.~P., \& Sutherland, J.~D. 2020, Science
  Advances, 6, eaax3419

\bibitem[{Schmider {et~al.}(2021)Schmider, Maier, \&
  Deutschmann}]{Schmider_2021}
Schmider, D., Maier, L., \& Deutschmann, O. 2021, Industrial \& Engineering
  Chemistry Research, 60, 5792

\bibitem[{Shemansky(1972)}]{Shemansky_1972}
Shemansky, D. 1972, The Journal of Chemical Physics, 56, 1582

\bibitem[{Sleep {et~al.}(1989)Sleep, Zahnle, Kasting, \& Morowitz}]{Sleep_1989}
Sleep, N.~H., Zahnle, K.~J., Kasting, J.~F., \& Morowitz, H.~J. 1989, Nature,
  342, 139

\bibitem[{Smith \& Posey(1957)}]{Smith_1957}
Smith, H.~A., \& Posey, J.~C. 1957, Journal of the American Chemical Society,
  79, 1310

\bibitem[{Stribling \& Miller(1987)}]{Stribling_1987}
Stribling, R., \& Miller, S.~L. 1987, Origins of Life and Evolution of the
  Biosphere, 17, 261

\bibitem[{Strobel {et~al.}(2009)Strobel, Atreya, B{\'e}zard, Ferri, Flasar,
  Fulchignoni, Lellouch, \& M{\"u}ller-Wodarg}]{Strobel_2009}
Strobel, D.~F., Atreya, S.~K., B{\'e}zard, B., {et~al.} 2009, Titan from
  Cassini-Huygens, 235

\bibitem[{Sutherland(2016)}]{Sutherland_2016}
Sutherland, J.~D. 2016, Angewandte Chemie International Edition, 55, 104

\bibitem[{Takahashi(1986)}]{Takahashi_1986}
Takahashi, E. 1986, Journal of Geophysical Research: Solid Earth, 91, 9367

\bibitem[{Thompson {et~al.}(2022)Thompson, Krissansen-Totton, Wogan, Telus, \&
  Fortney}]{Thompson_2022}
Thompson, M.~A., Krissansen-Totton, J., Wogan, N., Telus, M., \& Fortney, J.~J.
  2022, Proceedings of the National Academy of Sciences, 119, e2117933119

\bibitem[{Tian {et~al.}(2011)Tian, Kasting, \& Zahnle}]{Tian_2011}
Tian, F., Kasting, J., \& Zahnle, K. 2011, Earth and Planetary Science Letters,
  308, 417

\bibitem[{Todd {et~al.}(2022)Todd, Lozano, Kufner, Sasselov, \&
  Catling}]{Todd_2022}
Todd, Z.~R., Lozano, G.~G., Kufner, C.~L., Sasselov, D.~D., \& Catling, D.~C.
  2022, Geochimica et Cosmochimica Acta, 335, 1

\bibitem[{Todd \& {\"O}berg(2020)}]{Todd_2020}
Todd, Z.~R., \& {\"O}berg, K.~I. 2020, Astrobiology, 20, 1109

\bibitem[{Toner \& Catling(2019)}]{Toner_2019}
Toner, J.~D., \& Catling, D.~C. 2019, Geochimica et Cosmochimica Acta, 260, 124

\bibitem[{Toon {et~al.}(1989)Toon, McKay, Ackerman, \& Santhanam}]{Toon_1989}
Toon, O.~B., McKay, C., Ackerman, T., \& Santhanam, K. 1989, Journal of
  Geophysical Research: Atmospheres, 94, 16287

\bibitem[{Trainer {et~al.}(2006)Trainer, Pavlov, DeWitt, Jimenez, McKay, Toon,
  \& Tolbert}]{Trainer_2006}
Trainer, M.~G., Pavlov, A.~A., DeWitt, H.~L., {et~al.} 2006, Proceedings of the
  National Academy of Sciences, 103, 18035

\bibitem[{Urey(1952)}]{Urey_1952}
Urey, H.~C. 1952, Proceedings of the National Academy of Sciences, 38, 351

\bibitem[{Vardavas \& Carver(1984)}]{Vardavas_1984}
Vardavas, I., \& Carver, J.~H. 1984, Planetary and Space Science, 32, 1307

\bibitem[{Vuitton {et~al.}(2006)Vuitton, Yelle, \& Anicich}]{Vuitton_2006}
Vuitton, V., Yelle, R., \& Anicich, V. 2006, The Astrophysical Journal, 647,
  L175

\bibitem[{White(1976)}]{White_1976}
White, H.~B. 1976, Journal of molecular evolution, 7, 101

\bibitem[{Wogan {et~al.}(2022)Wogan, Catling, Zahnle, \& Claire}]{Wogan_2022}
Wogan, N., Catling, D., Zahnle, K., \& Claire, M. 2022, Proceedings of the
  National Academy of Sciences

\bibitem[{Wogan {et~al.}(2020)Wogan, Krissansen-Totton, \&
  Catling}]{Wogan_2020}
Wogan, N., Krissansen-Totton, J., \& Catling, D.~C. 2020, The Planetary Science
  Journal, 1, 58

\bibitem[{Wolf \& Toon(2010)}]{Wolf_2010}
Wolf, E., \& Toon, O. 2010, Science, 328, 1266

\bibitem[{Wolf {et~al.}(2022)Wolf, Kopparapu, Haqq-Misra, \&
  Fauchez}]{Wolf_2022}
Wolf, E.~T., Kopparapu, R., Haqq-Misra, J., \& Fauchez, T.~J. 2022, The
  Planetary Science Journal, 3, 7

\bibitem[{Wordsworth {et~al.}(2017)Wordsworth, Kalugina, Lokshtanov, Vigasin,
  Ehlmann, Head, Sanders, \& Wang}]{Wordsworth_2017}
Wordsworth, R., Kalugina, Y., Lokshtanov, S., {et~al.} 2017, Geophysical
  Research Letters, 44, 665

\bibitem[{Yadav {et~al.}(2020)Yadav, Kumar, \& Krishnamurthy}]{Yadav_2020}
Yadav, M., Kumar, R., \& Krishnamurthy, R. 2020, Chemical reviews, 120, 4766

\bibitem[{Zahnle {et~al.}(2016)Zahnle, Marley, Morley, \& Moses}]{Zahnle_2016}
Zahnle, K., Marley, M.~S., Morley, C.~V., \& Moses, J.~I. 2016, The
  Astrophysical Journal, 824, 137

\bibitem[{Zahnle(1986)}]{Zahnle_1986}
Zahnle, K.~J. 1986, Journal of Geophysical Research: Atmospheres, 91, 2819

\bibitem[{Zahnle {et~al.}(2019)Zahnle, Gacesa, \& Catling}]{Zahnle_2019}
Zahnle, K.~J., Gacesa, M., \& Catling, D.~C. 2019, Geochimica et Cosmochimica
  Acta, 244, 56

\bibitem[{Zahnle {et~al.}(2020)Zahnle, Lupu, Catling, \& Wogan}]{Zahnle_2020}
Zahnle, K.~J., Lupu, R., Catling, D.~C., \& Wogan, N. 2020, The Planetary
  Science Journal, 1, 11

\bibitem[{Zahnle \& Marley(2014)}]{Zahnle_2014}
Zahnle, K.~J., \& Marley, M.~S. 2014, The Astrophysical Journal, 797, 41

\end{thebibliography}
\bibliographystyle{aasjournal}

\end{document}